\documentclass{raa}

\usepackage{graphicx,times}
\usepackage{natbib}
\usepackage{amssymb,amsmath}
\usepackage{array}
\usepackage{url}
\usepackage{xcolor}
\bibpunct{(}{)}{;}{a}{}{,}

\newcommand{\raatablestyle}{\footnotesize\setlength{\extrarowheight}{1.5pt}\renewcommand{\arraystretch}{1.15}}
\makeatletter
\def\volnopage#1{}
\makeatother

\begin{document}

\title{Shigatse Astronomical Site Testing. I. Cloud-cover Climatology and Selected Local Meteorological Conditions}

\volnopage{}
\setcounter{page}{1}


\author{Baiyu Zhang
   \inst{1}
\and Hejun Yang
   \inst{2,3}
\and Xiaojun Dong
   \inst{1}
\and Lingling Wang
   \inst{1}
\and Juean Luobu
   \inst{1}
\and Minfeng Gu
   \inst{1}
\and Xiyan Peng
   \inst{1}
\and Hao Luo
   \inst{1}
\and Yindun Mao
   \inst{1}
\and ZhaoXiang Qi
   \inst{1,4}
\and Basangzeren
   \inst{5}
\and Qihang He
   \inst{6}
\and Guojie Feng
   \inst{7}
\and Chunhai Bai
   \inst{7}
\and Ali Esamdin
   \inst{7}
\and Wenbo Gu
   \inst{7,4}
\and Siqi Wang
   \inst{7,4}
\and Zihuang Cao
   \inst{8,4,*}\footnotetext{$*$Corresponding author: Zihuang Cao, zhcao@nao.cas.cn}
}

\institute{Shanghai Astronomical Observatory, Chinese Academy of Sciences, Shanghai 200030, China\\
\and
College of Computer Science, Beijing University of Technology, Beijing, China, 100124\\
\and
Beijing Artificial Intelligence Institute, Beijing University of Technology, Beijing, China, 100124\\
\and
University of Chinese Academy of Sciences, Beijing 100049, China\\
\and
Xizang Autonomous Region Meteorological Information Network Center, Lhasa 850000, China\\
\and
Xizang University, Lhasa 850000, China\\
\and
Xinjiang Astronomical Observatory, Chinese Academy of Sciences, Urumqi 830011, China\\
\and
National Astronomical Observatories, Chinese Academy of Sciences, Beijing 100101, China\\
\vs\no
{\small Author-created preprint submitted to Research in Astronomy and Astrophysics.}}

\abstract{As the first paper in a Shigatse astronomical site-testing series, we present a multi-source assessment of cloud cover and selected local meteorological conditions at the Shigatse 40 m site on the southern Tibetan Plateau. The study combines CALIPSO-GOCCP active-lidar climatology, ISCCP HXG passive-satellite cloud fields, conventional total-cloud-amount observations from the Shigatse Meteorological Station, and on-site Weather Station measurements. Together, these records characterize Shigatse as a southern-plateau monsoon-transition cloud regime: the active-lidar climatology gives a moderate-to-low annual cloud fraction, and the cloudier months are concentrated in the June--September monsoon interval. In GOCCP, the annual mean cloud fraction is 42.1\%, while the October--May low-cloud season has a mean cloud fraction of 26.3\%, compared with 73.7\% during the June--September monsoon interval. ISCCP gives higher absolute cloud fractions but supports the same seasonal phase and local spatial placement. The aligned 1988--2013 meteorological-station record gives a total-cloud-amount $\leq 40\%$ fraction of 80.7\% during October--May, rising to 90.7\% in the November--January core, and decreasing to 39.9\% during June--September. The 2024--2025 Weather Station archive further shows high fractions of valid samples satisfying the adopted meteorological criteria during the low-cloud months: 92.6\% for the October--May night-time proxy and 94.6\% for the corresponding 24 h samples. These results identify Shigatse as a measured lower-latitude southern-plateau cloud-cover reference within China's site-testing network, with a well-defined October--May low-cloud observing period and a Shigatse--Ali low-cloud corridor for subsequent regional site testing.
}

\authorrunning{B. Zhang et al.}
\titlerunning{Shigatse site testing. I. Cloud climatology}
\date{2026 May 30}

\maketitle

\section{Introduction}\label{introduction}

The performance of a ground-based astronomical facility depends on the atmosphere above the site as well as on the telescope and instrumentation. For optical and near-infrared observations, cloud cover determines the available observing time, while seeing, precipitable water vapor (PWV), sky brightness, and local meteorology determine the quality and operational use of that time. Large site-selection programs therefore treat site selection, site testing, and site assessment as staged tasks. The Thirty Meter Telescope (TMT) campaign used co-located meteorological sensors, seeing monitors, water-vapor measurements, sky-brightness monitoring, and high-cadence cloud monitoring to compare candidate sites \citep{Schoeck2009TMTOverview,Skidmore2008TMTASC}. Similar multi-parameter approaches have been used in European Extremely Large Telescope site-assessment work and in statistical comparisons of established observatory climates \citep{Varela2014ELTGroundMeteorology,Hellemeier2019WeatherSites}. In this framework, cloud-cover climatology provides the regional basis for estimating available observing time before site-specific instruments quantify seeing, PWV, sky background, and other facility-dependent parameters. Satellite cloud products are therefore useful for regional site assessment and for placing local monitoring results in a wider geographic context \citep{Carrasco2017SatelliteSurvey,Chepfer2010GOCCP,Stubenrauch2013GlobalCloudDatasets}.

Western China has become a major region for astronomical site testing over the past two decades. The Ali region of Xizang was identified as a high-altitude and dry candidate area, and subsequent work added radiosonde, turbulence, and cloud-cover constraints \citep{Ye2016AliAtacama,Qian2018AliRadiosonde,Hickson2020AliTurbulence,Qian2024AliCloud}. Lenghu on Saishiteng Mountain has become the most extensively documented domestic optical-site benchmark, with published evidence of a high fraction of clear nights, good seeing, low PWV, and suitable meteorological conditions \citep{Deng2021Lenghu,Li2024LenghuCloud}. The western-China Large Optical/infrared Telescope site-testing campaign further established a practical framework for homogeneous instrumentation, data processing, and quality control in high-plateau candidate-site studies \citep{Feng2020LOTOverview,Cao2020DataProducts}. In this context, the long-term satellite analysis by \citet{Cao2020ClearNights} is a direct methodological predecessor of the present study. It showed that multi-year satellite cloud data can identify regional clear-night structures before expensive local campaigns are narrowed to specific ridges or peaks.

The Shigatse region occupies a different part of this site-testing parameter space. A site near \(29.2^\circ\) N provides lower-latitude sky access than sites near \(38^\circ\)--\(39^\circ\) N, including higher elevations for southern targets and fields toward the Galactic bulge and southern time-domain sky. Geographically, Shigatse belongs to the southern Tibetan Plateau, near the transition between Indian monsoon influence and the drier interior plateau. Tibetan Plateau cloud studies show strong summer-monsoon enhancement, dry-season reduction, terrain-controlled gradients, diurnal variability, and product-dependent satellite retrieval uncertainty over complex terrain and high-albedo surfaces \citep{DuanWu2006TPCloudClimate,Zhang2007TPCloud,Liu2021TPSatelliteCloud,Shang2018TPDiurnalCloud,Liu2019TPCirrusGOCCP,Zhao2023TPCloudDiurnal,Wu2024TPCloudReview}. A previous total-sky-image study in the Shigatse area also reported an autumn--winter low-cloud regime and a summer maximum \citep{Yang2018ShigatseSkyImages}. These results motivate an independent cloud-cover reference for the southern Tibetan Plateau.

The existing Shigatse 40 m radio site provides the physical reference point for this first assessment. Access, power, communications, and routine local operations support sustained environmental monitoring and repeat field campaigns. The present work addresses two linked cloud-climatology questions. The first is what the annual and monthly cloud-cover regime is measured at the 40 m site and in its surrounding southern-plateau region. The second is whether the seasonal organization of low-cloud months identifies months prioritized for continued local monitoring and provides a basis for later regional site testing toward western Xizang. The cloud analysis and on-site Weather Station statistics therefore provide the first atmospheric-availability assessment for Shigatse-based site testing. Subsequent optical and near-infrared site characterization will require high-cadence cloud monitoring, direct seeing and PWV measurements, calibrated sky-brightness monitoring, and aerosol measurements. For radio and multi-wavelength development, radio-frequency interference, wind loading, humidity, and long-term operational stability also remain essential site parameters \citep{Umar2014RFISiteSelection,Li2020ReanalysisRadioSiteTesting,Wang2023QitaiRadioTelescope}.

The analysis combines data sets with different spatial scales, temporal sampling, and retrieval physics, allowing regional cloud climatology, product-dependent differences, and local fixed-time cloud observations to be examined separately. CALIPSO-GOCCP is adopted as the primary active-lidar cloud product because CALIOP provides vertically resolved cloud detection, including optically thin clouds, and its lidar retrieval is less dependent on surface reflectance than passive visible-infrared cloud retrievals over bright highland terrain \citep{Chepfer2010GOCCP}. ISCCP HXG provides a finer horizontal grid and a long climate-data context, while retaining the known limitations of passive visible-infrared retrievals over the Tibetan Plateau, including low-cloud, multilayer-cloud, and high-albedo retrieval issues \citep{Rossow1999ISCCP,Young2018ISCCPHSeries,NaudChen2010ISCCPTP,Liu2021TPSatelliteCloud}. Conventional total-cloud-amount observations from the Shigatse Meteorological Station provide an independent, long-term local check of the monthly phase of cloud amount; their four fixed observing times constrain monthly recurrence and low-order diurnal behavior. On-site Weather Station measurements characterize local wind, humidity, dew point, and precipitation conditions during the low-cloud months. A VIIRS/DNB-derived artificial sky-brightness model sequence is used later as an ancillary long-term indicator of artificial-light evolution.

This paper is the first contribution in a Shigatse astronomical site-testing series and focuses on the annual-to-monthly cloud-cover climatology and selected local meteorological conditions. Using these data sets, it quantifies the annual and monthly cloud-cover regime at the Shigatse 40 m site, tests the seasonal phase with long-term ground-based cloud observations, and examines whether the satellite-defined low-cloud months are supported by local meteorological conditions. The goal is to identify Shigatse as a measured southern-plateau cloud-cover reference and to define the low-cloud observing period and westward low-cloud corridor for subsequent site testing. A follow-up paper in the same series, led by Qihang He at Xizang University, will use all-sky-camera data to resolve continuous 24 h cloud evolution, night-time cloud duration, and cloud-gap statistics.

The manuscript is organized as follows. Section~\ref{site-and-data-sets} describes the Shigatse site and the data sets. Sections~\ref{goccp-cloud-climatology} and \ref{isccp-cloud-climatology} present the GOCCP and ISCCP cloud-cover climatologies in parallel. Section~\ref{ground-based-cloud-amount-analysis-at-shigatse} analyzes the long-term ground-based cloud amount at Shigatse. Section~\ref{consistency-among-cloud-data-sets-and-observing-conditions} examines consistency among the cloud data sets and connects the low-cloud season with local meteorological conditions. Section~\ref{discussion} discusses the selected Chinese site-testing context, the Shigatse--Ali low-cloud corridor, the artificial sky-brightness trend, multi-decadal stability in cloud amount, and the scale translation from satellite cloud climatology to site-level observing statistics. Section~\ref{conclusions} summarizes the main conclusions and their implications for subsequent Shigatse-based site testing.

\section{Site and Data Sets}\label{site-and-data-sets}

\subsection{Shigatse 40 m Site and Regional Setting}\label{shigatse-site-and-regional-setting}

The Shigatse 40 m radio site analyzed here is located on the southern Tibetan Plateau, approximately 30 km west of Shigatse city, at \(88.6322^\circ\) E, \(29.2056^\circ\) N, and an elevation of approximately 4060 m. The site lies in the southern-plateau transition zone between Indian-monsoon-influenced terrain and the drier interior plateau. For optical and near-infrared site testing, cloud cover first sets the available observing time; seeing, sky brightness, precipitable water vapor, and local meteorology then determine its scientific and operational value \citep{Schoeck2009TMTOverview,Skidmore2008TMTASC,Varela2014ELTGroundMeteorology}. Cloud-cover climatology is therefore adopted as the first environmental diagnostic for Shigatse, with emphasis on the annual cloud level, monthly cycle, and candidate low-cloud periods at the existing 40 m site and in the surrounding southern-plateau region.

The timing of the low-cloud months is as important as the annual mean cloud amount. A recurring non-monsoon low-cloud season defines calendar periods for monitoring, maintenance, instrument work, and field campaigns during cloudier months. Previous Tibetan Plateau and Shigatse studies show summer-monsoon cloud enhancement, dry-season reduction, diurnal variability, and terrain-controlled gradients \citep{DuanWu2006TPCloudClimate,Zhang2007TPCloud,Shang2018TPDiurnalCloud,Wu2024TPCloudReview,Yang2018ShigatseSkyImages}. With the existing 40 m facility as the monitoring base, the present analysis evaluates whether this seasonal organization is sufficiently coherent to guide continued Shigatse site testing and westward exploration toward the northern Himalayas within the staged site-testing framework described above.

\begin{figure}[htbp]
\centering
\includegraphics[width=0.95\linewidth,keepaspectratio]{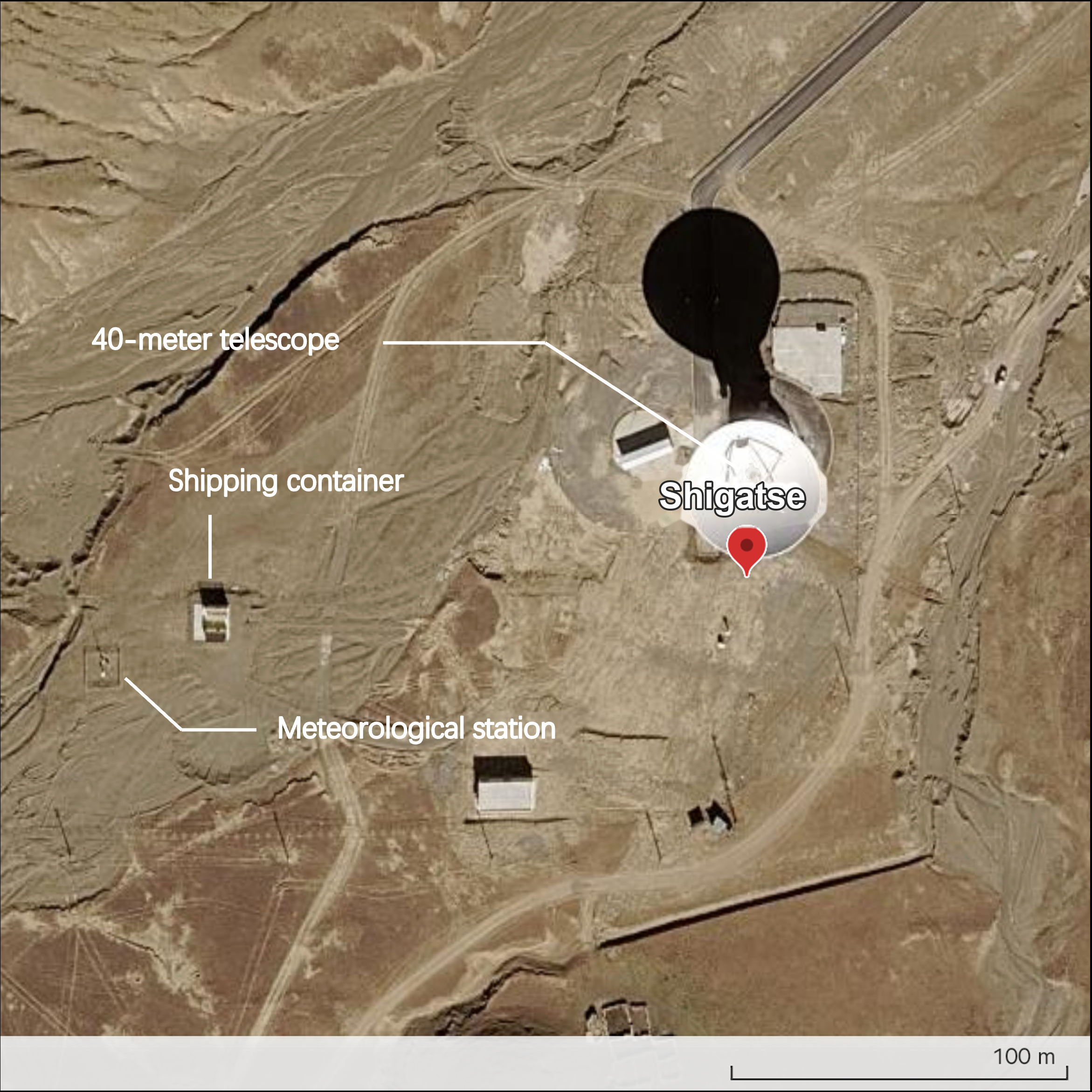}
\caption{Satellite view of the Shigatse 40 m radio site. The 40 m telescope, on-site Weather Station, and shipping container used for instruments are identified; the scale bar shows 100 m. The figure shows the local relationship among the telescope, the point meteorological measurements, and the nearby instrument container. Base image: Google Maps satellite imagery.}
\label{fig:2-1}
\end{figure}

\subsection{Data Sets}\label{data-sets}

The four data sources provide complementary observational constraints for the Shigatse cloud-cover assessment. CALIPSO-GOCCP provides the regional active-lidar cloud-cover climatology, ISCCP HXG gives finer passive-satellite spatial context, and conventional total-cloud-amount observations from the Shigatse Meteorological Station supply a multi-decade fixed-time ground cloud record. After the low-cloud months are identified, the on-site Weather Station data evaluate wind speed, humidity, dew-point depression, and precipitation at the 40 m site under the adopted meteorological criteria. A VIIRS/DNB-derived artificial sky-brightness sequence is retained as an ancillary indicator for dark-sky preservation. This separation of cloud cover, local meteorology, and sky-background information follows major site-testing work, where cloud cover, local meteorology, seeing, water vapor, sky background, and operational constraints are measured as distinct observational quantities \citep{Schoeck2009TMTOverview,Skidmore2008TMTASC,Varela2014ELTGroundMeteorology,Deng2021Lenghu,Li2024LenghuCloud,Qian2024AliCloud}. Table~\ref{tab:2-1} summarizes the data sets used in this paper.

\begin{table}[htbp]
\centering
\caption{Data Sets in the Shigatse Cloud-cover and Selected Local Meteorological Condition Assessment.}
\label{tab:2-1}
\begingroup
\raatablestyle
\setlength{\tabcolsep}{1.5pt}
\renewcommand{\arraystretch}{1.18}
\begin{tabular}{@{}>{\raggedright\arraybackslash}p{0.135\linewidth} >{\raggedright\arraybackslash}p{0.145\linewidth} >{\raggedright\arraybackslash}p{0.175\linewidth} >{\raggedright\arraybackslash}p{0.145\linewidth} >{\raggedright\arraybackslash}p{0.22\linewidth} >{\raggedright\arraybackslash}p{0.14\linewidth}@{}}
\hline
Data set & Spatial reference & Period and cadence & Primary variables & Function in the analysis & Limitation \\
\hline
CALIPSO-GOCCP & $2^\circ \times 2^\circ$ satellite grid & 2007--2016 monthly climatology & Cloud fraction and clear-sky fraction & Primary long-baseline satellite cloud-cover climatology & Coarse grid; lidar sampling is sparse in time \\
ISCCP HXG & $0.1^\circ \times 0.1^\circ$ satellite grid & 2007--2016 corrected daily product aggregated to monthly mean fields & Cloud fraction and clear-sky fraction & Finer-grid passive satellite check on spatial placement and monthly phase & Passive retrievals may be biased over snow or bright highland surfaces \\
Shigatse Meteorological Station cloud data & Met. station no. 55578; Shigatse city & 1988--2019, conventional observing times retained as 02, 08, 14, and 20 & Total cloud amount (\%, from the 0--10 tenths scale) & Multi-decade ground-based cloud-cover evidence for monthly recurrence and low-order fixed-time diurnal indications & Station is lower and about 30 km east of the Shigatse 40 m radio site; no direct cross-site comparison \\
On-site Weather Station & Shigatse site & 2024--2025, nominal 10-minute engineering data summarized to astronomical-use intervals & Wind, gust, wind direction, temperature, humidity, dew point, pressure, precipitation & Operational meteorological constraints & Short data span; surface meteorology only \\
\hline
\end{tabular}
\endgroup
\end{table}

\subsection{Satellite Cloud Products and Sampling}\label{satellite-cloud-products}

The satellite analysis uses two cloud products with different observing principles and spatial scales. CALIPSO-GOCCP is the primary long-baseline active-lidar cloud-cover climatology, while ISCCP HXG is a finer-grid passive satellite product for local spatial context and monthly consistency checks. Satellite cloud intercomparison studies show that products differ in grid size, illumination dependence, cloud-detection threshold, cloud-top retrieval method, and sensitivity to the lower atmosphere and surface background \citep{Stubenrauch2013GlobalCloudDatasets,Young2018ISCCPHSeries}. For a high-elevation site on the southern Tibetan Plateau, GOCCP provides the active-lidar climatological reference, and ISCCP is used to examine finer-scale spatial placement. Product completeness is evaluated at the file, nearest-grid-cell, and local-box levels before the fields are interpreted.

New-generation geostationary imagers, including FY-4/AGRI, provide high-cadence regional cloud monitoring over China and are well suited to cloud-continuity, diurnal-cycle, and event-scale studies \citep{NSMC2026AGRI,WMO2026AGRI}. The present analysis has a different requirement: it seeks a traceable annual-to-monthly cloud-cover climatology based on decade-scale records that can be compared across Shigatse, Ali, Lenghu, Xinglong, and the Shigatse--Ali low-cloud corridor. Baseline length is therefore a primary data-selection constraint. Shorter recent geostationary records are valuable for process and monitoring studies, but they are more sensitive to interannual variability, recent climate anomalies, and possible trends when used alone for climatological site comparison. For this reason, GOCCP and ISCCP are adopted as the primary satellite products in the present work: GOCCP provides documented multi-year active-lidar cloud climatology with vertically resolved detection, including optically thin layers, while ISCCP provides a long passive-satellite climate-data framework \citep{Chepfer2010GOCCP,Young2018ISCCPHSeries,Cao2020ClearNights}. FY-4/AGRI and comparable geostationary products are reserved for follow-up high-cadence studies of cloud continuity, diurnal behavior, and event-scale cloud evolution, including planned work with Xinjiang Astronomical Observatory.

For satellite cloud products, data completeness and temporal representativeness are distinct concepts. A monthly product may be complete in terms of file availability and valid grid cells, while its temporal support is set by the sampling cycle of the observing system. CALIPSO-GOCCP is constrained by the periodic local sampling of a nadir-viewing lidar orbit, whereas ISCCP HXG is constrained by the synoptic sampling and multi-satellite merging strategy of a passive climate data product \citep{Winker2010CALIPSO,Young2018ISCCPHSeries}. Accordingly, Table~\ref{tab:2-2} distinguishes between product/grid completeness and orbit- or sampling-cycle completeness. The former can be audited directly from the NetCDF files used here; the latter is characterized from the satellite design and product metadata, because exact daily or orbit-level counts over Shigatse are not stored in the monthly files.

\begin{table}[htbp]
\centering
\caption{Satellite Sampling Cycle and Completeness at Shigatse.}
\label{tab:2-2}
\begingroup
\raatablestyle
\setlength{\tabcolsep}{1.5pt}
\renewcommand{\arraystretch}{1.18}
\begin{tabular}{@{}>{\centering\arraybackslash}p{0.14\linewidth} >{\raggedright\arraybackslash}p{0.45\linewidth} >{\raggedright\arraybackslash}p{0.38\linewidth}@{}}
\hline
\multicolumn{1}{>{\centering\arraybackslash}p{0.14\linewidth}}{Product} &
\multicolumn{1}{>{\centering\arraybackslash}p{0.45\linewidth}}{Nominal sampling cycle near Shigatse} &
\multicolumn{1}{>{\centering\arraybackslash}p{0.38\linewidth}}{Completeness supported by current files} \\
\hline
CALIPSO-GOCCP & Nadir active lidar; approximately 99 min orbital period and a 16-day repeating ground-track pattern. A calendar-month climatology over 2007--2016 spans approximately 17.7--19.4 nominal 16-day repeat cycles, depending on month length. & All 12 monthly climatology files are present. The nearest Shigatse grid cell and the local box are valid in all months. Daily, orbit-level, and per-grid-cell sample counts are not stored. \\
ISCCP HXG & Passive multi-satellite product; not governed by a single repeat track over Shigatse. Native H-Series products are organized at synoptic intervals, while the local files used here are monthly means derived from corrected daily data. A 2007--2016 calendar-month data set contains 283--310 daily opportunities. & All 12 monthly products are present. The nearest Shigatse grid cell and the local box are valid in all months. Metadata give 9--10 valid years and at least 15 daily samples per valid year, but exact daily or synoptic counts are not stored. \\
\hline
\end{tabular}
\endgroup
\end{table}

Satellite data quality is evaluated using traceable file, grid-cell, and local-box completeness metrics. For both GOCCP and ISCCP, all 12 monthly products are present, the nearest Shigatse grid cell is valid in every month, and the local Shigatse box has valid grid cells throughout the annual cycle. The temporal sample counts behind the monthly fields are documented differently in the two products: GOCCP monthly files do not store the exact local orbit or profile counts, and the ISCCP monthly files retain valid-year and minimum-day metadata rather than exact daily or synoptic sample counts at the Shigatse grid cell. GOCCP supports monthly regional cloud-cover climatology, and ISCCP supports monthly spatial and month-group consistency checks. These products are therefore used to test annual cloud level, monthly phase, and spatial placement, while the meteorological-station cloud data test fixed-time local behavior over multiple decades.

\subsubsection{CALIPSO-GOCCP Cloud Product}\label{calipso-goccp-active-lidar-cloud-product}

CALIPSO is a NASA--CNES satellite mission designed to provide a global three-dimensional view of clouds and aerosols with an active lidar system \citep{Winker2010CALIPSO}. It flies in a near-polar, sun-synchronous orbit as part of the A-Train constellation, and its main cloud instrument, CALIOP, is a two-wavelength polarization lidar operating at 532 and 1064 nm \citep{Winker2010CALIPSO}. For high-elevation site studies, this active-lidar geometry is important because it detects optically thin cloud layers and provides vertically resolved cloud occurrence. These properties reduce the ambiguity associated with passive visible-infrared retrievals over bright terrain, snow-covered surfaces, and regions with weak thermal contrast \citep{Chepfer2010GOCCP,Stubenrauch2013GlobalCloudDatasets}.

GOCCP is the GCM-Oriented CALIPSO Cloud Product derived from CALIOP lidar profiles \citep{Chepfer2010GOCCP}. It was developed for simulator-consistent model evaluation, and its uniform lidar-based cloud detection provides a regional cloud-cover product for astronomical site studies where long homogeneous ground records are limited. In the GOCCP processing concept, instantaneous scattering-ratio profiles are diagnosed at the original CALIOP along-track scale and then accumulated into monthly gridded diagnostics; the \(2^\circ \times 2^\circ\) grid used here is therefore a climatological aggregation of active-lidar cloud occurrence. The Shigatse analysis uses 12 multi-year monthly climatology files for 2007--2016 and extracts \texttt{cltcalipso}, the total cloud-fraction variable reported as a unitless fraction from 0 to 1. The numerical extraction uses total cloud fraction only. In this paper, GOCCP quantifies the regional cloud-cover regime, the persistence of the non-monsoon low-cloud season, and the placement of Shigatse relative to other western China astronomical regions within the same active-lidar framework \citep{Cao2020ClearNights}.

The GOCCP archive is complete at the monthly-product level. All 12 monthly climatology products are present, the nearest GOCCP grid cell to the Shigatse site is valid in every month, and the valid-grid-cell fraction is 100\% in the local box used for regional context (\(27^\circ\)--\(31^\circ\) N, \(86^\circ\)--\(91^\circ\) E). The monthly fields therefore do not require spatial interpolation or month-dependent masking around Shigatse. Because of the CALIPSO repeating ground-track cycle, each calendar-month climatology over 2007--2016 pools samples from approximately 17.7--19.4 nominal 16-day repeat cycles, depending on month length. This sampling scale supports monthly climatology and month-group interpretation; individual-night cloud continuity is left to local high-cadence sky monitoring in the subsequent site-testing program. GOCCP is consequently interpreted as a complete monthly climatological constraint on the regional cloud regime. This product choice is also consistent with Tibetan Plateau cloud studies that have used CALIPSO-GOCCP, together with AIRS or other data sets, to diagnose high-cloud and cirrus seasonality over the plateau \citep{Liu2019TPCirrusGOCCP}. Thus, GOCCP provides the active-lidar regional cloud-cover reference against which the Shigatse monthly regime is evaluated.

\subsubsection{ISCCP HXG Cloud Product}\label{isccp-hxg-passive-satellite-cloud-product}

ISCCP, the International Satellite Cloud Climatology Project, was established under the World Climate Research Programme to construct a global, long-term satellite cloud climatology from calibrated visible and infrared radiances \citep{Schiffer1983ISCCP,Rossow1999ISCCP}. The H-Series Climate Data Record uses improved calibration, higher-resolution geostationary and polar-orbiting inputs, updated ancillary atmospheric information, and NetCDF product delivery relative to earlier ISCCP products \citep{Young2018ISCCPHSeries,NOAA2022ISCCPHUserGuide}. The retrieval is passive: cloudy pixels are identified from departures of observed radiances from estimated clear-sky radiances, and cloud-top temperature, cloud-top pressure, optical thickness, and related cloud properties are then retrieved from radiative-transfer calculations \citep{Rossow1999ISCCP,Young2018ISCCPHSeries}. This passive, multi-satellite design provides ISCCP with a long temporal baseline and dense spatial coverage, both of which are useful for placing a point-like site inside a regional cloud field.

The ISCCP HXG files used here are 2007--2016 corrected multi-year monthly mean cloud-fraction fields from the ISCCP H-Series Climate Data Record \citep{Young2018ISCCPHSeries}. In the H-Series product family, HXG is the \(0.1^\circ\) equal-angle product, distinct from the coarser HGS/HGG/HGH/HGM products and from the swath-level HXS product described in the ISCCP-H documentation \citep{Young2018ISCCPHSeries,NOAA2022ISCCPHUserGuide}. The local NetCDF files contain an 1800 by 3600 latitude--longitude grid with \(0.1^\circ \times 0.1^\circ\) spacing, and the cloud-fraction variable is stored in percent. ISCCP is used to test the finer-scale spatial placement of Shigatse within the broader southern-plateau cloud pattern. At the same time, ISCCP absolute values over the Tibetan Plateau require caution because passive cloud retrievals over high terrain can be affected by snow cover, high surface albedo, strong surface temperature gradients, multilayer clouds, optically thin clouds, and weak contrast between low clouds and the underlying surface \citep{Li2006TPCloudTypes,NaudChen2010ISCCPTP,Liu2021TPSatelliteCloud,Wu2024TPCloudReview}. In a CloudSat-CALIPSO assessment over the Tibetan Plateau, \citet{NaudChen2010ISCCPTP} found that ISCCP cloud cover was underestimated in their test case and identified missed low-level nighttime clouds and cloud-top pressure biases as important limitations; this result motivates the product-specific interpretation adopted here.

The ISCCP archive is also complete at the monthly-product level. All 12 corrected monthly products are present, the nearest Shigatse grid cell is valid in every month, and the valid-grid-cell fraction is 100\% in the same local box used for GOCCP. Unlike CALIPSO, ISCCP does not provide a single repeat-track overpass sequence above Shigatse. Its native H-Series sampling is organized around synoptic time slots in a passive multi-satellite system, while the local files used here are monthly mean fields derived from corrected daily data. The current files retain metadata on the number of valid years and the minimum daily samples required for each monthly mean. In the Shigatse grid cell, the monthly means have 9--10 valid years, with a minimum threshold of 15 daily samples per valid year; this corresponds to a metadata-guaranteed lower bound of 135--150 daily grid-cell samples and a calendar upper bound of 283--310 possible daily samples over 2007--2016. As with GOCCP, the monthly products do not preserve the exact daily or synoptic sample counts entering each climatological value. ISCCP is therefore interpreted as a monthly spatial and month-group cloud-field product.

\begin{figure}[htbp]
\centering
\begin{minipage}{0.48\linewidth}
\includegraphics[width=\linewidth,keepaspectratio]{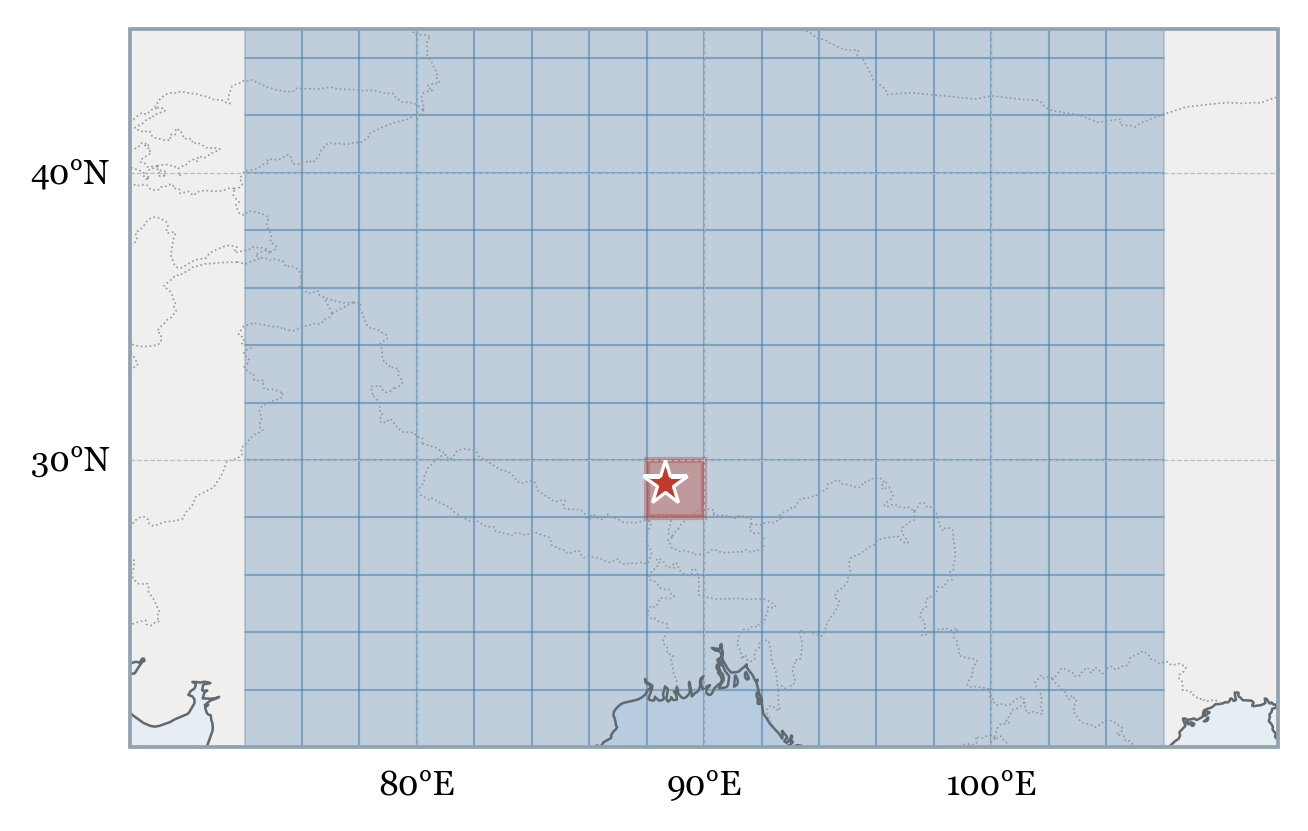}
\end{minipage}\hfill
\begin{minipage}{0.48\linewidth}
\includegraphics[width=\linewidth,keepaspectratio]{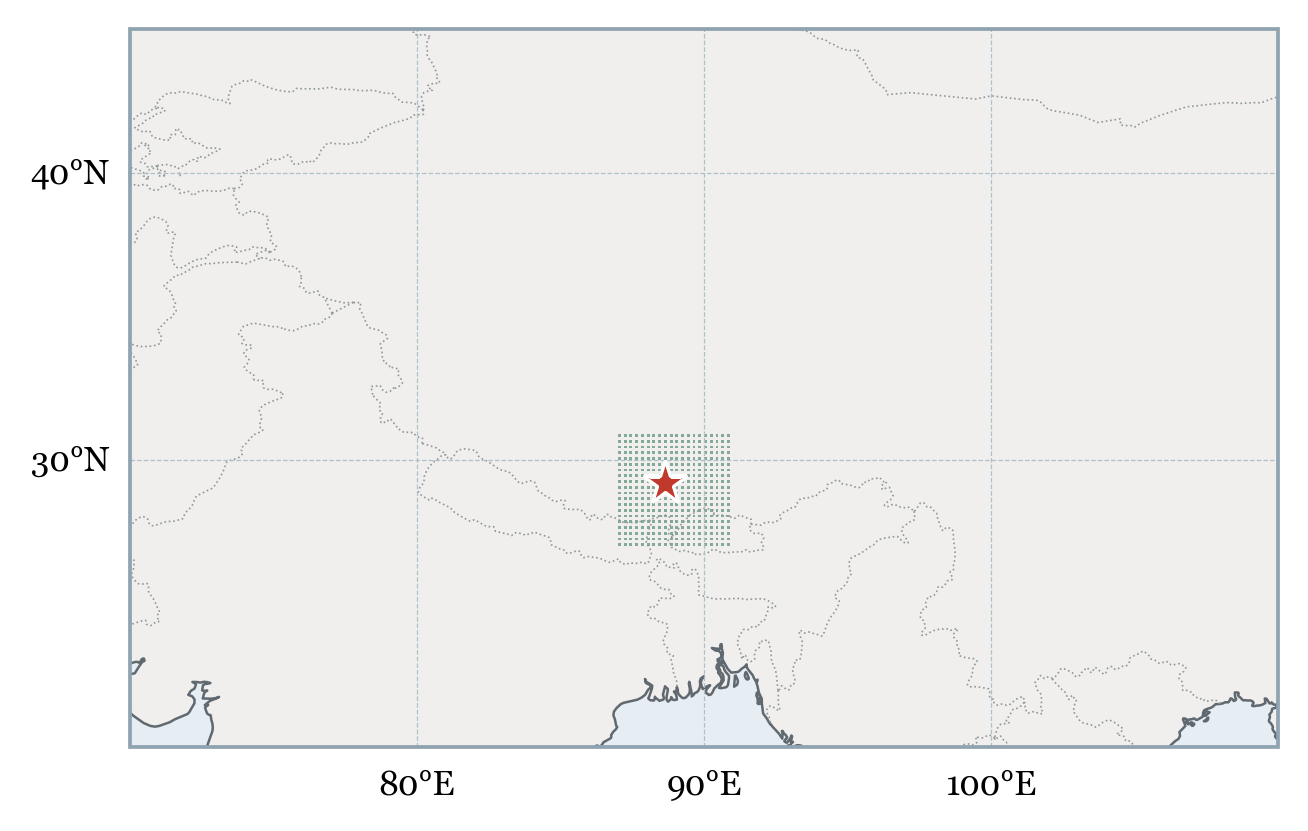}
\end{minipage}
\caption{Native grid support of the two satellite cloud products used at Shigatse. The left panel shows the GOCCP $2^\circ \times 2^\circ$ regional active-lidar cloud-climatology grid, and the right panel shows the ISCCP HXG $0.1^\circ \times 0.1^\circ$ passive-satellite grid.}
\label{fig:2-2}
\end{figure}

Consistent with this data strategy, the joint satellite interpretation emphasizes agreement in monthly timing and broad spatial structure while preserving product-specific absolute cloud fractions. Differences in absolute cloud fraction are interpreted through observing physics: GOCCP has active-lidar sensitivity and a coarse grid, whereas ISCCP has finer spatial sampling and passive-retrieval sensitivity to surface and illumination conditions. This complementarity sets the comparison strategy: GOCCP defines the regional cloud-cover reference, ISCCP is used to examine the local spatial placement of the site within that reference, and the long meteorological-station cloud data then check whether the satellite-inferred local low-cloud season is also present in ground observations. The native spatial supports of the two satellite cloud products are shown in Figure~\ref{fig:2-2}.

\subsection{Meteorological-station Cloud-amount Data}\label{long-term-meteorological-station-cloud-data}

The ground-based cloud component is provided by conventional cloud observations from the Shigatse Meteorological Station. The station is the national meteorological station no. 55578, located near Shigatse city at approximately \(29^\circ 15'\) N, \(88^\circ 53'\) E, and an elevation of approximately 3836 m. It is approximately 30 km east of the Shigatse 40 m radio site and lower by roughly 200 m. This distance and elevation difference preclude interpreting the meteorological-station data as co-located telescope-site data. Its value is climatological: it provides a multi-decade, locally relevant ground data set of total cloud amount in the Shigatse basin and adjacent southern-plateau environment.

The source table spans 1988 January 1 to 2019 May 31 and contains 11,474 daily entries for each of the four conventional observing-time labels: 02:00, 08:00, 14:00, and 20:00, giving 45,896 time-labeled entries in total. Total cloud amount is reported on the conventional 0--10 tenths scale; valid 0--10 entries are expressed as 0--100\% cloud amount by multiplying by 10. Blank entries and source values larger than 10 are treated as missing/invalid samples and excluded from cloud-amount statistics. After this validity rule is applied, 36,626 entries remain valid, with 8418, 9620, 10,198, and 8390 valid samples at 02:00, 08:00, 14:00, and 20:00, respectively. Because the 02:00 column contains valid data only through 2013 December 31, the formal four-time analyses are restricted to the common period from 1988 January 1 to 2013 December 31. Within this aligned period, the valid samples used in the main analysis are 8418, 7884, 8411, and 6799 at 02:00, 08:00, 14:00, and 20:00, respectively. The later valid samples at 08:00, 14:00, and 20:00 are retained for completeness accounting and all-available sensitivity checks, but are not used in the aligned four-time cloud-amount analysis. This treatment keeps the four observing-time samples comparable and prevents the post-2013 absence of 02:00 data from biasing fixed-time comparisons.

For optical night-time astronomy, the two night-time proxy samples are the most directly relevant. However, the site is not restricted to optical use. The existing 40 m-class radio facility and possible future multi-wavelength operations motivate retention of all four fixed cloud-observation times together with the separate 24 h Weather Station data. The four cloud-observation times are therefore retained in the data description, while the night-time subset is isolated later when cloud-limited optical observing conditions are discussed. This design does not provide continuous 24 h cloud monitoring. The distinction is important because Tibetan Plateau cloud cover has a documented diurnal component and because different satellite, reanalysis, and model data sets can represent that diurnal cycle differently \citep{Shang2018TPDiurnalCloud,Zhao2023TPCloudDiurnal}.

\begin{figure}[htbp]
\centering
\includegraphics[width=\linewidth,keepaspectratio]{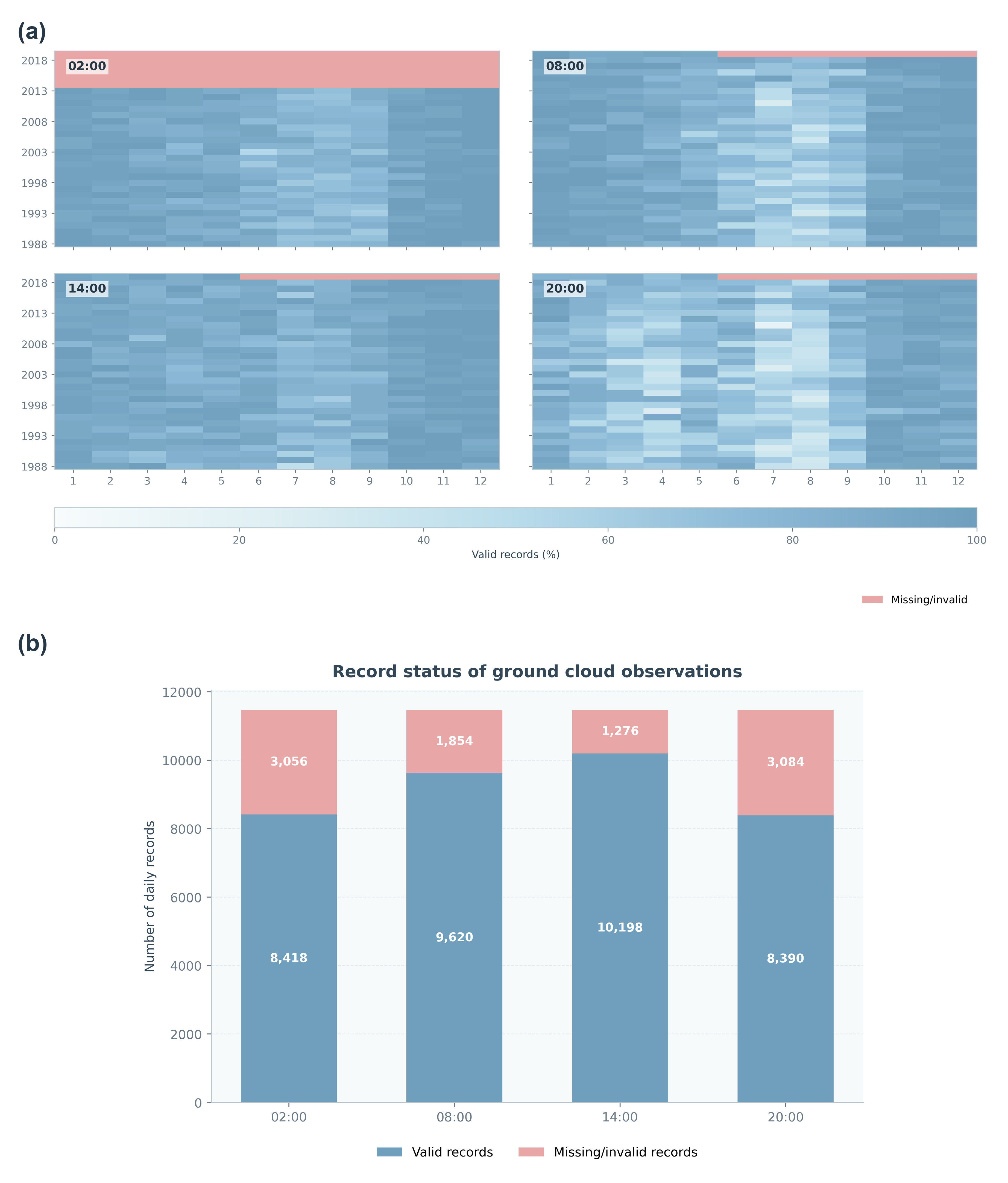}
\caption{Data-quality diagnostics for the Shigatse Meteorological Station total-cloud-amount data. Panel (a) gives monthly valid-sample completeness for the four conventional observing-time labels. Panel (b) gives the full-source valid and missing/invalid record counts for each observing time as vertical stacked bars. Valid total-cloud-amount entries are reported on the conventional 0--10 tenths scale and are expressed as 0--100\% cloud amount in the analysis. Missing/invalid samples include blank entries, out-of-range values larger than 10, and terminal intervals after the last valid entry of a given observing-time column.}
\label{fig:2-3}
\end{figure}

Figure~\ref{fig:2-3} defines the valid sampling domain for the meteorological-station cloud data. The meteorological-station cloud data provide an independent ground-based test of the satellite-inferred low-cloud season: they test whether the Shigatse region has a persistent, locally observed, multi-decade low-cloud season. Both all-available and strictly complete-day versions are retained as sensitivity checks. The 1988--2013 common-period aligned record is therefore the ground-based cloud data set used to test the monthly recurrence of the Shigatse low-cloud season.

\subsection{On-site Weather Station Data}\label{on-site-weather-station-data}

The on-site Weather Station data provide the local meteorological layer at the Shigatse 40 m site. The station was deployed by Shanghai Astronomical Observatory and operated during 2024--2025. The archive used here covers 2024 January to 2025 December; the 2025 December file in the manuscript data package ends at 2025 Dec 9 12:20 and is retained as a partial monthly archive because no later Weather Station data were added by the mid-2026 writing stage. The data include near-surface wind speed, wind direction, gusts, temperature, relative humidity, dew-point temperature, pressure, and precipitation.

Table~\ref{tab:2-3} gives representative measurement ranges and stated precision for high-precision automatic meteorological instruments used in astronomical site testing and Chinese meteorological observation networks \citep{Liu2019SnowAblationSensors,XuLi2020DaliMetObs,WMO2023CIMOGuide}. The ranges cover the Shigatse environment, including the low-pressure conditions near 4060 m, dry-season low humidity, weak-to-moderate night-time winds, and the observed temperature range. The precision is adequate for the monthly and night-time statistics used here, especially the wind-speed, relative-humidity, dew-point-depression, and precipitation-occurrence thresholds adopted in Section~\ref{local-meteorological-assessment-during-the-satellite-defined-low-cloud-season}. Temperature and pressure are retained as environmental descriptors, while gust speed and wind direction are documented in the archive but are not formal classification variables in the final analysis.

\begin{table}[htbp]
\centering
\caption{Reference Measurement Ranges and Precision for the On-site Weather Station.}
\label{tab:2-3}
\raatablestyle
\setlength{\tabcolsep}{2.6pt}
\begin{tabular}{@{}p{0.22\linewidth}p{0.25\linewidth}p{0.18\linewidth}p{0.30\linewidth}@{}}
\hline
Value & Range & Resolution & Precision \\
\hline
Temperature & \(-50\) to \(+60~^\circ\)C & 0.01 \(^\circ\)C & \(\pm 0.1~^\circ\)C \\
Relative humidity & 0--100\% RH & 0.1\% RH & \(\pm 2\%\) RH \\
Pressure & 500--1100 hPa & 0.1 hPa & \(\pm 0.2\) hPa \\
Wind speed & 0--75 m/s & 0.1 m/s & \(\pm 0.3\) m/s \\
Wind direction & 0--360\(^\circ\) & 3\(^\circ\) & \(\pm 5^\circ\) \\
Precipitation rate & 0--4 mm/min & 0.1 mm & \(\pm 0.4\) mm for totals \(\leq 10\) mm \\
\hline
\end{tabular}
\end{table}

The Weather Station raw archive consists of monthly CSV files sampled at a nominal 10-minute cadence, corresponding to 144 expected samples per day. Rows are sorted by observation time and de-duplicated by timestamp, and completeness is evaluated by nominal monthly file. The key variables used in this paper are also checked against broad physical limits to separate sampling gaps from physically invalid measurements.

\begin{figure}[htbp]
\centering
\includegraphics[width=\linewidth,keepaspectratio]{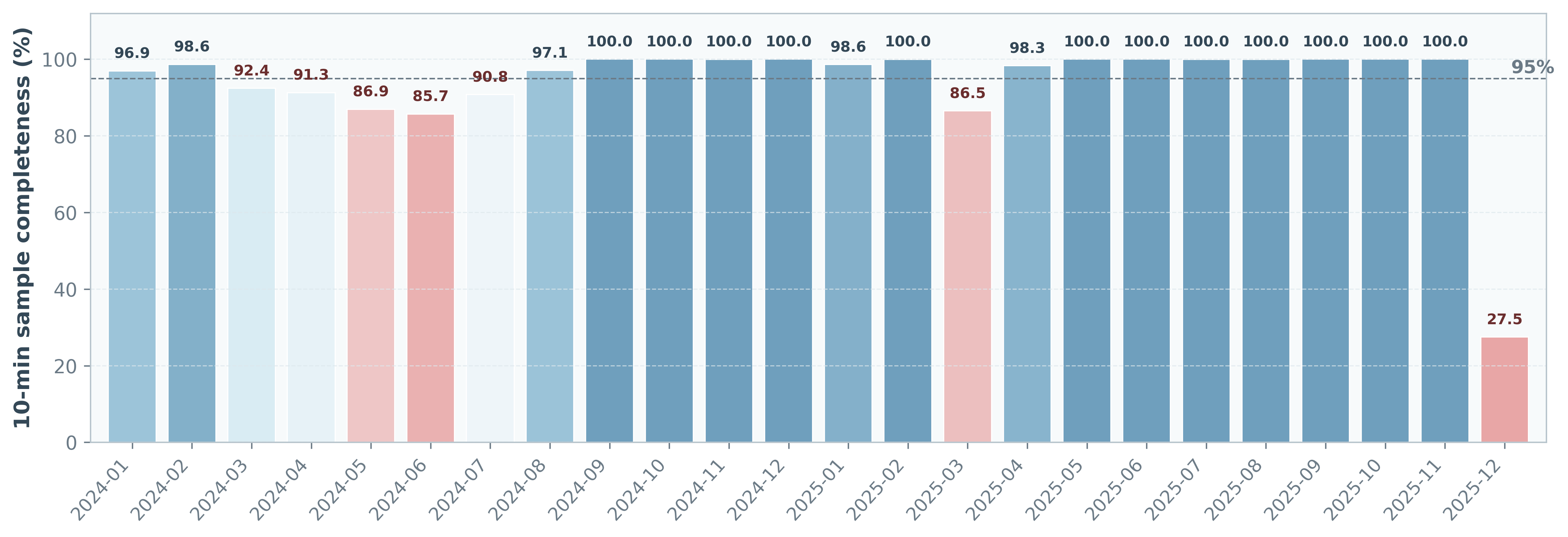}
\caption{Monthly completeness of the on-site Weather Station archive. Completeness is defined as $100 \times N_{\rm unique}/N_{\rm expected}$, where $N_{\rm expected}$ assumes 144 samples per day. The dashed line marks 95\% completeness; blue bars indicate relatively complete months, and muted rose bars mark lower-completeness months.}
\label{fig:2-6}
\end{figure}

Figure~\ref{fig:2-6} shows that missing time stamps are the main Weather Station data-quality limitation. Over January 2024--December 2025, the archive contains 98,623 unique 10-minute observation times out of 105,264 expected samples, or 93.69\%. Excluding the partial 2025 December file, the completeness is 97,397 out of 100,800 expected samples, or 96.62\%. Reduced-completeness intervals are concentrated mainly in early 2024, March 2025, and the 2025 December archive cutoff, while most months after September 2024 are nearly complete at the monthly scale. The early-2024 lower-completeness months correspond to the initial operation stage, when data acquisition was affected by interruptions including unexpected power outages. The 2025 December file is partial because the data package used here ends at 2025 Dec 9 12:20 and was not supplemented with later data by the mid-2026 writing stage. No duplicate timestamp rows or physically invalid values are found under the quality-control bounds used for this data-description section.

The Weather Station archive provides the local wind, humidity, dew point, and precipitation layer for the low-cloud month groups identified independently by GOCCP, ISCCP, and the meteorological-station cloud data.

\subsection{VIIRS/DNB Artificial Sky-brightness Data}\label{ancillary-viirsdnb-night-sky-background-data}

The VIIRS/DNB-derived sequence is used as an ancillary, model-based data set for dark-sky preservation. The data set consists of annual VIIRS/DNB-derived Sky Brightness model values from the lightpollutionmap.info framework for 2012--2025, extracted at Shigatse and the comparison sites. This approach follows global artificial night-sky-brightness mapping, in which satellite night-light radiance is combined with atmospheric propagation models to estimate artificial sky brightness and SQM-equivalent quantities \citep{Cinzano2001WorldAtlas,Falchi2016WorldAtlas}. The Shigatse sequence is complete at the annual-product level (14/14 years).

The values are model estimates derived from satellite night-light data and atmospheric propagation assumptions. Annual VIIRS products apply background-removal thresholds, and the lightpollutionmap.info layer is derived from VIIRS/Black Marble data and calibrated against reference measurements \citep{Elvidge2021AnnualVIIRS,GoogleEarthEngineVIIRSAnnualV22,LightPollutionMapHelp2026}. For very dark sites, values can be close to the reporting threshold or model floor; any small plotting floor used later is a visualization convention for floor-level dark-regime indicators. Local SQM/SQM-LE-class monitoring is the appropriate follow-up measurement for calibrated, high-cadence night-sky brightness statistics \citep{Hanel2018NightSkyMethods,Plauchu2017SPM,Cavazzani2020SQMSatellite}. Here, the annual VIIRS/DNB-derived sequence traces artificial-light evolution at Shigatse and along the Shigatse--Ali low-cloud corridor.

\subsection{Comparison Sites}\label{sites-for-comparison}

The comparison sample places Shigatse within the existing Chinese site-testing framework. Three reference sites are retained: Lenghu on Saishiteng Mountain, Ali Observatory, and Xinglong Observatory. They represent distinct geographic, climatic, and site-development settings within Chinese astronomical site testing. These settings provide reference points for interpreting the Shigatse cloud-cover cycle and observing season. Their basic coordinates and elevations are summarized in Table~\ref{tab:2-4}.

Lenghu is the key domestic optical benchmark, with published measurements of clear nights, seeing, PWV, cloud cover, and local meteorology \citep{Deng2021Lenghu,Li2024LenghuCloud}. It is also the retained comparison site closest in longitude to Shigatse, but lies nearly \(9.4^\circ\) farther north. The Shigatse--Lenghu pair therefore contrasts a northern-plateau optical benchmark with a lower-latitude southern-plateau site, a difference relevant to sky access because latitude affects the elevation of low-declination and southern targets.

Ali is the key western-Xizang high-altitude reference. It was included in the western-China LOT site-testing campaign, with homogeneous meteorological, cloud, seeing, sky-background, and PWV data products collected from 2017 to 2019, and it has since accumulated dedicated radiosonde, turbulence, PWV, and cloud studies \citep{Feng2020LOTOverview,Cao2020DataProducts,Qian2018AliRadiosonde,Hickson2020AliTurbulence,Qian2024AliCloud}. Ali connects Shigatse to the Xizang high-plateau literature background and to the westward regional cloud-climatology framework developed in \citet{Cao2020ClearNights}. Here, Ali-A denotes the specific Ali reference point used in that framework, located at \(32.32573^\circ\) N, \(80.02671^\circ\) E and about 5040 m elevation \citep{Cao2020DataProducts}. The Shigatse--Ali-A separation is used later as a reference radius for examining the Shigatse--Ali low-cloud corridor at the satellite cloud-product scale.

Xinglong is retained as a reference eastern observatory with a long operational history and published observing-condition statistics for 2007--2014 \citep{Zhang2016XinglongObservingConditions}. Its lower elevation, eastern longitude, and monsoon-influenced setting provide a contrast to the high-plateau cases.

\begin{table}[htbp]
\centering
\caption{Basic Site Parameters for Shigatse and the Comparison Sites Ali, Lenghu, and Xinglong.}
\label{tab:2-4}
\raatablestyle
\setlength{\tabcolsep}{4.0pt}
\begin{tabular*}{\linewidth}{@{\extracolsep{\fill}}lccc@{}}
\hline
Site & Latitude & Longitude & Elevation \\
\hline
Shigatse site \citep{Yang2018ShigatseSkyImages} & \(29.2056^\circ\) N & \(88.6322^\circ\) E & about 4060 m \\
Lenghu \citep{Deng2021Lenghu,Li2024LenghuCloud} & \(38.6068^\circ\) N & \(93.8961^\circ\) E & about 4200 m \\
Ali \citep{Cao2020DataProducts,Qian2024AliCloud} & \(32.32573^\circ\) N & \(80.02671^\circ\) E & about 5040 m \\
Xinglong \citep{Zhang2016XinglongObservingConditions} & \(40^\circ23'26''\) N & \(117^\circ34'39''\) E & about 900 m \\
\hline
\end{tabular*}
\end{table}

The quantitative claims of the paper are limited to cloud-cover climatology, selected local meteorological conditions, and comparison-site context. The main result is a cloud-cover climatology and a non-monsoon observing season for Shigatse; the supporting result is a Weather Station surface-meteorological analysis that identifies potential weather-related constraints during that season. The comparison sites therefore provide reference points for interpreting Shigatse in latitude, longitude, cloud-season timing, and western-China site-testing context.

\section{GOCCP Cloud-cover Climatology}\label{goccp-cloud-climatology}

GOCCP is used to define the regional active-lidar cloud-cover reference for Shigatse. The product is derived from CALIPSO/CALIOP lidar observations and applies a consistent cloud-detection framework to long-term satellite profiles, which is valuable over complex terrain where passive retrievals may be affected by bright surfaces, snow cover, and weak cloud--surface thermal contrast \citep{Chepfer2010GOCCP,Winker2010CALIPSO,Stubenrauch2013GlobalCloudDatasets}. The \(2^\circ \times 2^\circ\) GOCCP grid provides a regional description of the southern-plateau cloud environment. The Tibetan Plateau literature provides additional product context: CALIPSO-GOCCP has been used with AIRS to analyze the seasonality of high clouds and cirrus over the plateau, including the association of southern-plateau summer high clouds with Asian monsoon moisture and deep convection \citep{Liu2019TPCirrusGOCCP}. That result supports the physical interpretation of a summer cloud maximum in the Shigatse region. The analysis follows the same sequence used for ISCCP in Section~\ref{isccp-cloud-climatology}: annual spatial pattern, monthly cycle, monthly spatial evolution, Hovmöller diagnostics, and within-product site context.

\subsection{Annual Spatial Pattern of Cloud Fraction}\label{annual-cloud-fraction-pattern}

\begin{figure}[htbp]
\centering
\includegraphics[width=\linewidth,keepaspectratio]{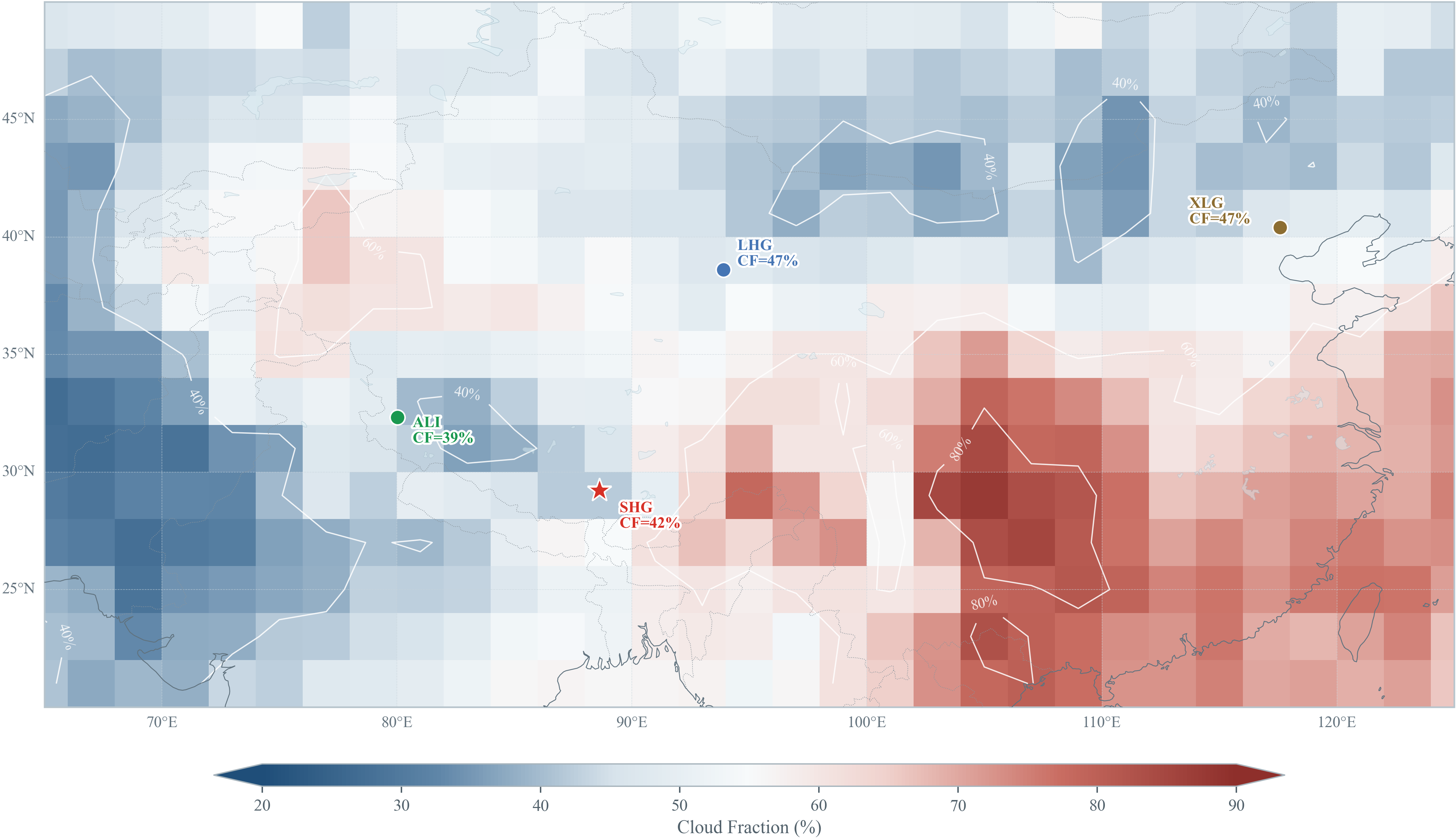}
\caption{Annual mean cloud fraction from CALIPSO-GOCCP. The map defines the regional active-lidar cloud-climatology context for Shigatse at the GOCCP grid-cell scale. Site codes in this and subsequent satellite cloud maps are SHG (Shigatse), LHG (Lenghu), ALI (Ali), and XLG (Xinglong).}
\label{fig:3-1}
\end{figure}

The annual mean GOCCP cloud-fraction map places Shigatse in a relatively lower-cloud southern-plateau region near the transition between monsoon-influenced terrain and the drier interior plateau (Figure~\ref{fig:3-1}). The extracted Shigatse grid cell has an annual mean cloud fraction of 42.1\%, corresponding to an annual clear-sky fraction of 57.9\%. This value is lower than the cloud fractions typical of the eastern monsoon-influenced part of the domain and is consistent with the broader Tibetan Plateau pattern, in which cloud cover is controlled by monsoon moisture transport, terrain-related gradients, and dry-season circulation \citep{DuanWu2006TPCloudClimate,Zhang2007TPCloud,Shang2018TPDiurnalCloud,Wu2024TPCloudReview}.

The annual value is the first-order cloud-climatology quantity. For astronomical use, the temporal concentration of low-cloud conditions is equally important because a moderate annual cloud fraction can still support a productive observing program if low-cloud months recur during a stable and sufficiently long season. This point is central to the Shigatse assessment: the existing facility provides an operational reference point, while the regional cloud climate identifies the low-cloud observing period for systematic monitoring at the present site and for further exploration toward the northern Himalayan side.

\subsection{Monthly Cloud-fraction Cycle at Shigatse}\label{monthly-cloud-fraction-cycle-at-shigatse}

\begin{figure}[htbp]
\centering
\includegraphics[width=0.90\linewidth,keepaspectratio]{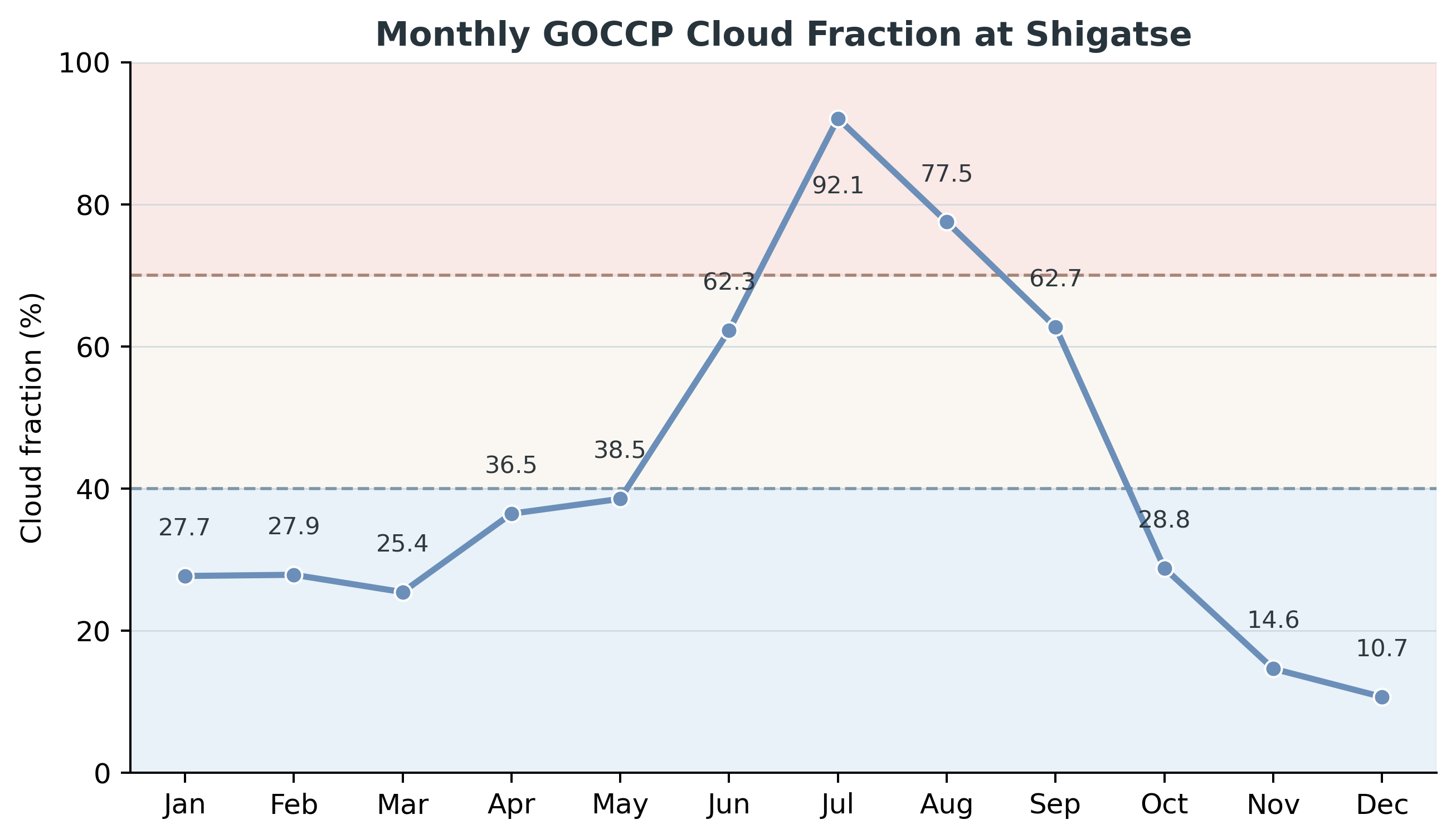}
\caption{Monthly GOCCP cloud fraction for the Shigatse grid cell. Numerical labels give the monthly values used in the text and in the cross-product monthly-value table in Section~\ref{goccpisccp-monthly-phase-and-product-dependent-offset}. The 40\% and 70\% cloud-fraction guide levels correspond to cloud-cover classes used as less-cloudy and observable conditions in recent western-China cloud-cover studies \citep{Qian2024AliCloud,Li2024LenghuCloud}; in this paper, they define monthly satellite cloud-fraction reference levels for the cloud-climatology analysis.}
\label{fig:3-2}
\end{figure}

The monthly GOCCP data show a sharply defined annual cycle (Figure~\ref{fig:3-2}). Cloud fraction is lowest from late autumn through winter, with 14.6\% in November and 10.7\% in December. January--March remain low, at 25.4\%--27.9\%, and the spring transition months of April and May remain below 40\%. Cloud fraction increases rapidly after the onset of the summer monsoon, reaching 62.3\% in June, 92.1\% in July, 77.5\% in August, and 62.7\% in September.

For Shigatse, the most useful temporal description is a set of locally meaningful month groups, rather than a conventional four-season average alone. The November--January core has a GOCCP mean cloud fraction of 17.7\%, corresponding to a clear-sky fraction of 82.3\%. The broader October--May low-cloud season has a mean cloud fraction of 26.3\% and contains all eight months with GOCCP cloud fraction below 40\%. The June--September interval has a mean cloud fraction of 73.7\% and defines the main monsoon-affected period. These month groups follow the observed monthly structure and match the expected separation between the dry non-monsoon regime and the summer-monsoon cloud maximum over the southern Tibetan Plateau \citep{DuanWu2006TPCloudClimate,Zhang2007TPCloud,Shang2018TPDiurnalCloud,Liu2019TPCirrusGOCCP,Wu2024TPCloudReview,Yang2018ShigatseSkyImages}.

This monthly structure defines the calendar groups used in the subsequent analysis. The November--January core is the strongest initial optical interval. October--May is the broader candidate observing and monitoring season. June--September is the monsoon-affected interval in which optical scheduling should be treated conservatively, while non-observing engineering work and instrument maintenance can be concentrated when possible.

\subsection{Monthly Spatial Patterns and Month-to-month Evolution}\label{monthly-spatial-patterns-and-month-to-month-evolution}

\begin{figure}[htbp]
\centering
\includegraphics[width=0.985\linewidth,keepaspectratio]{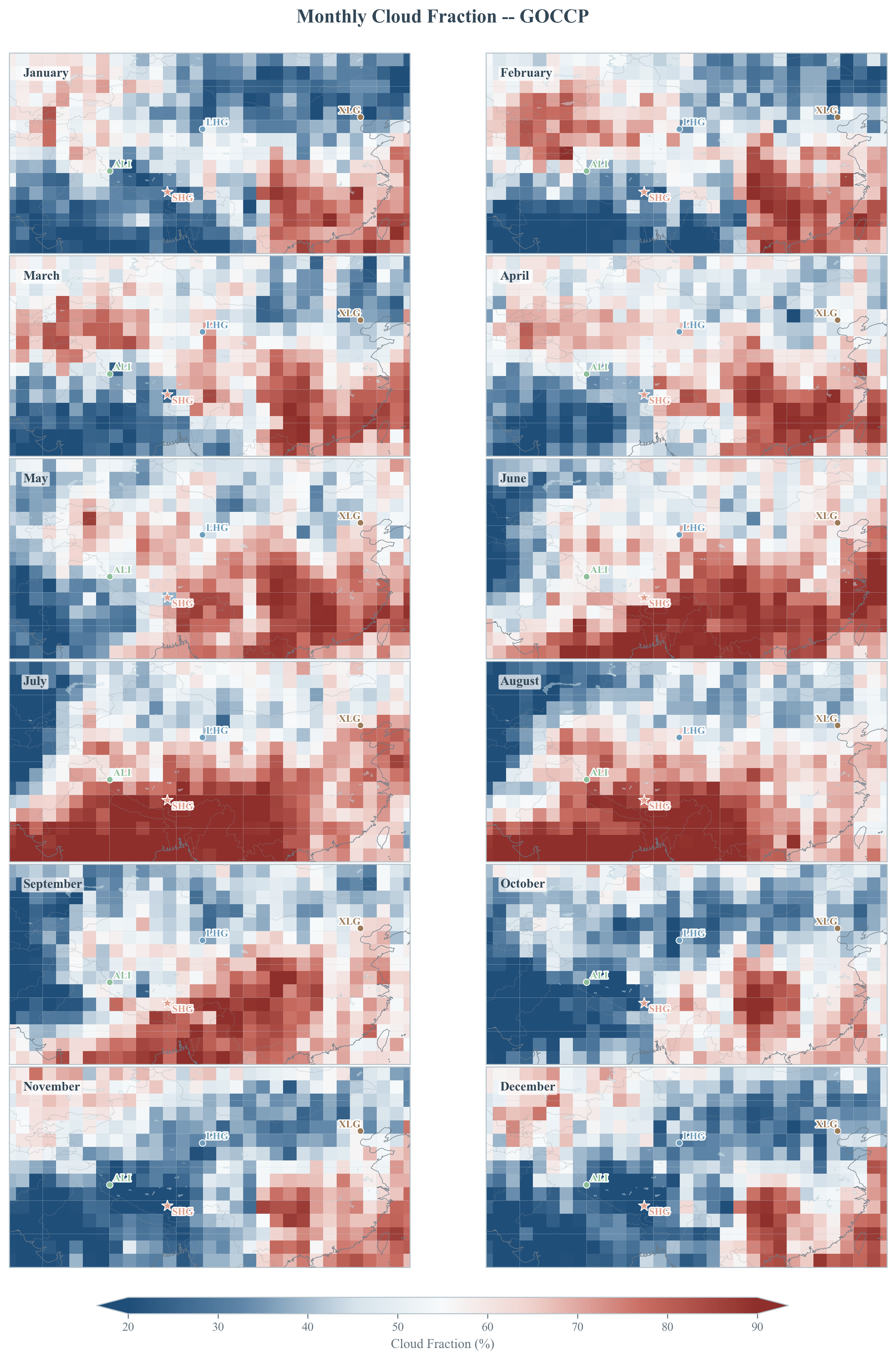}
\caption{Monthly GOCCP cloud-fraction maps from January to December in a two-column, six-row layout. The panels show the active-lidar cloud field at $2^\circ \times 2^\circ$ support with a common 20--90\% color scale. Site codes: SHG = Shigatse, LHG = Lenghu, ALI = Ali, and XLG = Xinglong.}
\label{fig:3-3}
\end{figure}

The monthly GOCCP maps test whether the grid-cell low-cloud signal identified in Section~\ref{monthly-cloud-fraction-cycle-at-shigatse} is embedded in a coherent regional cloud field (Figure~\ref{fig:3-3}). From October through May, lower cloud fractions extend across a broader part of the southern and western plateau. During June--September, cloud fraction increases over the Himalayan foreland and southern plateau, with the strongest expansion in July and August. This month-by-month view provides the main spatial support for interpreting the Shigatse monthly cycle as part of a regional seasonal regime.

\subsection{Regional Longitude--Month and Latitude--Month Structure}\label{regional-hovmuxf6ller-diagnostics}

\begin{figure}[htbp]
\centering
\includegraphics[width=0.82\linewidth,keepaspectratio]{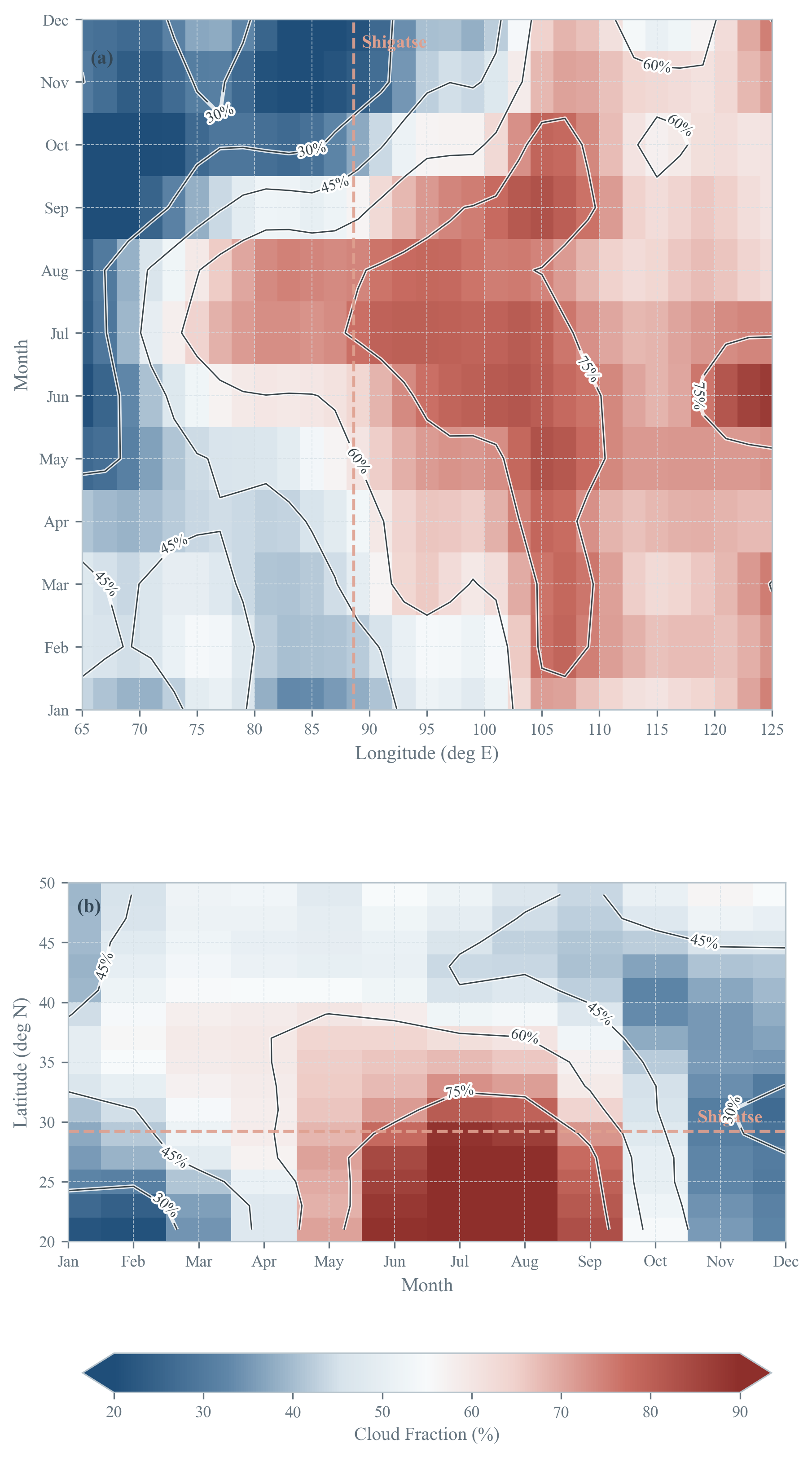}
\caption{GOCCP Hovmöller diagnostics for the Shigatse region. The upper panel shows longitude--month structure along $28^\circ$--$36^\circ$ N, and the lower panel shows latitude--month structure along $85^\circ$--$100^\circ$ E. The panels diagnose the regional seasonal phase and the spatial embedding of the Shigatse low-cloud period in the active-lidar product.}
\label{fig:3-5}
\end{figure}

The GOCCP Hovmöller diagrams provide the regional active-lidar diagnostic corresponding to the monthly maps (Figure~\ref{fig:3-5}). They connect the monthly Shigatse curve with the seasonal displacement of the southern-plateau cloud field, rather than simply repeating the single-grid time series. Along the \(28^\circ\)--\(36^\circ\) N belt, the Shigatse longitude lies within a low-cloud zone during October--May, while high cloud fraction expands across much of the belt during June--September. Along the \(85^\circ\)--\(100^\circ\) E belt, the monsoon-related cloud maximum reaches its greatest northward extent in July. The latitude of Shigatse is therefore most strongly affected during the four-month monsoon interval.

The Hovmöller diagnostics add the regional context missing from a single-grid annual statistic. A single-grid annual cloud fraction cannot distinguish a stable seasonal pattern from a local anomaly or a spatially drifting feature. Cross-site monthly summaries from the same GOCCP extraction are discussed alongside their ISCCP counterparts in Section~\ref{shigatse-in-the-cloud-climatology-context-of-selected-chinese-sites}. The longitude--month and latitude--month diagnostics place the Shigatse low-cloud period within a coherent regional seasonal regime: the low-cloud period follows the withdrawal of monsoon cloud and recurs across the longitude and latitude bands relevant to further site exploration. These patterns identify the calendar and spatial domain in which higher-altitude or more remote Himalayan-side candidates can be prioritized for ground testing.

\section{ISCCP Cloud-cover Climatology}\label{isccp-cloud-climatology}

ISCCP HXG is used to define the finer-grid passive-satellite cloud-climatology context for Shigatse. ISCCP is a long-standing international cloud data record built from geostationary and polar-orbiting radiance observations \citep{Schiffer1983ISCCP,Rossow1999ISCCP,Young2018ISCCPHSeries}. Its retrieval physics and cloud definitions differ from those of CALIPSO-GOCCP, and the product is affected by known uncertainties over high terrain, snow or bright surfaces, multilayer clouds, optically thin clouds, and weak cloud--surface thermal contrast \citep{NaudChen2010ISCCPTP,Li2006TPCloudTypes,Liu2021TPSatelliteCloud,Wu2024TPCloudReview}. These effects are especially relevant for the Tibetan Plateau, where CloudSat-CALIPSO comparisons show that ISCCP can underestimate cloud cover and misplace cloud-top pressure in product- and season-dependent ways \citep{NaudChen2010ISCCPTP}. The \(0.1^\circ \times 0.1^\circ\) ISCCP grid tests the local spatial placement and monthly phase of Shigatse within the southern-plateau cloud field, while absolute cloud fractions are interpreted on a product-by-product basis. The section mirrors the GOCCP sequence in Section~\ref{goccp-cloud-climatology}.

\subsection{Annual Spatial Pattern of Cloud Fraction}\label{annual-cloud-fraction-pattern-1}

\begin{figure}[htbp]
\centering
\includegraphics[width=\linewidth,keepaspectratio]{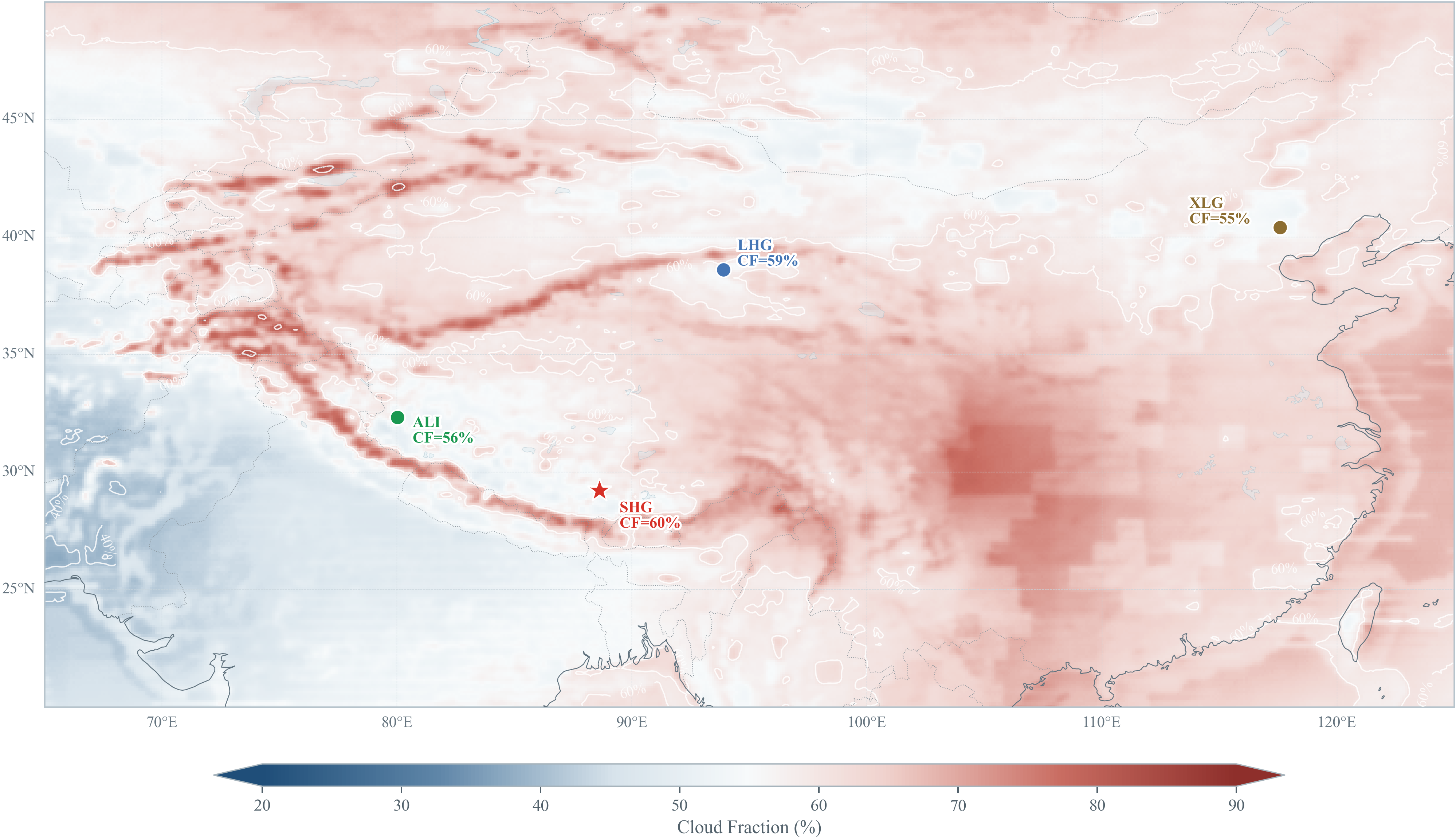}
\caption{Annual mean cloud fraction from ISCCP HXG. The map shows the finer-grid passive-satellite cloud-climatology context for Shigatse and its local spatial placement within the southern-plateau cloud field. Site codes: SHG = Shigatse, LHG = Lenghu, ALI = Ali, and XLG = Xinglong.}
\label{fig:4-1}
\end{figure}

The annual ISCCP cloud-fraction field places Shigatse within a relatively low-cloud part of the southern Tibetan Plateau embedded in a coherent local cloud gradient (Figure~\ref{fig:4-1}). Compared with the GOCCP grid, the finer ISCCP grid can resolve stronger local gradients associated with river valleys, high terrain, and the northern Himalayan slope. ISCCP therefore provides a finer-grid passive-satellite check on the spatial placement of the GOCCP-identified low-cloud regime around Shigatse.

The extracted Shigatse ISCCP cloud fraction is 59.7\% in the annual mean. This value is higher than the corresponding GOCCP cloud fraction, and a 0.1\(^\circ\) passive-satellite grid cell still averages over terrain and retrieval conditions that differ from on-site ground data. At the ISCCP product scale, however, the Shigatse area lies within a coherent cloud gradient on the southern plateau, consistent with the regional GOCCP interpretation and supporting the month-by-month spatial test below.

\subsection{Monthly Cloud-fraction Cycle at Shigatse}\label{monthly-cloud-fraction-cycle-at-shigatse-1}

\begin{figure}[htbp]
\centering
\includegraphics[width=0.90\linewidth,keepaspectratio]{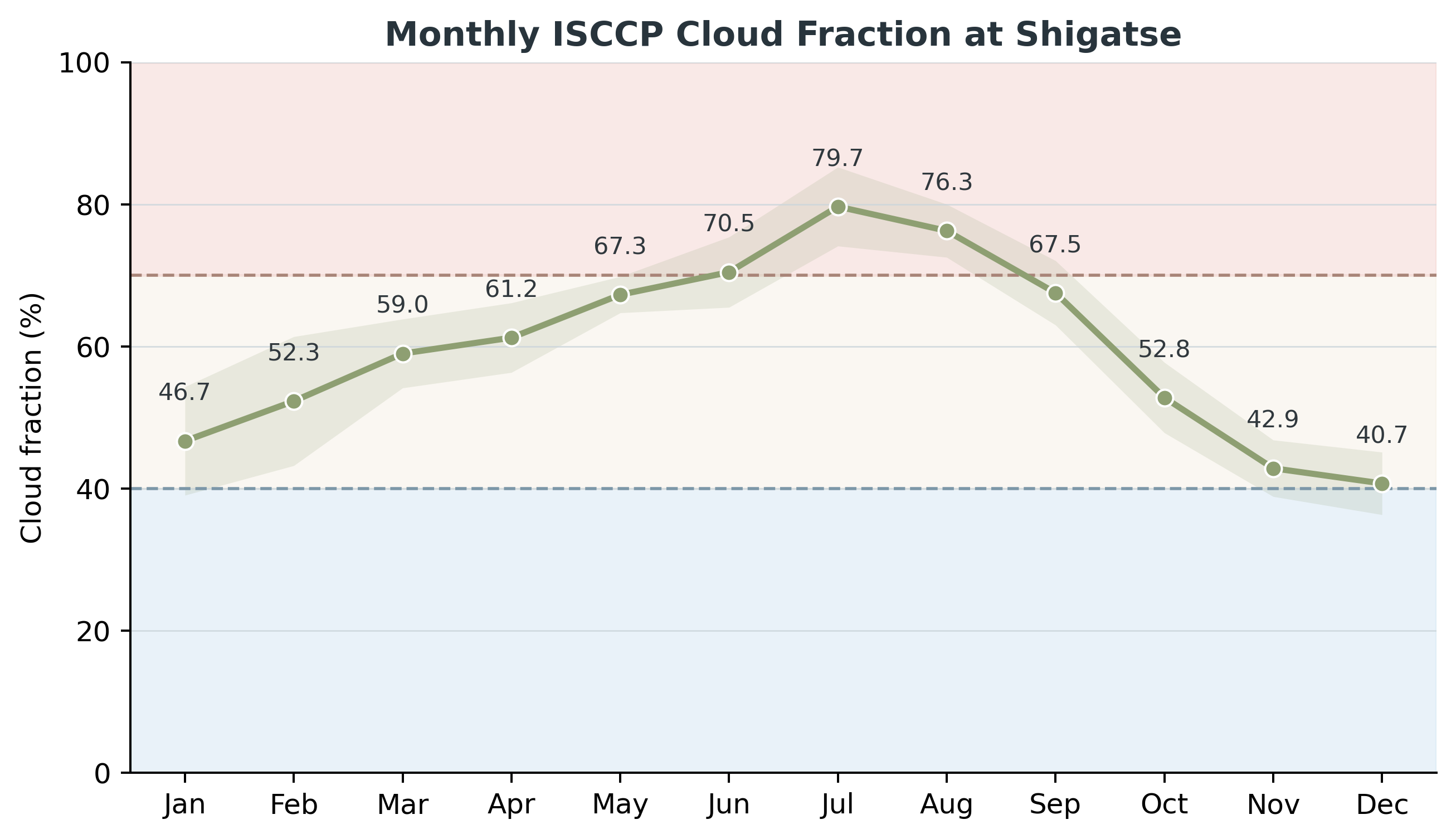}
\caption{Monthly ISCCP cloud fraction for the Shigatse grid cell. Numerical labels give the monthly mean cloud-fraction values used in the text and in the cross-product monthly-value table in Section~\ref{goccpisccp-monthly-phase-and-product-dependent-offset}; the shaded band shows the interannual $1\sigma$ spread. The 40\% and 70\% cloud-fraction guide levels correspond to the same cloud-cover classes shown in Figure~\ref{fig:3-2} \citep{Qian2024AliCloud,Li2024LenghuCloud}.}
\label{fig:4-2}
\end{figure}

The monthly ISCCP cycle shows a comparable seasonal phase to GOCCP (Figure~\ref{fig:4-2}). The Shigatse cloud fraction is lowest in November and December, at 42.9\% and 40.7\%, respectively, and remains relatively low in January and February, at 46.7\% and 52.3\%, respectively. It increases through the pre-monsoon transition and reaches the highest values during the summer monsoon, with 79.7\% in July and 76.3\% in August. For the same month groups used in the GOCCP and ground-data comparison, the November--January core has an ISCCP mean cloud fraction of 43.4\%, the broader October--May low-cloud season has 52.9\%, and the June--September monsoon interval has 73.5\%.

The higher ISCCP absolute cloud fractions reflect the different sensors, viewing geometries, cloud detection methods, and spatial averaging used by the two products. Over the Tibetan Plateau, passive cloud retrievals are additionally complicated by snow cover, bright surfaces, strong elevation gradients, low-cloud detection at night, and variable cloud vertical structure \citep{NaudChen2010ISCCPTP,Li2006TPCloudTypes,Liu2021TPSatelliteCloud}. The most important ISCCP result for the Shigatse assessment is the monthly ordering: a lower-cloud late-autumn-to-winter interval, a broader October--May low-cloud season, and a June--September maximum.

\subsection{Monthly Spatial Patterns and Month-to-month Evolution}\label{monthly-spatial-patterns-and-month-to-month-evolution-1}

\begin{figure}[htbp]
\centering
\includegraphics[width=0.985\linewidth,keepaspectratio]{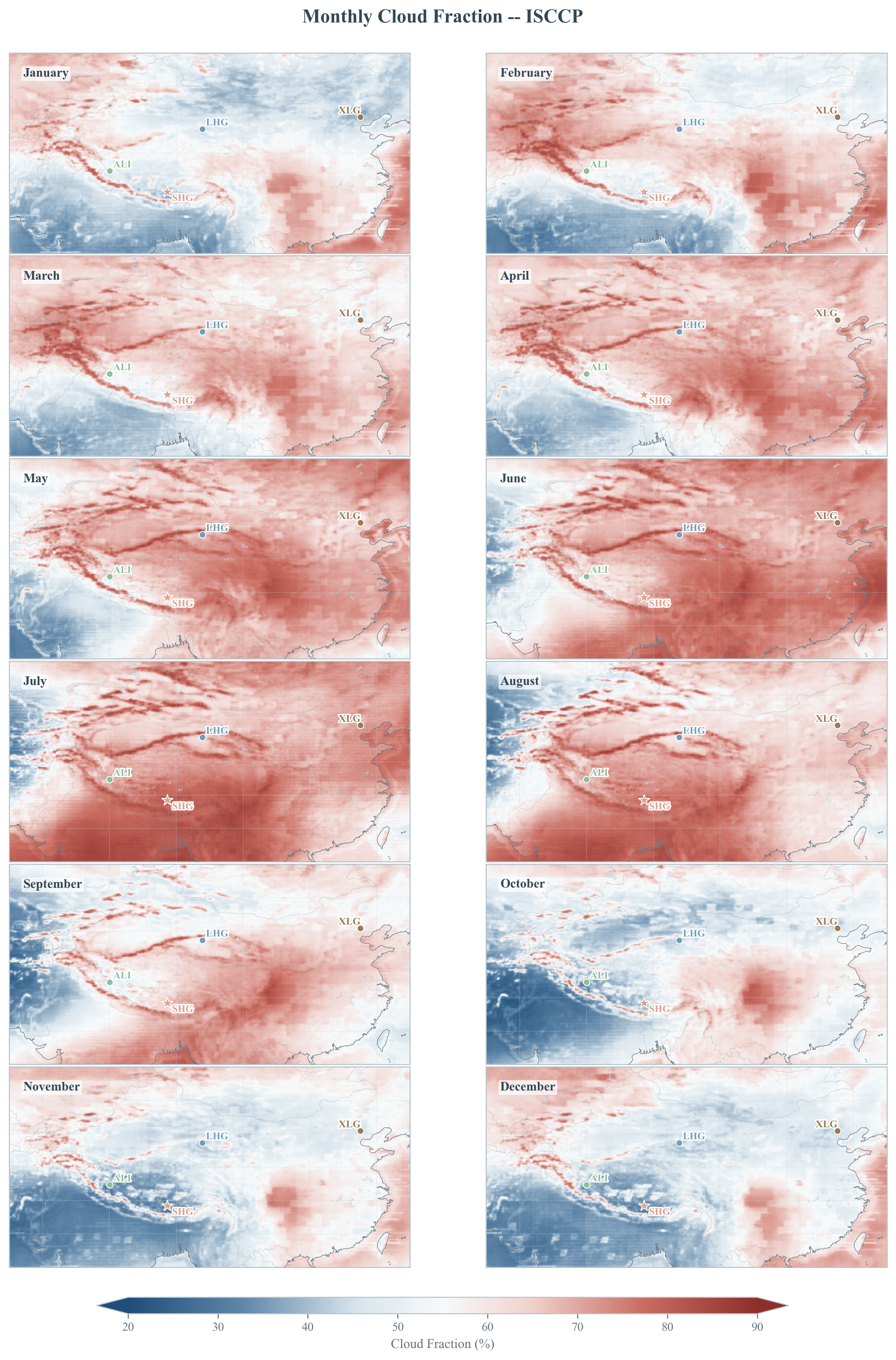}
\caption{Monthly ISCCP cloud-fraction maps from January to December. The two-column sequence shows the passive-satellite field at $0.1^\circ \times 0.1^\circ$ support with the same 20--90\% color scale and site codes as Figure~\ref{fig:3-3}.}
\label{fig:4-3}
\end{figure}

The monthly ISCCP maps test whether the relatively lower-cloud Shigatse environment persists through the low-cloud season and how it is placed within the finer-grid passive-satellite cloud field (Figure~\ref{fig:4-3}). The full twelve-month sequence follows the same two-column layout as the GOCCP maps in Figure~\ref{fig:3-3}. The Shigatse region remains in a lower-cloud southern-plateau setting during the non-monsoon months, while July and August show the broad monsoon cloud influence across the southern plateau.

Spatial persistence is required for interpreting a low-cloud annual or monthly signal: if the lower-cloud center shifts away from the Shigatse site, a site-extracted monthly cycle can overstate local relevance. ISCCP indicates that the Shigatse region is part of a recurrent southern-plateau low-cloud structure during the non-monsoon season. The summer maps identify monsoon cloudiness as a regional seasonal constraint for Shigatse, while the non-monsoon maps support its recurrent low-cloud placement.

\subsection{Regional Longitude--Month and Latitude--Month Structure}\label{regional-hovmuxf6ller-diagnostics-1}

\begin{figure}[htbp]
\centering
\includegraphics[width=0.82\linewidth,keepaspectratio]{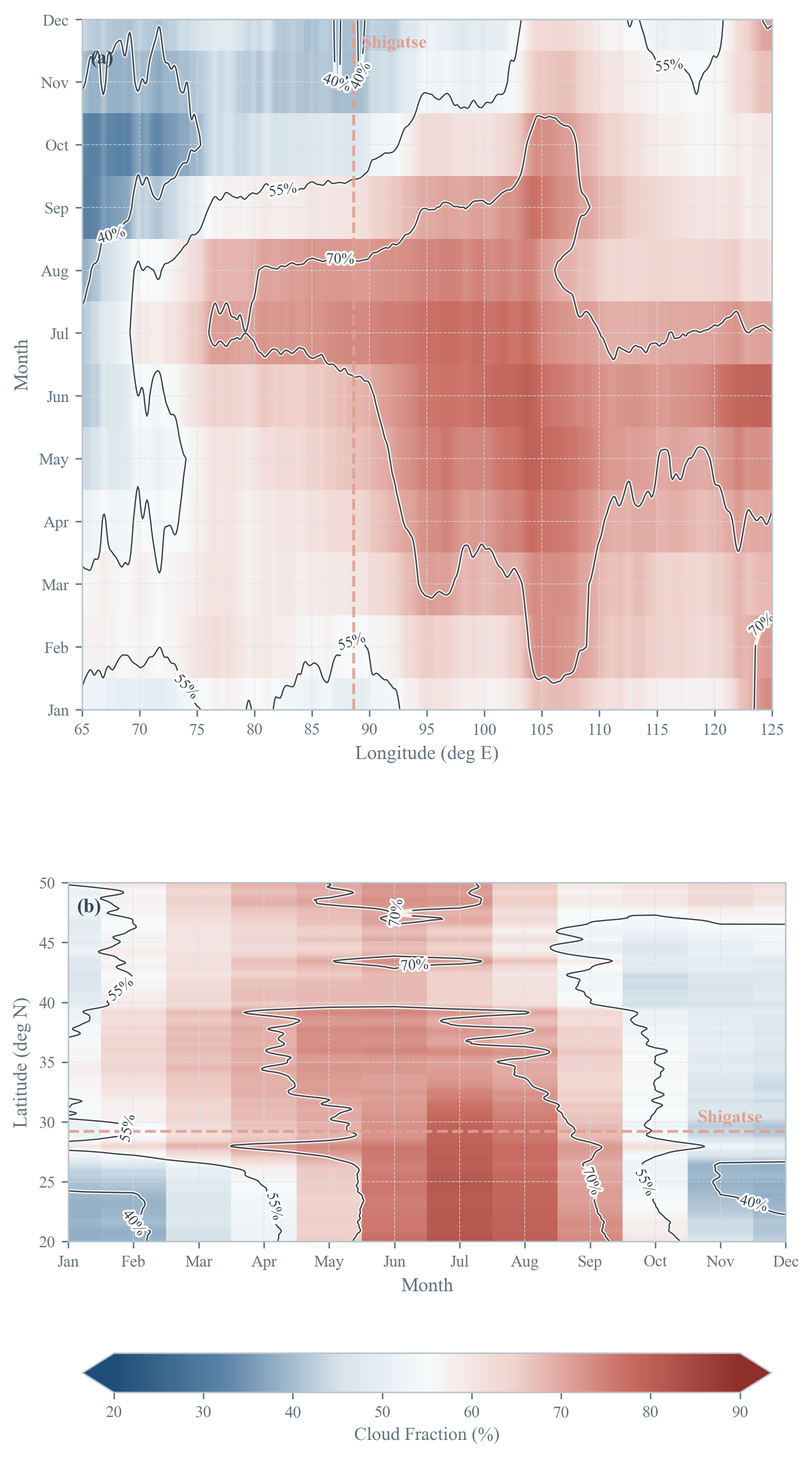}
\caption{ISCCP Hovmöller diagnostics for the Shigatse region. The upper panel shows longitude--month structure along $28^\circ$--$36^\circ$ N, and the lower panel shows latitude--month structure along $85^\circ$--$100^\circ$ E. The panels are the ISCCP counterpart of Figure~\ref{fig:3-5} and diagnose seasonal phase and fine-grid spatial placement in the passive satellite product.}
\label{fig:4-5}
\end{figure}

The ISCCP Hovmöller diagrams provide the finer-grid passive-satellite diagnostic corresponding to the GOCCP sections in Figure~\ref{fig:3-5} (Figure~\ref{fig:4-5}). The longitude--month section averages cloud fraction across \(28^\circ\)--\(36^\circ\) N, and the latitude--month section averages across \(85^\circ\)--\(100^\circ\) E. This pair of sections places the Shigatse monthly curve within the surrounding regional structure of the passive product.

The ISCCP sections reproduce the main seasonal phase identified by GOCCP: lower cloud fractions occur during the non-monsoon part of the year, whereas the summer monsoon produces a broad cloud maximum. The longitude--month section shows that the Shigatse longitude remains within a comparatively lower-cloud region during the late-autumn and winter interval, while high cloud fraction expands across much of the southern-plateau belt in June--September. The latitude--month section shows the same summer enhancement near the Shigatse latitude and a sharper north--south gradient than is resolved by the coarser GOCCP grid.

The product differences are also informative. ISCCP gives higher absolute cloud fractions and more small-scale texture than GOCCP, particularly over high terrain and along strong surface-contrast gradients. This behavior is consistent with the product role defined in Section~\ref{satellite-cloud-products}: ISCCP is a passive multi-satellite cloud product with finer spatial sampling, but its retrievals over the Tibetan Plateau can be affected by snow, bright surfaces, complex terrain, and weak cloud--surface thermal contrast \citep{NaudChen2010ISCCPTP,Young2018ISCCPHSeries,Liu2021TPSatelliteCloud,Wu2024TPCloudReview}. The four-site monthly context for both satellite products is discussed together in Section~\ref{shigatse-in-the-cloud-climatology-context-of-selected-chinese-sites}. At finer grid spacing, ISCCP therefore places Shigatse in the same seasonal cloud regime as GOCCP, with a recurrent non-monsoon lower-cloud interval and a monsoon-limited summer.

\section{Ground-based Cloud-amount Statistics at Shigatse}\label{ground-based-cloud-amount-analysis-at-shigatse}

The long-term meteorological-station cloud data set is the third cloud data source used in this paper. It comes from conventional cloud observations at the Shigatse meteorological station, near Shigatse city and approximately 30 km east of the Shigatse 40 m radio site. The station is roughly 200 m lower than the telescope site. This basin-scale record provides a multi-decade, locally grounded test of whether the satellite-inferred monthly cloud pattern is also present in traditional total-cloud-amount data from the Shigatse region.

\subsection{Data Cleaning and Common-period Alignment}\label{data-cleaning-and-common-period-alignment}

The source table contains four conventional observing times: 02:00, 08:00, 14:00, and 20:00. Total cloud amount is stored on the conventional 0--10 tenths scale. Values from 0 to 10 are retained as valid cloud amounts and converted to 0--100\% by multiplying by 10; value 11 and values greater than 10 are treated as invalid or missing values. The full source interval extends from 1988 January 1 to 2019 May 31, but the main four-time analysis uses the common valid period from 1988 January 1 to 2013 December 31. This alignment is required because the 02:00 data series contains valid values only through 2013 December 31, while the 08:00, 14:00, and 20:00 columns continue into 2014--2019.

Within the aligned period, the valid sample counts are 8418 at 02:00, 7884 at 08:00, 8411 at 14:00, and 6799 at 20:00. The later 2014--2019 valid samples at 08:00, 14:00, and 20:00 are retained in the data archive and completeness diagnostics, but they are not used in the main four-time climatological comparison. This treatment keeps the observing-time samples temporally matched and avoids mixing a shorter night-time data span with longer daytime and evening data.

\begin{figure}[htbp]
\centering
\includegraphics[width=\linewidth,keepaspectratio]{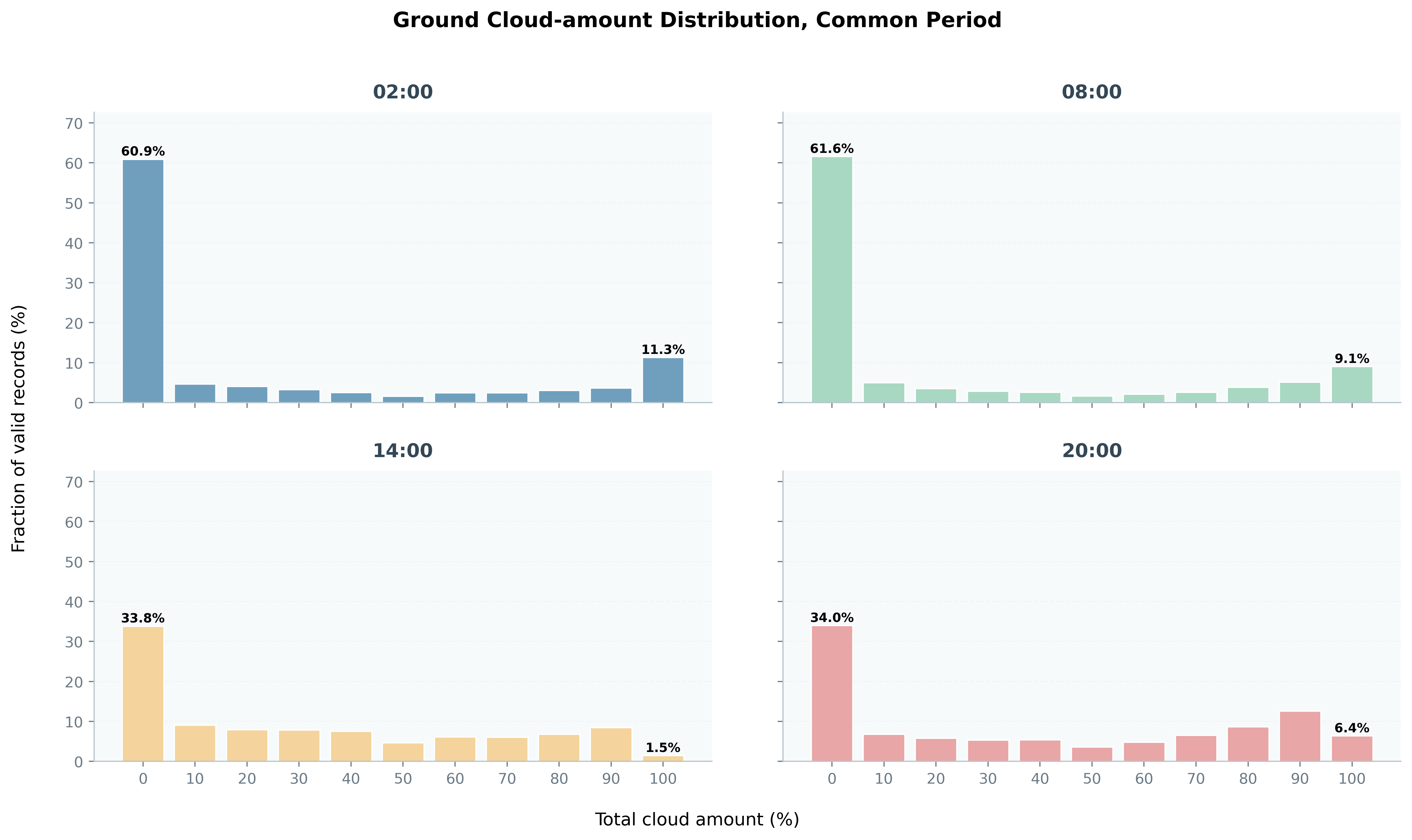}
\caption{Distribution of valid total cloud amount at the four conventional observing times during the common 1988--2013 period. The original 0--10 tenths values are expressed as 0--100\% cloud amount for plotting.}
\label{fig:5-1}
\end{figure}

The distribution of total cloud amount indicates that low-cloud observations are frequent in the aligned data. The mean cloud amounts are 24.3\%, 23.8\%, 32.7\%, and 39.5\% at 02:00, 08:00, 14:00, and 20:00, respectively. For the classification analysis below, an observation with total cloud amount \(\leq 40\%\) is treated as the conventional-cloud analogue of a clear or less-cloudy condition. This threshold is tied to the upper boundary of the less-cloudy class used in recent Ali cloud-cover statistics \citep{Qian2024AliCloud}. Under this criterion, the fractions of observations with cloud amount \(\leq 40\%\) are 75.4\%, 75.5\%, 66.3\%, and 57.5\% at 02:00, 08:00, 14:00, and 20:00. The larger low-cloud fractions at the two night-time proxy samples provide the first fixed-time indication that the clearest ground-observed cloud conditions occur preferentially during the night-side part of the observing cycle.

\subsection{Monthly Mean Cloud Amount}\label{monthly-mean-cloud-amount}

\begin{figure}[htbp]
\centering
\includegraphics[width=0.95\linewidth,keepaspectratio]{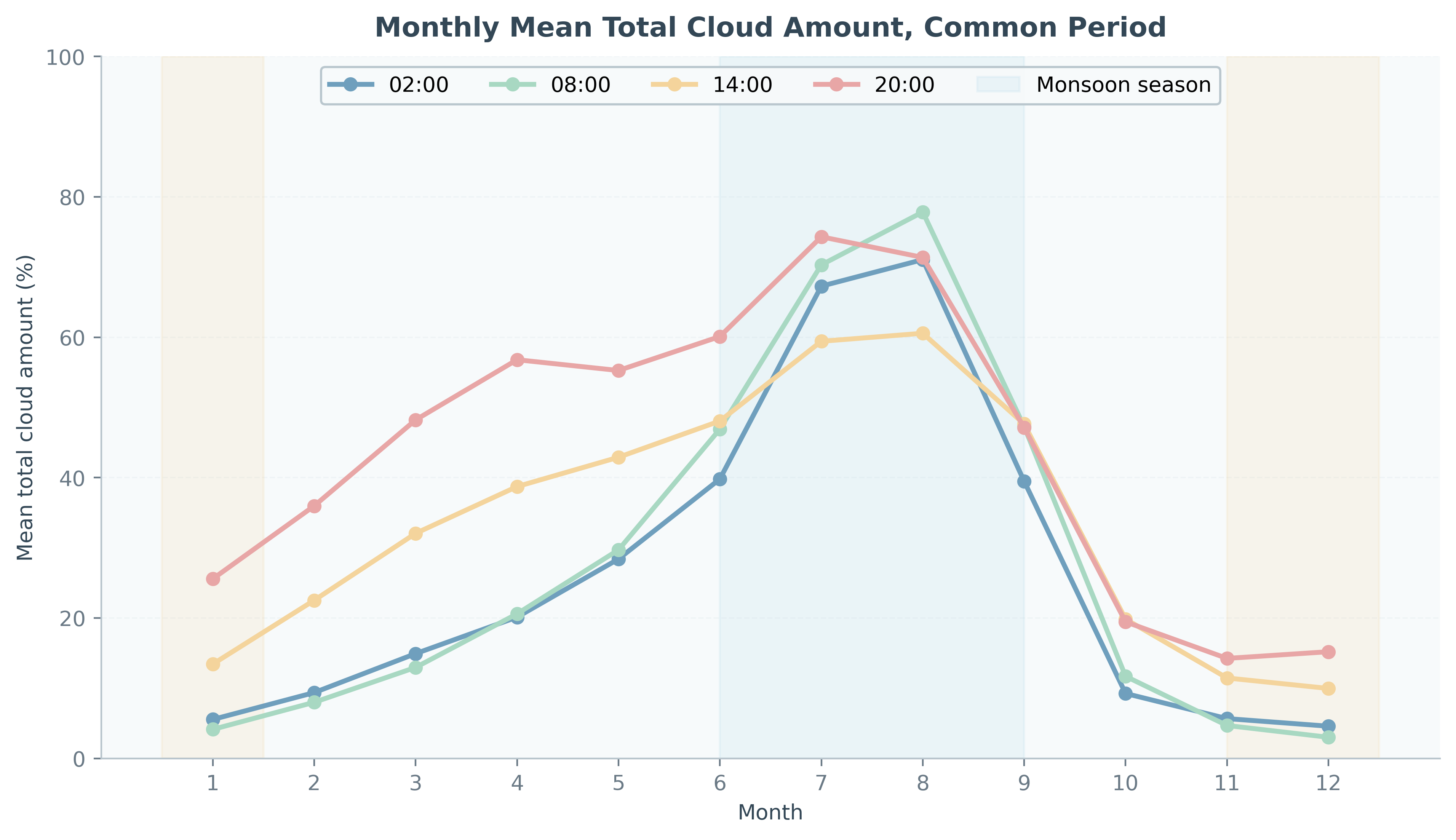}
\caption{Monthly mean total cloud amount from the common-period ground data. The curve shows the late-autumn/winter minimum and the summer maximum in the fixed-time conventional observations.}
\label{fig:5-2}
\end{figure}

The monthly mean ground cloud amount shows a seasonal cycle closely aligned with the satellite results (Figure~\ref{fig:5-2}). At 02:00, the mean cloud amount is 4.6\% in December, 5.6\% in January, and 5.7\% in November, then rises to 67.3\% in July and 71.1\% in August. At 08:00, the corresponding winter values are similarly low, with 3.0\% in December, 4.1\% in January, and 4.7\% in November, while July and August reach 70.3\% and 77.9\%. The 14:00 and 20:00 time series show higher non-monsoon cloud amounts, yet retain the same broad annual phase.

Figure~\ref{fig:5-2} establishes the phase of the ground-observed seasonal cycle. The meteorological-station record supplies a long-duration local climatological check on the satellite-inferred monthly structure. GOCCP identifies a low-cloud period from October to May and a cloud maximum in June--September; the ground station independently shows the same late-autumn to winter minimum and summer maximum. This agreement supports the ground-based consistency check on the satellite-inferred monthly cloud phase.

\subsection{Low-cloud Fraction at Four Fixed Observing Times}\label{low-cloud-fraction-and-fixed-time-diurnal-indications}

\begin{figure}[htbp]
\centering
\includegraphics[width=0.95\linewidth,keepaspectratio]{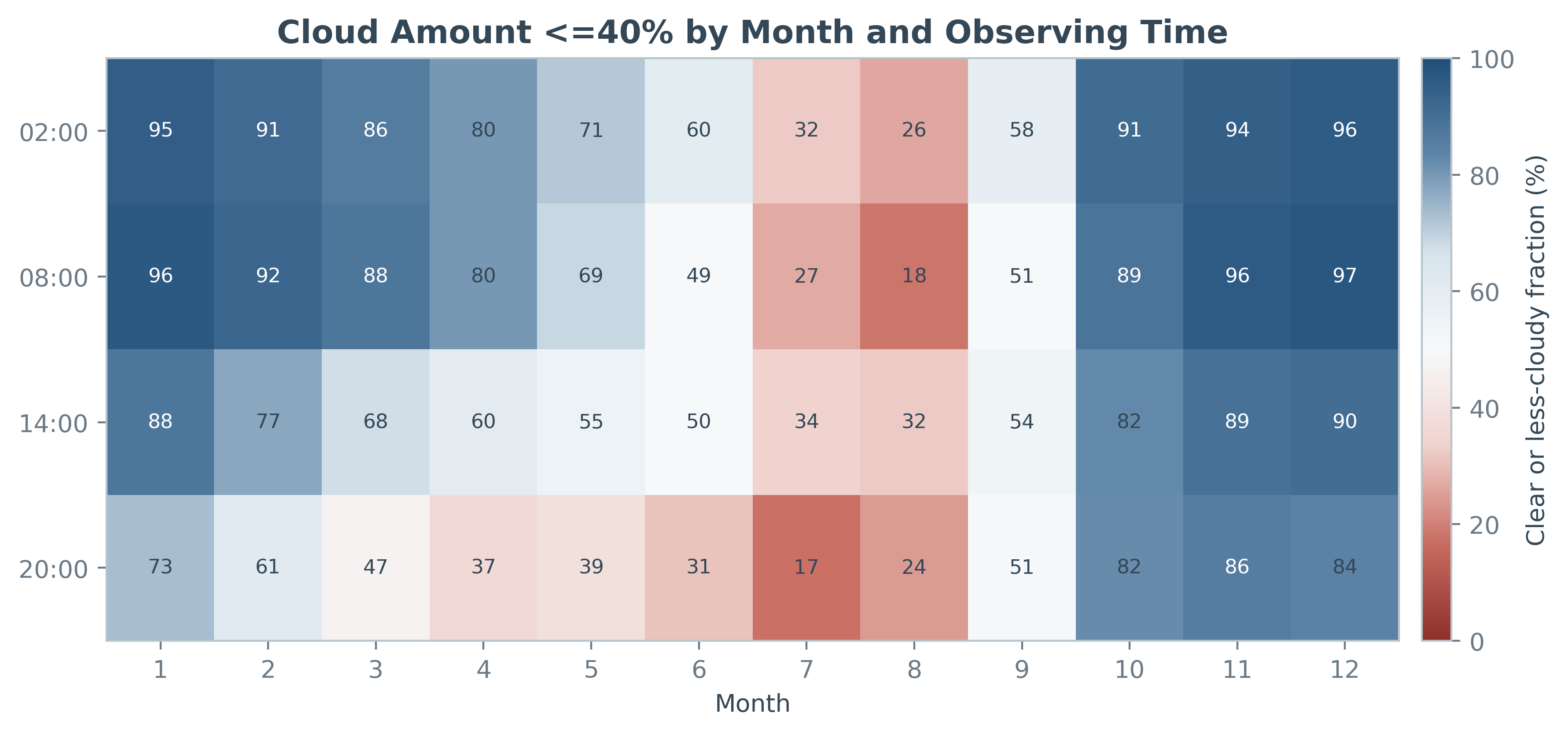}
\caption{Monthly fraction of ground observations with total cloud amount $\leq 40\%$ during the common period. This threshold is adopted as the ground clear-or-less-cloudy boundary for fixed-time conventional cloud observations. Blue colors indicate a higher clear or less-cloudy fraction on ground, while red colors indicate a lower fraction; the color scale spans 0--100\%.}
\label{fig:5-3}
\end{figure}

The thresholded statistic in Figure~\ref{fig:5-3} converts the mean seasonal cycle into the frequency of low-cloud observations. Using total cloud amount \(\leq 40\%\) as the conventional-cloud analogue of a clear or less-cloudy condition, the two night-time proxy samples show very high fractions during the November--January core and a sharp decline during the monsoon maximum. At 02:00 and 08:00, the November--January values are generally above 94\%, whereas July--August falls below about one-third at 02:00 and below 30\% at 08:00. This statistic is closer to a fixed-time cloud-availability metric than the monthly mean, while remaining limited to four fixed conventional observing times.

\begin{figure}[htbp]
\centering
\includegraphics[width=0.92\linewidth,keepaspectratio]{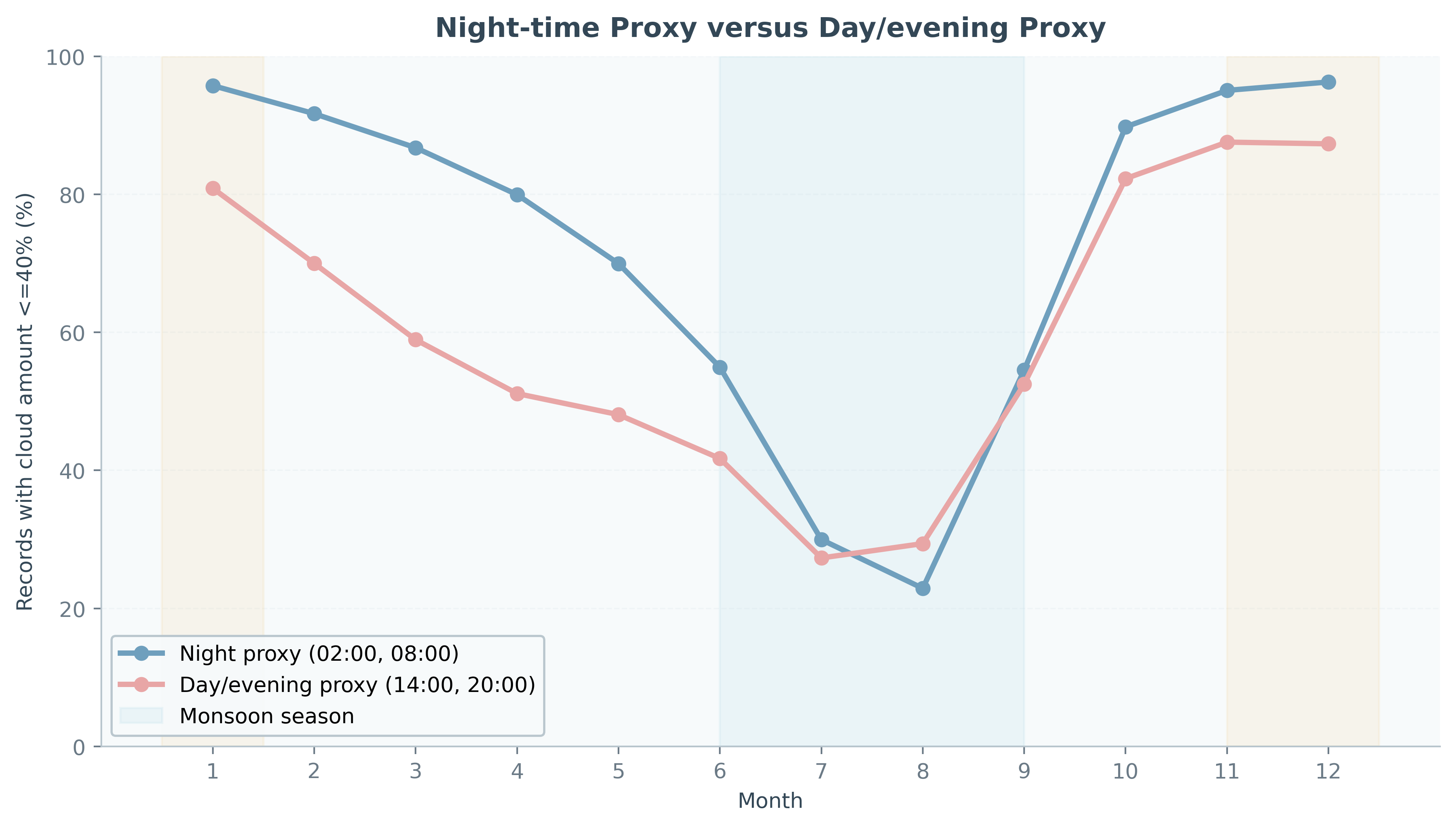}
\caption{Clear or less-cloudy fraction on ground for night-time proxy samples and day/evening proxy samples in the aligned 1988--2013 common period. The fraction is defined by total cloud amount $\leq 40\%$. The four fixed conventional observing times show a stronger night-side low-cloud signal during the non-monsoon months.}
\label{fig:5-4}
\end{figure}

The night--day comparison requires care for two reasons. First, conventional observing times do not coincide with astronomical night, which is commonly associated with the Sun being below \(-18^\circ\) \citep{NOAA2026SolarCalculatorGlossary}. Second, visual cloud estimates at night depend on available illumination from the Moon, stars, and artificial light; cloud cover can be underestimated when illumination is inadequate, especially for middle and high clouds \citep{WMO2017CloudAtlas,Hahn1995MoonlightCloudCover}. Based on four fixed observing times from the ground meteorological station, the analysis is therefore restricted to low-order evidence for the diurnal variation of cloud amount. It cannot measure continuous 24 h cloud evolution, night-time cloud-duration statistics, or cloud-gap structure, which require high-cadence sky monitoring. This restraint is consistent with Tibetan Plateau cloud studies showing that diurnal cloud behavior is physically meaningful but data-set dependent \citep{Shang2018TPDiurnalCloud,Zhao2023TPCloudDiurnal}. In these fixed-time data, 02:00 and 08:00 are treated as night-time proxy samples, while 14:00 and 20:00 are treated as day/evening proxy samples. The night-time proxy group has a higher clear or less-cloudy fraction on ground during most of the non-monsoon season, while the contrast weakens during summer when all four observing times are affected by monsoon cloud (Figure~\ref{fig:5-4}).

The paired 02:00--08:00 statistics provide an additional check on night-time persistence within the limits of the conventional data. Among days with valid values at both night-time proxy samples in the aligned period (\(N=7275\)), both fixed observations have cloud amount \(\leq 40\%\) in 72.8\% of cases. The paired fraction rises to 94.0\% during the November--January core and 83.9\% during the October--May low-cloud season, compared with 36.0\% during June--September. This paired statistic shows that the winter low-cloud signal is present in most paired conventional night-time proxy samples, providing fixed-time evidence for night-side persistence.

The ground data therefore connect the satellite seasonal phase to conventional local cloud observations. The low-cloud season is evident both in monthly satellite cloud fields and in conventional total-cloud observations at the local basin scale. The concentration of low-cloud conditions into a recurrent non-monsoon interval is relevant to observing operations, because continuous campaigns and instrument maintenance can be planned around a predictable annual structure.

\subsection{Interannual Recurrence of Low-cloud Months}\label{interannual-recurrence-of-the-low-cloud-season}

\begin{figure}[htbp]
\centering
\includegraphics[width=0.95\linewidth,keepaspectratio]{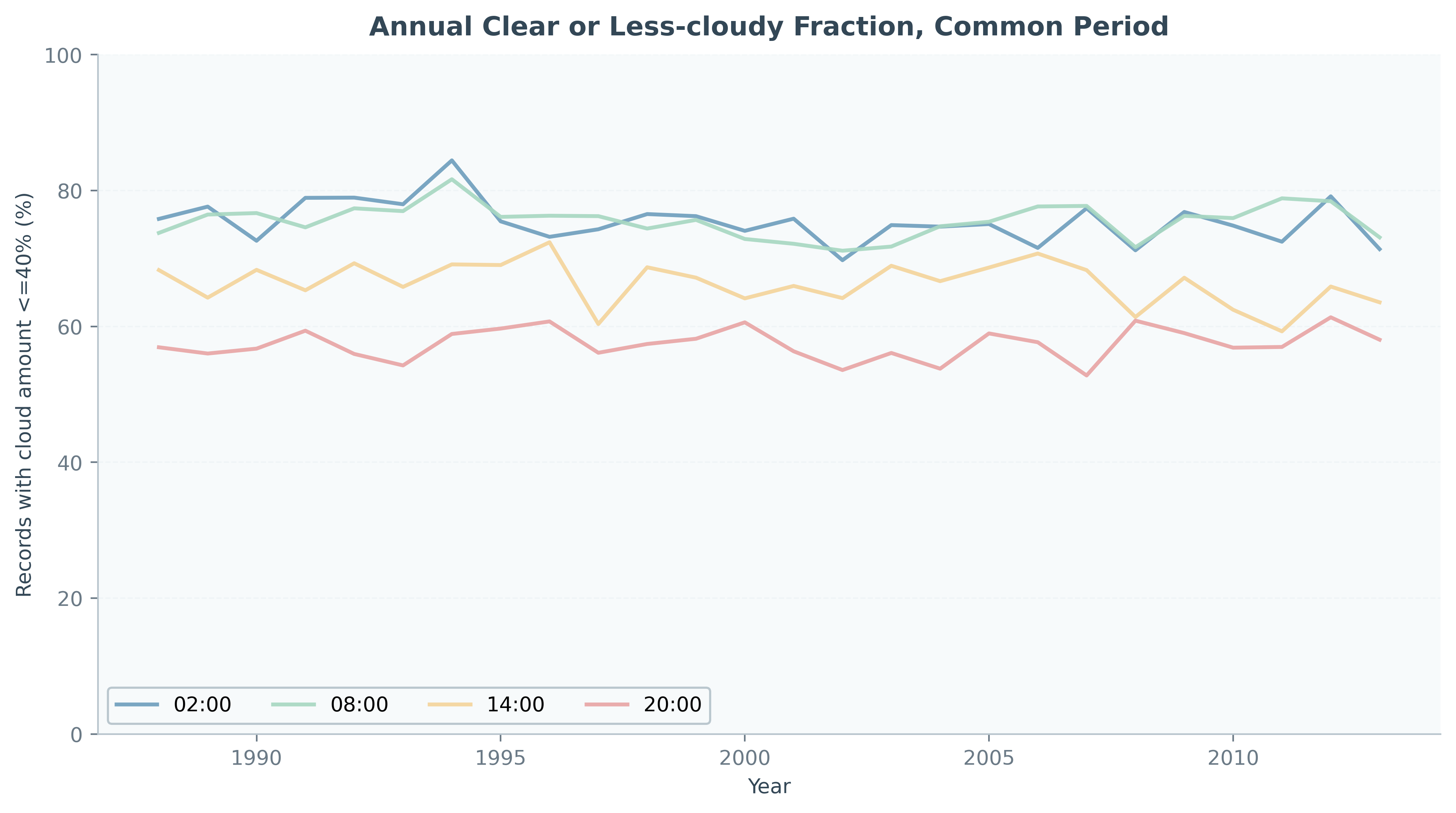}
\caption{Annual fraction of ground observations with total cloud amount $\leq 40\%$ during the common period. Interannual variability is present, while the non-monsoon low-cloud regime recurs throughout the aligned ground record.}
\label{fig:5-5}
\end{figure}

The annual clear-or-less-cloudy fraction confirms that the low-cloud season recurs throughout the aligned ground data (Figure~\ref{fig:5-5}). Interannual variability is evident, especially during monsoon-affected months, but the non-monsoon low-cloud regime remains the stable component of the data set. Aggregated over all four observing times, the October--May fraction is 80.7\%, and the November--January core is 90.7\%. By contrast, the June--September fraction is 39.9\%.

The low-cloud season appears repeatedly in the multi-decade ground data and has the same broad timing as the satellite products. The multi-decade ground data therefore establish the recurrence of the Shigatse low-cloud season and provide the local climatological basis for translating the satellite result into site-testing priorities.

\section{Multi-source Consistency and Local Meteorological Conditions}\label{consistency-among-cloud-data-sets-and-observing-conditions}

The three cloud data sets used in this paper differ in retrieval physics, spatial scales, and observing practices. GOCCP provides a coarse but homogeneous active-lidar climatology. ISCCP provides finer passive-satellite spatial sampling with known product-dependent behavior over high terrain. The Shigatse meteorological-station cloud data provide a long conventional total-cloud-amount series from the local basin environment. The comparison keeps the native support of each data set explicit: GOCCP defines the regional active-lidar cloud-cover climatology, ISCCP checks finer spatial placement and monthly phase, and the meteorological-station data test whether a corresponding seasonal structure appears in a local long-term observation. The on-site Weather Station subsequently provides a local meteorological-condition layer for observing conditions.

The analysis first compares the monthly phase and month-group structure of cloud amount, and then examines the local meteorological conditions during the low-cloud months. The satellite products cover 2007--2016 monthly climatological fields, the aligned ground-based cloud analysis uses the 1988--2013 common period, and the Weather Station archive samples 2024--2025 local surface conditions. This separation allows the section to test three quantities at their proper support: satellite phase agreement, ground-based seasonal recurrence, and local meteorological conditions during the low-cloud months.

\subsection{GOCCP--ISCCP Monthly Phase and Product Differences}\label{goccpisccp-monthly-phase-and-product-dependent-offset}

\begin{figure}[htbp]
\centering
\includegraphics[width=\linewidth,keepaspectratio]{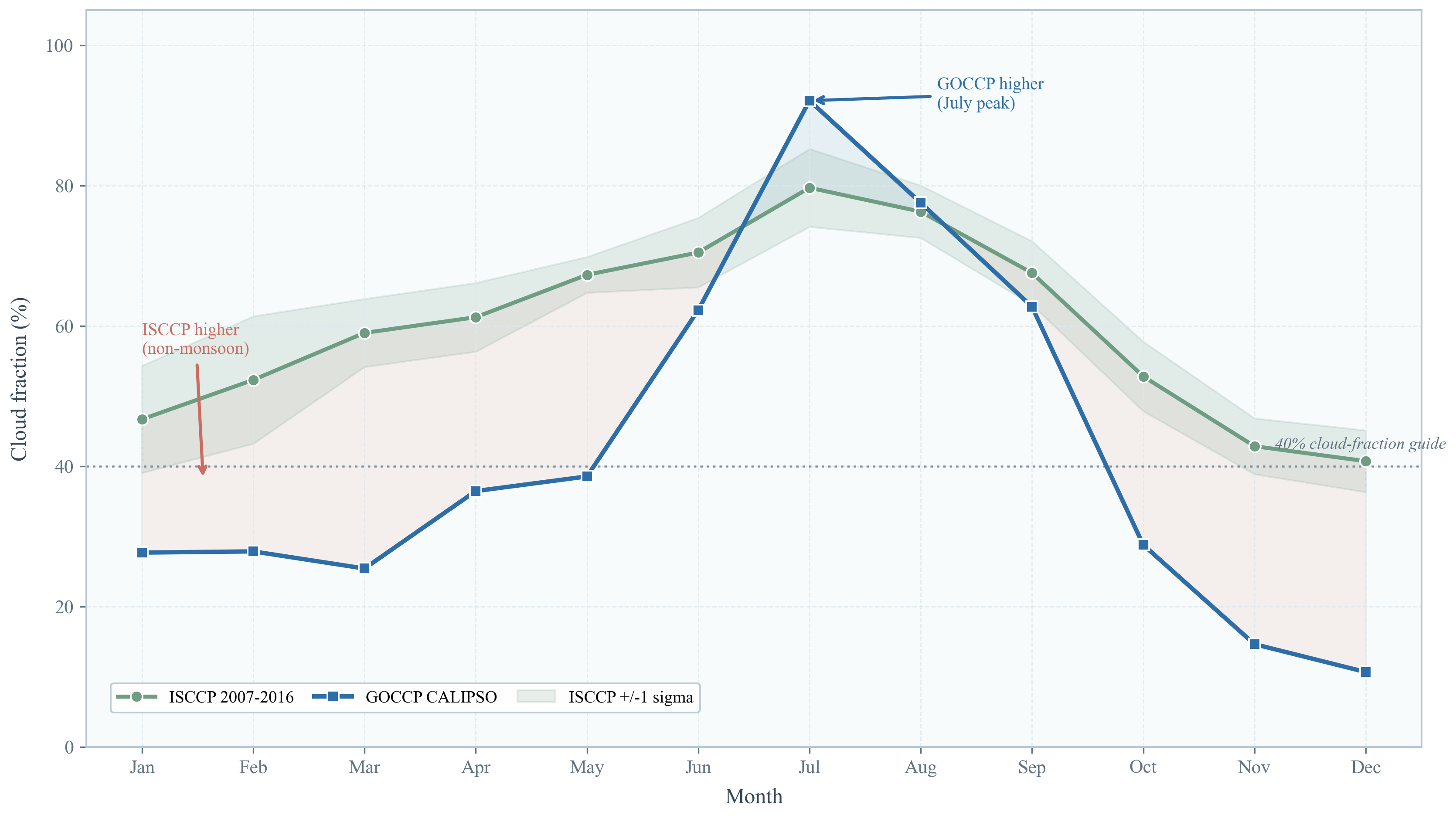}
\caption{Monthly GOCCP--ISCCP cloud-fraction comparison for Shigatse, regenerated from the original monthly NetCDF products. The ISCCP curve includes its interannual $1\sigma$ envelope; the shaded gap between the curves marks the product-dependent monthly offset, and the 40\% cloud-fraction guide is shown for reference. Both products show a late-autumn/winter cloud minimum and a summer cloud maximum.}
\label{fig:6-1}
\end{figure}

GOCCP and ISCCP exhibit comparable seasonal phases in both the lower-cloud and higher-cloud parts of the year (Figure~\ref{fig:6-1}). Both products give their lowest cloud fractions in the late-autumn to winter interval and their highest values in July and August. The four-site monthly summaries in Section~\ref{shigatse-in-the-cloud-climatology-context-of-selected-chinese-sites} provide the within-product perspective for this interpretation: Shigatse is evaluated through the timing and continuity of its low-cloud months, with product-specific annual values retained for reference. The repeated monthly ordering is the satellite result least affected by the product-dependent absolute offset because the two products have different cloud definitions, spatial averaging, sensor physics, and sampling histories \citep{Stubenrauch2013GlobalCloudDatasets,Young2018ISCCPHSeries}.

\begin{table}[htbp]
\centering
\caption{Monthly Satellite Cloud-fraction Values for the Shigatse Grid Cell. The values are the same as those labeled in Figures~\ref{fig:3-2} and \ref{fig:4-2} and provide the numerical basis for the GOCCP--ISCCP monthly phase comparison.}
\label{tab:6-monthly-satellite-cf}
\raatablestyle
\setlength{\tabcolsep}{4.4pt}
\begin{tabular}{lrrrrrr}
\hline
Product & Jan & Feb & Mar & Apr & May & Jun \\
\hline
GOCCP CF (\%) & 27.7 & 27.9 & 25.4 & 36.5 & 38.5 & 62.3 \\
ISCCP CF (\%) & 46.7 & 52.3 & 59.0 & 61.2 & 67.3 & 70.5 \\
\hline
Product & Jul & Aug & Sep & Oct & Nov & Dec \\
\hline
GOCCP CF (\%) & 92.1 & 77.5 & 62.7 & 28.8 & 14.6 & 10.7 \\
ISCCP CF (\%) & 79.7 & 76.3 & 67.5 & 52.8 & 42.9 & 40.7 \\
\hline
\end{tabular}
\end{table}

The absolute values differ substantially. In the Shigatse extraction, GOCCP gives an annual cloud fraction of 42.1\%, while ISCCP gives 59.7\%. The difference is largest during the non-monsoon season, when snow or bright highland surfaces and weak thermal contrast can affect passive cloud detection over the plateau \citep{NaudChen2010ISCCPTP,Li2006TPCloudTypes,Liu2021TPSatelliteCloud,Wu2024TPCloudReview}. This offset is treated as a product-dependent difference tied to measurement physics and spatial support. The main satellite result is that both products show a comparable seasonal phase despite their different absolute scales.

\subsection{Spatial Correspondence between GOCCP and ISCCP}\label{spatial-scale-consistency}

Spatial correspondence is evaluated relative to the spatial support of each product. The GOCCP fields used here are reported on a \(2^\circ \times 2^\circ\) latitude--longitude grid derived from CALIPSO/CALIOP active-lidar cloud occurrence \citep{Chepfer2010GOCCP,Winker2010CALIPSO}. At the latitude of Shigatse (\(\sim 29.2^\circ\) N), this grid spacing corresponds to approximately 223 km in the north--south direction and 194 km in the east--west direction; a GOCCP value is therefore a regional cloud-climatology estimate for the southern Tibetan Plateau around Shigatse. The ISCCP HXG product used here is a passive multi-satellite visible--infrared cloud data record on a \(0.1^\circ \times 0.1^\circ\) grid \citep{Rossow1999ISCCP,Young2018ISCCPHSeries}, corresponding near Shigatse to approximately 11.1 km by 9.7 km. The ISCCP grid cell area is thus about 1/400 of the GOCCP grid-cell area. This scale separation is central to the interpretation: GOCCP is used to assess whether Shigatse lies inside a broad active-lidar low-cloud regime, whereas ISCCP is used to examine whether the site remains in a comparable lower-cloud part of the regional pattern at finer passive-satellite spacing.

The two products also differ in measurement physics. GOCCP is based on nadir active-lidar profiles and is sensitive to optically thin cloud layers and vertical cloud occurrence, while its coarse grid and orbit sampling smooth ridge-scale and basin-scale contrasts. ISCCP provides much denser horizontal coverage from passive radiances, but over the Tibetan Plateau its absolute cloud fraction can be affected by snow cover, high surface albedo, complex terrain, and weak cloud--surface thermal contrast \citep{Stubenrauch2013GlobalCloudDatasets,NaudChen2010ISCCPTP,Liu2021TPSatelliteCloud,Wu2024TPCloudReview}. The comparison therefore tests whether a broad active-lidar seasonal signal and a finer-grid passive-satellite spatial placement support a consistent climatological interpretation.

The processing strategy follows this definition. The monthly GOCCP and ISCCP fields are shown on their native grids in the regional maps, using the valid nearest-grid-cell extraction for Shigatse and the surrounding fields described in Sections~\ref{goccp-cloud-climatology} and \ref{isccp-cloud-climatology}. For the quantitative spatial test, ISCCP is area-averaged to the GOCCP grid over the western-China map domain, and the paired fields are evaluated with Spearman spatial rank correlation and lower-tercile low-cloud-area overlap. This preserves the product-scale interpretation while testing whether the two products place lower- and higher-cloud regions in comparable parts of the map. The approach follows the general practice of satellite cloud-product intercomparison and threshold-based spatial verification \citep{Stubenrauch2013GlobalCloudDatasets,NaudChen2010ISCCPTP,Liu2021TPSatelliteCloud,Ebert2008FuzzyVerification,Ahijevych2009SpatialVerification}. Appendix~\ref{appendix-spatial-correspondence-diagnostics} gives the full monthly correspondence diagnostics and the definitions of the low-cloud-area overlap metrics. The within-product Hovmöller diagnostics in Figures~\ref{fig:3-5} and \ref{fig:4-5} provide the regional basis for this comparison: GOCCP gives the active-lidar view of the broad seasonal displacement, while ISCCP shows a comparable seasonal phase and local spatial placement at finer grid spacing.

\begin{figure}[htbp]
\centering
\includegraphics[width=\linewidth,keepaspectratio]{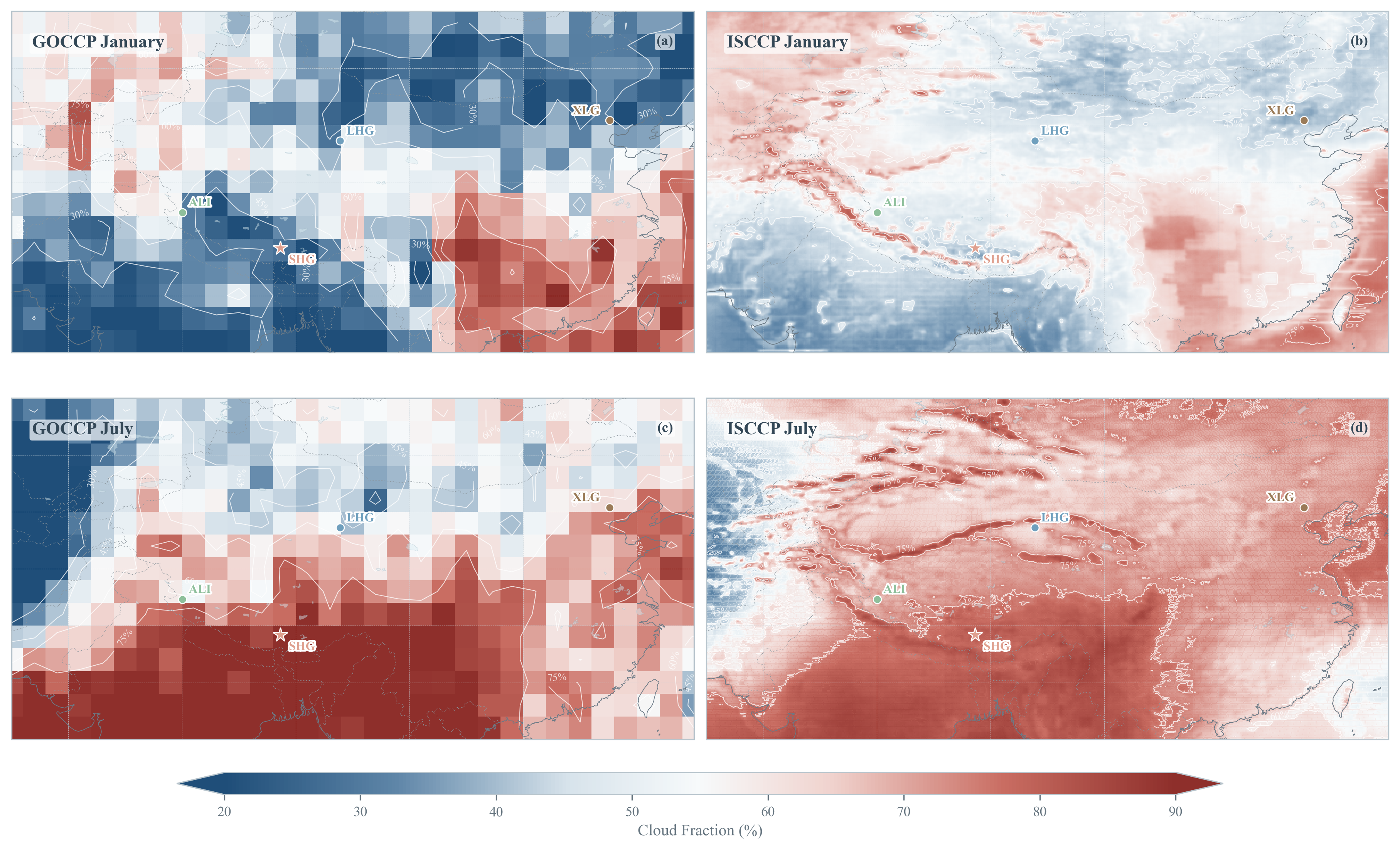}
\caption{Joint GOCCP and ISCCP spatial comparison for January and July. The panels show the contrast between the non-monsoon low-cloud environment and the monsoon-season cloud maximum. Site codes: SHG = Shigatse, LHG = Lenghu, ALI = Ali, and XLG = Xinglong.}
\label{fig:6-2}
\end{figure}

The January and July maps give the clearest spatial test (Figure~\ref{fig:6-2}). In January, both products place Shigatse within the clearer non-monsoon environment of the southern plateau, although their absolute cloud fractions differ. In July, both products show the expansion of monsoon cloud into the Shigatse region. After ISCCP is averaged to the GOCCP grid, the Spearman spatial rank correlation is 0.892 in January and 0.791 in July. The product-specific lower-tercile low-cloud masks have Jaccard overlaps of 0.632 and 0.469, respectively, and ISCCP recovers 77.4\% and 63.9\% of the GOCCP lower-tercile cells in those two months. Across the October--May low-cloud period, the median rank correlation is 0.846 and the median lower-tercile Jaccard overlap is 0.627. These values quantify the visual correspondence in Figure~\ref{fig:6-2}: the two products differ in absolute cloud fraction, but they preserve a common spatial ordering and a common placement of the low-cloud region. The cross-product comparison therefore supports product-scale spatial correspondence of the Shigatse low-cloud regime. In the standard hierarchy of astronomical site testing, regional satellite cloud-cover climatology identifies candidate low-cloud regimes, and local monitoring with all-sky cameras and co-located site instruments then converts those regimes into night-by-night usable-time statistics \citep{Schoeck2009TMTOverview,Skidmore2008TMTASC,Varela2014ELTGroundMeteorology}. The previous total-sky-image study in the Shigatse region also shows why local monitoring is needed: high-cadence sky images can resolve site-scale cloud evolution and seasonal cloud behavior \citep{Yang2018ShigatseSkyImages}. Continued ground monitoring can now translate this product-scale correspondence into ridge-scale, 24 h, and night-by-night observing statistics.

\subsection{Ground-based Comparison of the Satellite Seasonal Phase}\label{ground-based-consistency-check-of-the-satellite-seasonal-phase}

The long conventional cloud-amount data provide a ground-based consistency check on the satellite-defined seasonal cycle. Within the present data set, the meteorological-station observations supply the available multi-decade, ground-based cloud evidence for the Shigatse region. The comparison target is monthly recurrence and seasonal phase across data sets with different definitions, observing times, and spatial support. GOCCP and ISCCP cover 2007--2016 in the satellite analysis, while the aligned meteorological-station cloud data cover 1988--2013.

\begin{figure}[htbp]
\centering
\includegraphics[width=0.92\linewidth,keepaspectratio]{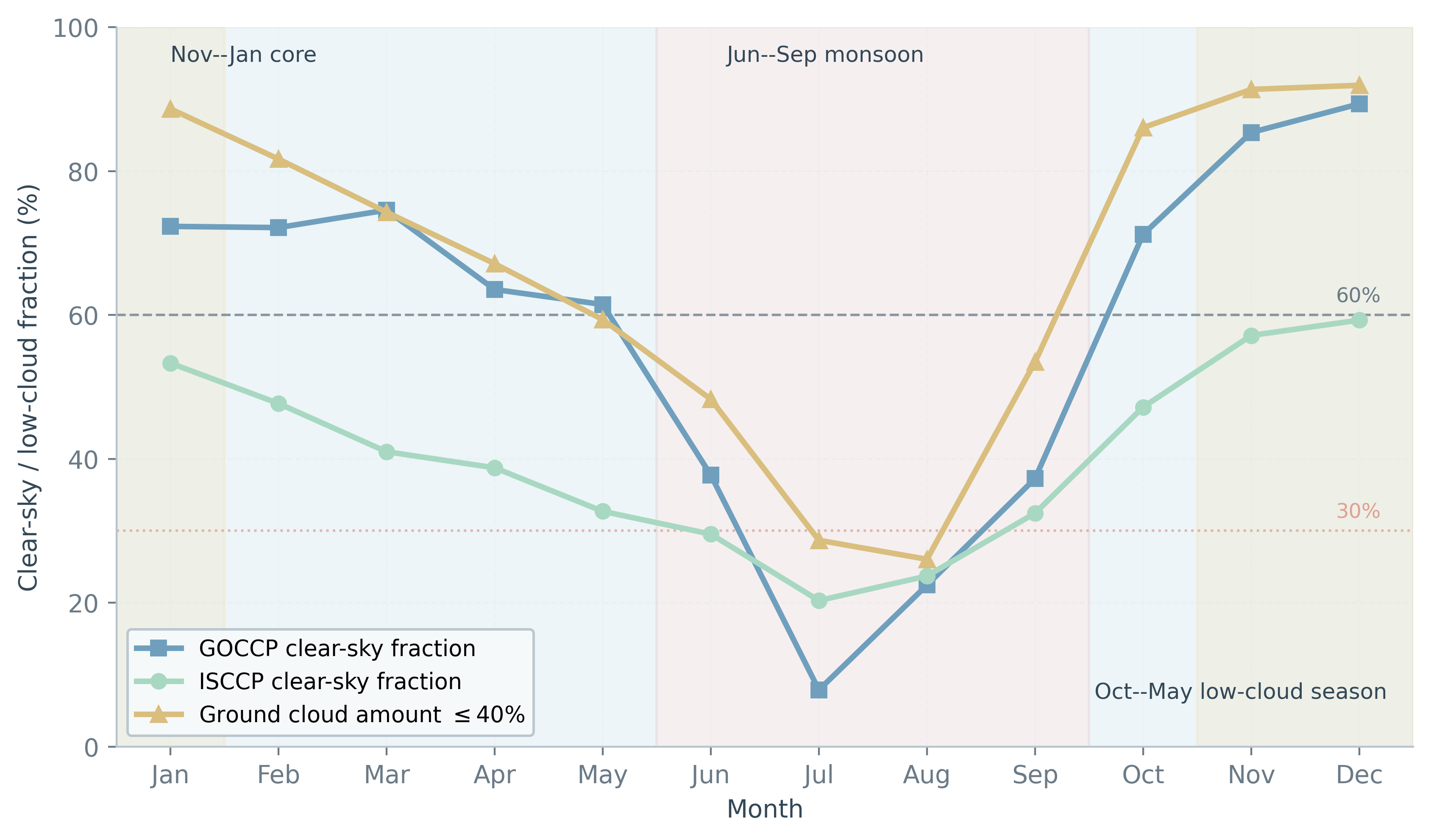}
\caption{Monthly ground-based consistency check of the satellite seasonal cycle at Shigatse. GOCCP and ISCCP are shown as clear-sky fractions, while the ground curve is the fraction of aligned conventional meteorological-station observations with total cloud amount $\leq 40\%$, weighted across the four fixed observing times. The three curves show the same late-autumn/winter low-cloud phase and summer cloud maximum.}
\label{fig:6-3}
\end{figure}

The comparison links the satellite-derived seasonal phase to the independent fixed-time ground cloud record (Figure~\ref{fig:6-3}; Table~\ref{tab:6-1}). All three data sets show the same annual cycle: lower cloud amount during the non-monsoon months and a pronounced cloud maximum in June--September. GOCCP has lower cloud fraction during October--May and higher values in June--September; ISCCP shows the corresponding non-monsoon minimum and monsoon maximum; and the meteorological-station data show lower total cloud amount and a higher clear or less-cloudy fraction on ground in late autumn and winter, followed by a strong summer maximum.

The agreement is also quantitative. For total cloud amount, the monthly ground mean correlates with GOCCP and ISCCP cloud fraction with Pearson \(r=0.96\) and \(0.97\), respectively; the corresponding Spearman \(\rho\) values are 0.94 and 0.99. For the complementary clear-sky metric, the clear or less-cloudy fraction on ground, defined by total cloud amount \(\leq 40\%\), correlates with GOCCP and ISCCP clear-sky fractions with Pearson \(r=0.96\) and \(0.97\). The standardized monthly-cycle RMSEs are 0.28--0.29 for the GOCCP--ground comparisons and 0.24--0.25 for the ISCCP--ground comparisons. These statistics measure similarity in normalized monthly phase among records with different spatial support, observing times, and cloud definitions; they are not conversion coefficients between products. In the aligned ground-based data, the fraction of fixed-time observations with total cloud amount \(\leq 40\%\) is 90.7\% in November--January, 80.7\% in October--May, and 39.9\% in June--September. The satellite and ground-based records therefore identify the same recurrent non-monsoon low-cloud season at Shigatse.

The same comparison was repeated for the 2007--2013 subset of the meteorological-station record, which is the portion closest in time to the 2007--2016 satellite climatologies. The subset retains the same monthly phase: the ground mean total-cloud amount is lowest in November and December, remains low in January, and reaches its maximum in July--August. Quantitatively, the 2007--2013 monthly ground mean cloud amount correlates with GOCCP and ISCCP cloud fraction with Pearson \(r=0.967\) and \(0.980\), respectively. The corresponding comparison between the ground \(\leq 40\%\) fraction and satellite clear-sky fraction gives Pearson \(r=0.966\) for GOCCP and \(0.976\) for ISCCP. Thus, the ground--satellite seasonal agreement is present both in the full aligned 1988--2013 record and in the satellite-era subset.

\begin{figure}[htbp]
\centering
\includegraphics[width=0.98\linewidth,keepaspectratio]{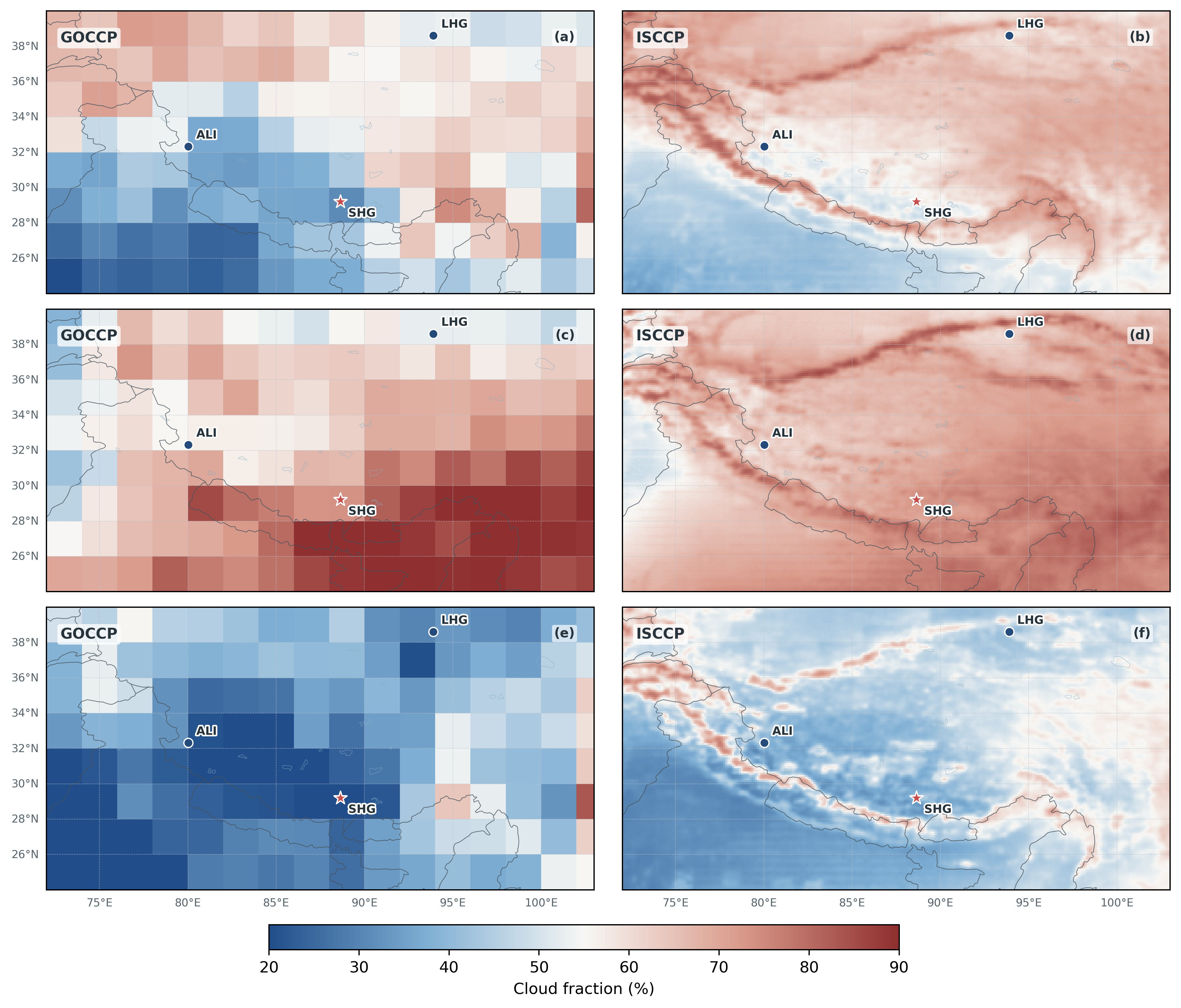}
\caption{Satellite cloud-fraction spatial patterns for the Shigatse-defined month groups. Rows show the January--May low-cloud period, the June--September monsoon-cloud period, and the October--December return to the low-cloud regime; columns show GOCCP and ISCCP. The grouping displays the low--high--low annual sequence relevant to the Shigatse cloud-cover interpretation. Site codes: SHG = Shigatse, ALI = Ali, and LHG = Lenghu.}
\label{fig:6-season-spatial}
\end{figure}

The same seasonal phase is spatially visible when the monthly satellite fields are grouped according to the Shigatse cloud regime rather than by the conventional four seasons (Figure~\ref{fig:6-season-spatial}). The January--May group represents the low-cloud period at the beginning of the calendar year, June--September isolates the monsoon-cloud period, and October--December shows the return to the low-cloud regime. GOCCP and ISCCP differ in absolute cloud fraction and spatial texture, but both products place Shigatse in the same low--high--low annual sequence. This figure replaces the conventional four-season summaries in the main text and gives the spatial counterpart to the monthly phase comparison.

\begin{table}[htbp]
\centering
\caption{Month-group Cloud Diagnostics for Shigatse. CF denotes cloud fraction. The satellite values are monthly climatological means for the Shigatse grid cell in each product. Ground mean cloud is the weighted mean total cloud amount across the four conventional observing times in the aligned 1988--2013 common period, and ground $\leq 40\%$ gives the corresponding clear or less-cloudy fraction on ground.}
\label{tab:6-1}
\raatablestyle
\setlength{\tabcolsep}{4.8pt}
\begin{tabular*}{\linewidth}{@{\extracolsep{\fill}}llrrrr@{}}
\hline
Month group & Months & GOCCP CF (\%) & ISCCP CF (\%) & Ground mean (\%) & Ground $\leq 40\%$ (\%) \\
\hline
Annual & All months & 42.1 & 59.7 & 29.7 & 69.1 \\
Nov--Jan core & Nov--Jan & 17.7 & 43.4 & 9.6 & 90.7 \\
Oct--May season & Oct--May & 26.3 & 52.9 & 19.0 & 80.7 \\
Jun--Sep cloudy & Jun--Sep & 73.7 & 73.5 & 56.8 & 39.9 \\
\hline
\end{tabular*}
\end{table}

\begin{table}[htbp]
\centering
\caption{Ground-cloud Threshold Sensitivity for the Aligned 1988--2013 Common Period. Values are fractions of valid four-time conventional cloud observations whose total cloud amount is at or below the listed threshold.}
\label{tab:6-threshold-sensitivity}
\raatablestyle
\setlength{\tabcolsep}{4.6pt}
\begin{tabular*}{\linewidth}{@{\extracolsep{\fill}}lrrrrr@{}}
\hline
Month group & \(N\) & Mean cloud (\%) & \(\leq 30\%\) & \(\leq 40\%\) & \(\leq 50\%\) \\
\hline
Annual & 31512 & 29.7 & 64.6 & 69.1 & 72.0 \\
Nov--Jan core & 9091 & 9.6 & 88.9 & 90.7 & 91.7 \\
Oct--May low-cloud & 22562 & 19.0 & 77.3 & 80.7 & 82.9 \\
Jun--Sep monsoon & 8950 & 56.8 & 32.8 & 39.9 & 44.6 \\
\hline
\end{tabular*}
\end{table}

Because the ground record contains four fixed observing times, it also constrains the monthly recurrence and low-order night--day behavior of the low-cloud regime. For the two conventional night-time proxy observations, 02:00 and 08:00, the clear or less-cloudy fractions on ground in November--January are generally near or above 94\% at each month, and paired 02:00--08:00 occurrence reaches 94.0\% in the November--January core and 83.9\% in the October--May low-cloud season. This paired statistic provides a fixed-time persistence indicator: the winter low-cloud condition is present in most paired conventional night-time proxy observations. Aggregated across all four fixed observing times, the clear or less-cloudy fraction on ground is 90.7\% for November--January and 80.7\% for October--May, compared with 39.9\% for June--September (Table~\ref{tab:6-1}). The threshold-sensitivity check in Table~\ref{tab:6-threshold-sensitivity} shows that this month-group ordering is stable under 30\%, 40\%, and 50\% total-cloud-amount thresholds. The 40\% threshold is retained for consistency with the cloud-cover class adopted above, while the low-cloud season itself does not depend on that single numerical choice.

This meteorological-station view complements the Weather Station assessment in the next section. The conventional cloud observations show that the low-cloud months recur over a long local record, while the Weather Station data quantify local meteorological conditions within the same calendar month groups. The fixed-time ground data provide low-order information on night--day behavior, and high-cadence sky monitoring can supply the continuous 24 h cloud evolution, night-time cloud-duration statistics, and cloud-gap structure. The fixed-time cloud record ties the satellite seasonal phase to a multi-decade local ground observation.

\subsection{Local Meteorological-condition Statistics during the Satellite-defined Low-cloud Season}\label{local-meteorological-assessment-during-the-satellite-defined-low-cloud-season}

The Weather Station archive was evaluated over the same calendar month groups to quantify local wind speed, relative humidity, dew-point depression, and precipitation at the Shigatse site. These statistics report how often valid 10-minute surface samples in the satellite-defined low-cloud months satisfy the adopted local meteorological criteria.

\begin{figure}[htbp]
\centering
\includegraphics[width=\linewidth,keepaspectratio]{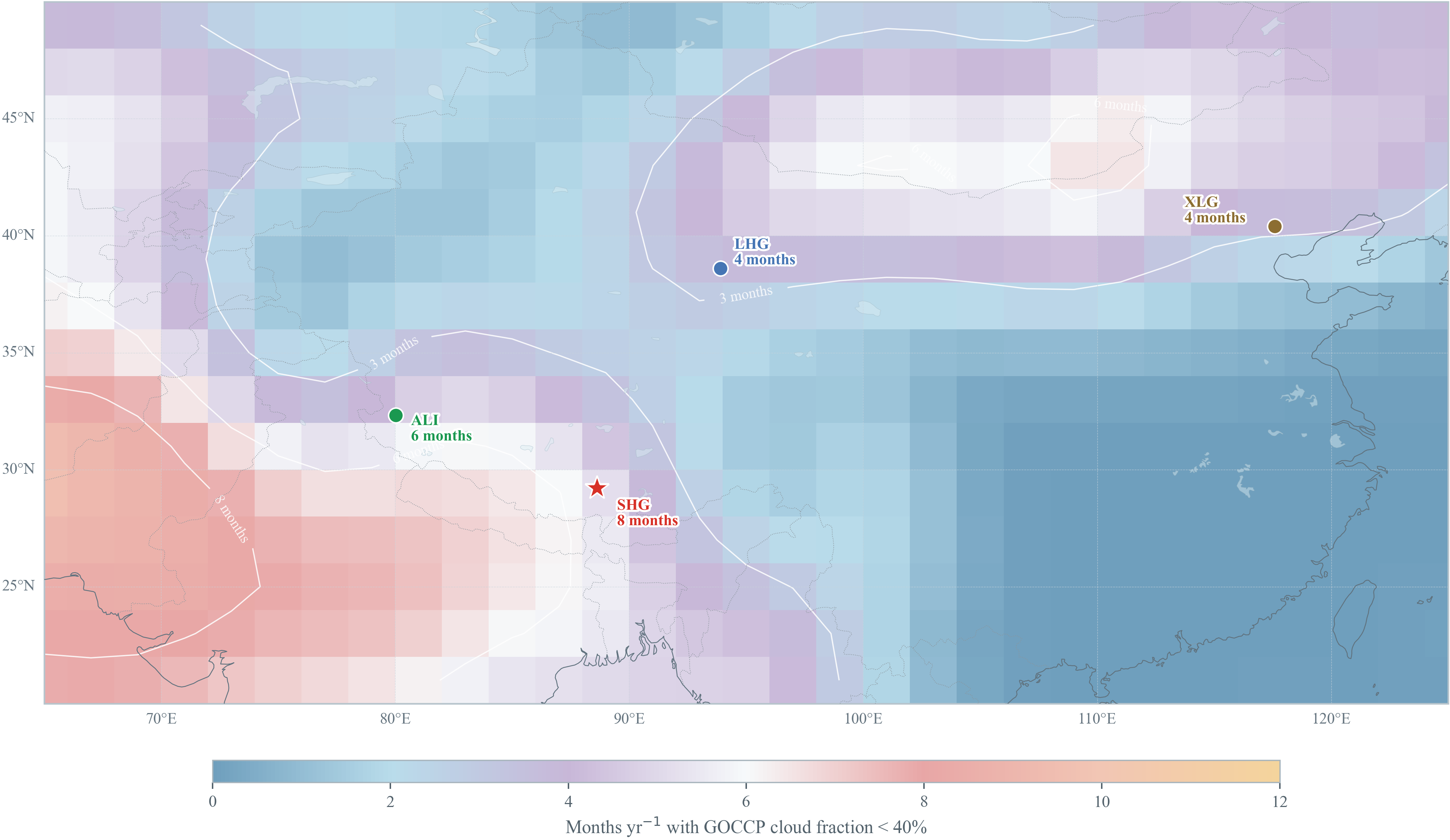}
\caption{Satellite-defined low-cloud month count in the cropped regional cloud-climatology map. A counted month has a GOCCP cloud fraction below 40\% in the monthly grid-cell climatology. Site codes and the corresponding month counts are labeled directly for Shigatse, Ali, Lenghu, and Xinglong.}
\label{fig:6-5}
\end{figure}

The month groups in this assessment are anchored to GOCCP because GOCCP provides the active-lidar regional cloud map used in this paper. This GOCCP anchoring is consistent with the shared seasonal phase shown in Section~\ref{ground-based-consistency-check-of-the-satellite-seasonal-phase}: ISCCP and the ground data show the same seasonal phase, while GOCCP offers a single, physically homogeneous active-lidar basis for the regional month-count map. A month is counted when the monthly GOCCP cloud fraction is below 40\%, the less-cloudy guide level adopted in recent western-China cloud-cover studies \citep{Qian2024AliCloud,Li2024LenghuCloud}. At the Shigatse grid cell, this criterion gives eight low-cloud months, corresponding to October--May, with November--January as the clearest core (Figure~\ref{fig:6-5}). The map is used as a regional cloud-climatology diagnostic, following the role of long-term satellite cloud statistics in western-China site testing \citep{Cao2020ClearNights}; it identifies candidate calendar periods for local meteorological assessment.

The adopted Weather Station condition classes are defined in Table~\ref{tab:6-2}. They use ESO Paranal/VLT wind, humidity, and dew-point limits as reference meteorological criteria and express them as Shigatse surface-meteorological classes for site testing \citep{ESOParanalAtTelescope}. Measurable hourly precipitation is added as a hard rejection term, because an optical telescope cannot operate safely during rain or snowfall even if the wind, humidity, and dew-point-depression thresholds are otherwise satisfied. Samples in the Good-condition class satisfy the VLT-based wind, humidity, and dew-point criteria, have no measurable hourly precipitation, and have wind speed \(V\leq 12\) m/s. Samples in the Normal-condition class also satisfy the hard rejection criteria but have \(12 < V \leq 18\) m/s, corresponding to the VLT wind-speed range associated with pointing-limited operation. Rejected samples have wind speed \(>18\) m/s, RH \(>70\%\), \(T-T_{\rm d}\leq 1.5~^\circ\)C, or hourly precipitation \(>0\). Telescope-surface temperature, enclosure state, cloud alarms, and instrument-specific limits are outside the Weather Station archive. These thresholds are stricter than the high-risk occurrence statistics used in the Muztagh-ata site-testing paper, where relative humidity \(>90\%\) and wind speed \(>15\) m/s were used to quantify humidity and strong-wind constraints \citep{Xu2020MuztaghMeteoSky}. The Muztagh-ata comparison shows that wind, humidity, and dew-point proximity are established meteorological diagnostics in high-altitude astronomical site testing, while the VLT-anchored criteria and the added precipitation rejection provide the conservative surface-meteorological condition classes adopted for the present Shigatse analysis.

\begin{table}[htbp]
\centering
\caption{VLT-anchored Meteorological-condition Classes Adopted for the Weather Station Statistics in This Paper. Here $V$ is the sustained 10 m wind speed, RH is relative humidity, and $T-T_{\rm d}$ is ambient-air dew-point depression. Hourly precipitation $>0$ is added as a hard local rejection term for optical observing.}
\label{tab:6-2}
\raatablestyle
\setlength{\tabcolsep}{3.0pt}
\begin{tabular}{@{}p{0.12\linewidth}p{0.38\linewidth}p{0.47\linewidth}@{}}
\hline
Condition class & Criterion applied to Weather Station data & Use in this paper \\
\hline
Good & $V\leq 12$ m/s, RH $\leq 70\%$, $T-T_{\rm d}>1.5~^\circ$C, and hourly precipitation $=0$ & Defines samples satisfying all adopted criteria with wind speed at or below 12 m/s. \\
Normal & $12 < V \leq 18$ m/s, RH $\leq 70\%$, $T-T_{\rm d}>1.5~^\circ$C, and hourly precipitation $=0$ & Defines the wind-limited local meteorological condition class, following the VLT wind-speed range associated with pointing-limited operation. \\
Rejected & Wind speed $>18$ m/s, or RH $>70\%$, or $T-T_{\rm d}\leq 1.5~^\circ$C, or hourly precipitation $>0$ & Defines samples excluded from the Good- and Normal-condition classes. \\
\hline
\end{tabular}
\end{table}

\newpage

\begin{table}[htbp]
\centering
\caption{Night-time Weather Station Meteorological-condition Statistics under the Adopted VLT-anchored Condition Classes. The night sample uses the fixed 20:00--06:00 local-time proxy. \(N\) is the number of valid 10-minute samples; all other columns are percentages. The diagnostic rejection-trigger columns are not mutually exclusive.}
\label{tab:6-3}
\raatablestyle
\setlength{\tabcolsep}{2.0pt}
\begin{tabular*}{\linewidth}{@{\extracolsep{\fill}}lrrrrrrr@{}}
\hline
Month group & $N$ & Good & Normal & RH $>70\%$ & $T-T_{\rm d}\leq 1.5~^\circ$C & Wind $>18$ m/s & Precip. $>0$ \\
\hline
Annual sample & 41045 & 82.2 & 0.0 & 17.4 & 3.0 & 0.0 & 1.3 \\
Nov--Jan core & 9571 & 96.2 & 0.0 & 2.8 & 1.3 & 0.0 & 1.4 \\
Oct--May low-cloud & 26817 & 92.6 & 0.0 & 6.9 & 1.3 & 0.0 & 1.2 \\
Jun--Sep monsoon & 14228 & 62.5 & 0.0 & 37.2 & 6.0 & 0.0 & 1.4 \\
\hline
\end{tabular*}
\end{table}

\begin{table}[htbp]
\centering
\caption{Full-day Weather Station Meteorological-condition Statistics under the Adopted VLT-anchored Condition Classes. \(N\) is the number of valid 10-minute samples; all other columns are percentages, and other column definitions follow Table~\ref{tab:6-3}.}
\label{tab:6-4}
\raatablestyle
\setlength{\tabcolsep}{2.0pt}
\begin{tabular*}{\linewidth}{@{\extracolsep{\fill}}lrrrrrrr@{}}
\hline
Month group & $N$ & Good & Normal & RH $>70\%$ & $T-T_{\rm d}\leq 1.5~^\circ$C & Wind $>18$ m/s & Precip. $>0$ \\
\hline
Annual sample & 98623 & 87.5 & 0.3 & 11.8 & 1.7 & 0.0 & 1.0 \\
Nov--Jan core & 23057 & 97.1 & 0.3 & 2.0 & 0.9 & 0.0 & 0.9 \\
Oct--May low-cloud & 64653 & 94.6 & 0.5 & 4.5 & 0.8 & 0.0 & 0.8 \\
Jun--Sep monsoon & 33970 & 73.9 & 0.0 & 25.7 & 3.5 & 0.0 & 1.3 \\
\hline
\end{tabular*}
\end{table}

Under the adopted VLT-anchored Shigatse meteorological-condition classes, the satellite-defined November--January core has high Good-condition fractions in the 2024--2025 Weather Station archive. The night-time Good-condition fraction is 96.2\% (\(N=9571\)), and the corresponding 24 h Good-condition fraction is 97.1\% (\(N=23057\)). For the broader October--May period, the Good-condition fractions are 92.6\% at night (\(N=26817\)) and 94.6\% over 24 h (\(N=64653\)). The 24 h statistics describe the surface-meteorological environment relevant to daytime operation, radio-related activities, maintenance, and equipment safety checks at the existing Shigatse facility. These values quantify the fraction of valid 10-minute local meteorological samples satisfying the adopted wind, humidity, dew-point-depression, and precipitation criteria. The Normal-condition fraction, representing otherwise acceptable samples with \(12 < V \leq 18\) m/s, is negligible at night and remains below 0.5\% in the 24 h statistics. Sustained wind above 18 m/s does not occur in the valid 10 m Weather Station samples used here, so the wind component does not dominate the adopted rejection statistics for Shigatse.

The dominant meteorological limiter is humidity. In the night-time proxy, RH \(>70\%\) occurs in 2.8\% of November--January samples, 6.9\% of October--May samples, and 37.2\% of June--September samples. The close-to-dew-point term, \(T-T_{\rm d}\leq 1.5~^\circ\)C, is smaller: 1.3\% in both November--January and October--May, rising to 6.0\% in June--September. Measurable hourly precipitation is rare in the low-cloud season but is retained as a hard rejection term: it occurs in 1.4\% of November--January night-time samples and 1.2\% of October--May night-time samples. Most November--January samples satisfy the Good-condition class under the adopted humidity, dew-point, wind, and precipitation criteria, and the broader October--May season retains high Good-condition fractions. The June--September monsoon interval is reduced mainly by moisture, consistent with the cloud maximum.

Under the adopted meteorological criteria, the low-cloud advantage is retained in the local surface record. GOCCP, ISCCP, and the long meteorological-station cloud record identify the recurring non-monsoon low-cloud season, and the Weather Station statistics show that most valid samples in these month groups also satisfy the wind, humidity, dew-point-depression, and precipitation criteria at the Shigatse site. The low-cloud season therefore becomes a practical low-cloud observing period for continued Shigatse site testing.

\section{Discussion}\label{discussion}

\subsection{Cloud-cover Cycles in Selected Chinese Site-testing Regions}\label{shigatse-in-the-cloud-climatology-context-of-selected-chinese-sites}

The comparison in this section is restricted to three Chinese reference cases with different site-testing roles. Lenghu represents the mature domestic optical benchmark, with published measurements of clear nights, seeing, PWV, cloud cover, and meteorological conditions \citep{Deng2021Lenghu,Li2024LenghuCloud}. Ali represents the western-Xizang high-altitude reference and the nearest established context for extending site testing westward from Shigatse \citep{Ye2016AliAtacama,Cao2020DataProducts,Qian2024AliCloud}. Xinglong represents a long-running eastern observatory at lower elevation, where observing conditions are shaped by a different monsoon and regional-climate setting \citep{Zhang2016XinglongObservingConditions}. This section compares the annual cloud-cover level, seasonal timing, and month-to-month evolution at the selected sites, placing the Shigatse result in the broader context of Chinese astronomical site testing.

The annual cloud fraction is the first comparison scale. In GOCCP, Shigatse has an annual cloud fraction of 42.1\%, compared with 39.5\% for Ali, 46.9\% for Lenghu, and 46.8\% for Xinglong. ISCCP gives higher absolute values, with 59.7\% at Shigatse, 55.8\% at Ali, 59.4\% at Lenghu, and 54.5\% at Xinglong. These annual values place Shigatse within the range of the selected Chinese reference sites. The observing value of a site also depends on the timing of low-cloud months and on geographic distribution. The selected sites span about \(37.6^\circ\) in longitude from Ali to Xinglong, corresponding to roughly 2.5 h in local sidereal-time phase, and about \(11.2^\circ\) in latitude between Shigatse and Xinglong. Shigatse, at \(29.2^\circ\) N, gives higher transit elevations for southern targets than Lenghu or Xinglong, whereas the higher-latitude sites have a different northern-sky emphasis. Distributed observing networks such as the Las Cumbres Observatory Global Telescope Network explicitly use geographic distribution to improve time-domain coverage \citep{Brown2013LCOGT}; large-telescope site-testing programs likewise treat cloud cover, local meteorology, seeing, water vapor, sky background, and field access as separate layers before a site can be characterized for a specific facility \citep{Schoeck2009TMTOverview,Skidmore2008TMTASC,Varela2014ELTGroundMeteorology}. In this context, Shigatse is a lower-latitude southern-plateau reference site that can complement established Chinese sites.

Table~\ref{tab:selected-site-cloud-context} therefore uses a site-specific monthly diagnostic. For each site and product, the table lists the three months with the lowest cloud fraction and the three months with the highest cloud fraction. The result shows that Shigatse and Ali share a late-autumn to winter low-cloud tendency and a summer cloud maximum, consistent with their location on the southern and western Tibetan Plateau. The Shigatse low-cloud months are especially concentrated in late autumn and winter: GOCCP gives the three lowest months as November, December, and March, with a mean cloud fraction of 16.9\%, while ISCCP gives November--January, with a mean of 43.4\%. The corresponding high-cloud months are July--September in GOCCP and June--August in ISCCP, reflecting the monsoon-season limitation already diagnosed in the main analysis.

Ali has a similar annual phase but a different balance between winter and summer. Its GOCCP annual cloud fraction is slightly lower than Shigatse, and the largest difference appears in the high-cloud part of the year: the three highest GOCCP months at Ali have a mean cloud fraction of 62.0\%, compared with 77.5\% at Shigatse. This supports the interpretation that the western-Xizang region has a weaker summer cloud penalty than the southern-plateau Shigatse region. The result is consistent with \citet{Cao2020ClearNights}, who interpreted the Ali--Payang--Mayum La region as a transition zone between monsoon-controlled cloud formation to the east and a drier western regime influenced more strongly by westerly circulation.

Lenghu and Xinglong provide two additional forms of context. Lenghu is a well documented optical-site benchmark: its site quality has been examined through clear-night, seeing, PWV, cloud-cover, and meteorological studies, and the region is now supporting large survey-facility development, including the 2.5 m Wide Field Survey Telescope and the 6.5 m Multiplexed Survey Telescope \citep{Deng2021Lenghu,Li2024LenghuCloud,Chen2024WFSTScheduling,Zhang2024MUSTDome}. Its satellite monthly cycle is clearly different from that of Shigatse. In GOCCP, the three lowest-cloud months at Lenghu are October, January, and November, while the highest-cloud months fall in February--April; in ISCCP, the lowest-cloud months shift to October--December and the highest-cloud months to May--July. This product-dependent behavior should be interpreted together with the published local Lenghu site-testing record. It also shows calendar complementarity in low-cloud months: months that are relatively cloudy at Lenghu in GOCCP, such as February--April with a mean of 59.7\%, remain much lower at Shigatse in the same product, where the February--April mean is about 29.9\%. Xinglong provides a different comparison. Its eastern, lower-elevation observing conditions are affected by cloud, humidity, dust, and related weather factors \citep{Zhang2016XinglongObservingConditions}. In GOCCP, Xinglong has a winter-to-early-spring low-cloud group but becomes cloudier in June, July, September, and November; the November contrast is particularly clear in GOCCP, with 57.2\% at Xinglong and 14.6\% at Shigatse. These examples show that cross-site value comes not only from annual cloud amount, but also from whether low-cloud months occur at different calendar times and at different longitudes.

These comparisons place Shigatse as a lower-latitude southern-plateau node with a strong non-monsoon low-cloud regime, a different sky-access geometry, and an existing 40 m facility that can support continued environmental monitoring. This result is aligned with the staged logic of major site-testing programs: satellite and long-term climatological assessment identifies promising regimes, while local instruments then determine whether cloud continuity, seeing, sky background, water vapor, wind, humidity, and operational constraints are adequate for specific facilities \citep{Schoeck2009TMTOverview,Skidmore2008TMTASC,Varela2014ELTGroundMeteorology}. The monthly contrast among Shigatse, Ali, Lenghu, and Xinglong therefore supports continued Shigatse monitoring and westward site testing along the Shigatse--Ali low-cloud corridor, with particular attention to the Payang--Huoerba--Mayum La region, where a related but drier cloud regime is expected.

\begin{table}[htbp]
\centering
\caption{Site-specific Monthly Cloud-fraction Context for Shigatse and Selected Chinese Observing or Site-testing Locations. Values are cloud fractions in percent, extracted at the nearest GOCCP or ISCCP grid cell used in this work. The low- and high-cloud entries list the three lowest and three highest monthly cloud-fraction values for each site and product separately.}
\label{tab:selected-site-cloud-context}
\raatablestyle
\setlength{\tabcolsep}{3.6pt}
\begin{tabular*}{\linewidth}{@{\extracolsep{\fill}}llrll@{}}
\hline
Site & Product & Annual CF & Lowest months (mean CF) & Highest months (mean CF) \\
\hline
Shigatse & GOCCP & 42.1 & Nov--Dec; Mar (16.9) & Jul--Sep (77.5) \\
         & ISCCP & 59.7 & Nov--Jan (43.4) & Jun--Aug (75.5) \\
Ali      & GOCCP & 39.5 & Nov--Dec; Mar (19.7) & Jun--Aug (62.0) \\
         & ISCCP & 55.8 & Oct--Dec (42.8) & Jun--Aug (66.2) \\
Lenghu   & GOCCP & 46.9 & Jan; Oct--Nov (26.8) & Feb--Apr (59.7) \\
         & ISCCP & 59.4 & Oct--Dec (47.6) & May--Jul (68.0) \\
Xinglong & GOCCP & 46.8 & Jan--Mar (33.1) & Jun--Jul; Sep (61.5) \\
         & ISCCP & 54.5 & Jan; Oct; Dec (43.7) & Jun--Aug (67.0) \\
\hline
\end{tabular*}
\end{table}

\begin{figure}[htbp]
\centering
\includegraphics[width=0.98\linewidth,keepaspectratio]{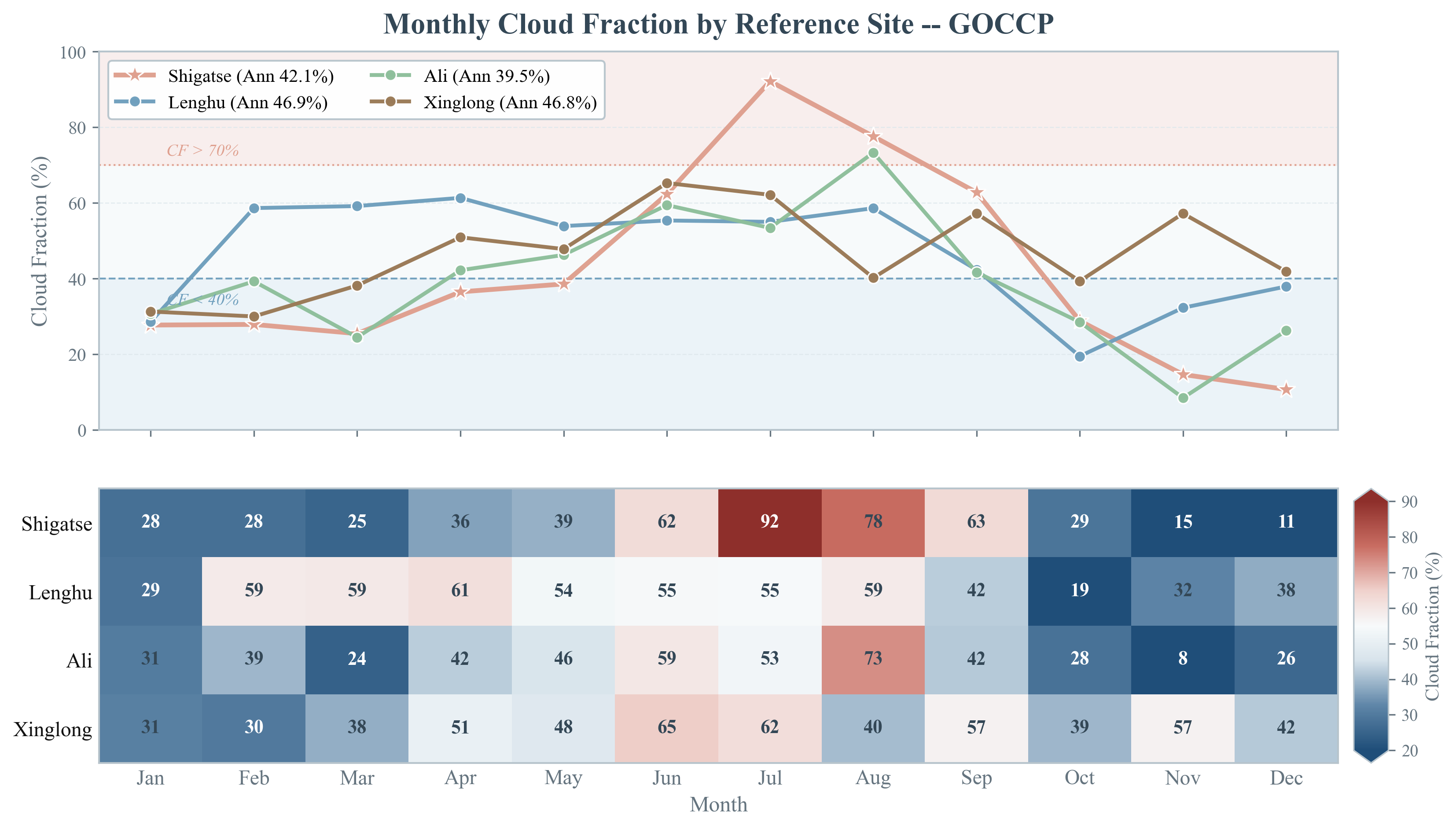}
\caption{Monthly GOCCP cloud-fraction summary for Shigatse and three Chinese reference observing or site-testing locations. The upper panel shows monthly cloud-fraction curves; the lower panel gives the same values as a site--month matrix. Values are percentages at the nearest GOCCP grid cell. The figure provides a within-product view of annual cloud level, site-specific low-cloud months, and month-to-month complementarity.}
\label{fig:goccp-site-monthly-summary}
\end{figure}

\begin{figure}[htbp]
\centering
\includegraphics[width=0.98\linewidth,keepaspectratio]{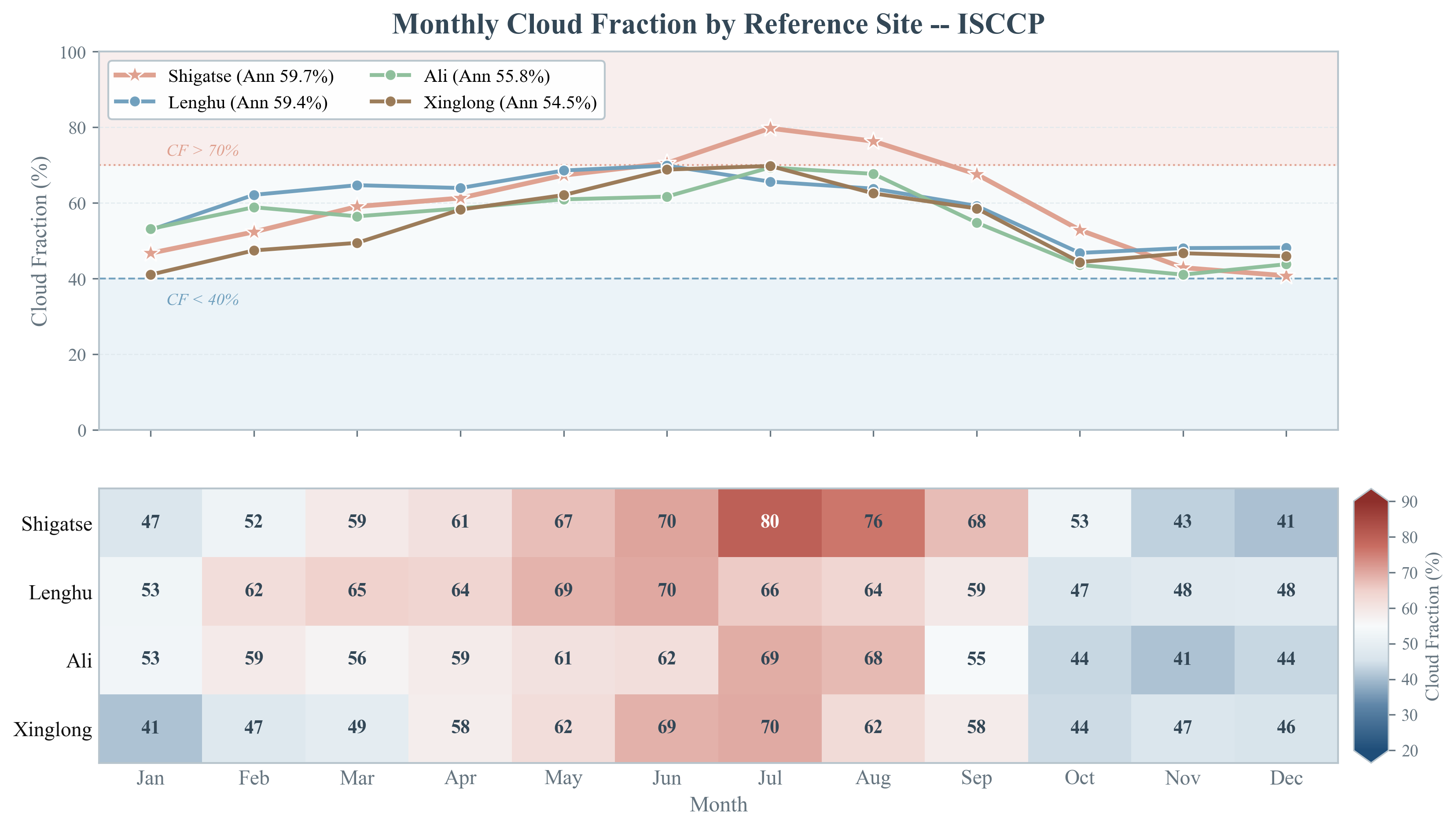}
\caption{Monthly ISCCP cloud-fraction summary for Shigatse and three Chinese reference observing or site-testing locations. The design follows Figure~\ref{fig:goccp-site-monthly-summary}, but the values are percentages at the nearest ISCCP HXG grid cell. Together with the GOCCP summary, the figure compares monthly phase, product-dependent differences, and site-specific observing-season complementarity.}
\label{fig:isccp-site-monthly-summary}
\end{figure}

The annual and monthly comparisons define the cloud-climatological position of Shigatse among the selected Chinese site-testing regions. Ali has the lower annual cloud fraction in both satellite products, while Shigatse contributes a different combination of lower-latitude sky access, an existing monitored 40 m facility, and a multi-source non-monsoon low-cloud season. The annual cloud fraction places Shigatse within the range of selected Chinese site-testing regions, and the monthly cycle indicates a persistent late-autumn-to-winter low-cloud period, strong summer monsoon limitation, and observing-season complementarity relative to sites with different longitude, latitude, and cloud-forming regimes. This cloud-cover result supports continued local site testing at Shigatse and motivates the Shigatse--Ali low-cloud corridor discussed below.

\subsection{Shigatse--Ali Low-cloud Corridor}\label{westward-low-cloud-corridor-from-shigatse-toward-ali}

The cloud-cover context above motivates examination of the Shigatse--Ali low-cloud corridor. The November--January core has the strongest local support at Shigatse: it combines low satellite cloud occurrence, high clear or less-cloudy fraction on ground in the aligned meteorological-station data, and high Good-condition fractions in the on-site meteorological statistics. The broader October--May period remains the main low-cloud season, while June--September is limited by the monsoon for optical night-time work. In the site-testing logic used for western China, long-baseline satellite cloud data identify where repeated field monitoring should be concentrated, and local instruments then convert that climatology into night-by-night observing statistics \citep{Cao2020ClearNights}. This section examines whether the Shigatse low-cloud regime connects westward with the Payang--Huoerba--Mayum La region identified in earlier western-Xizang cloud analyses.

\begin{figure}[htbp]
\centering
\includegraphics[width=\linewidth,keepaspectratio]{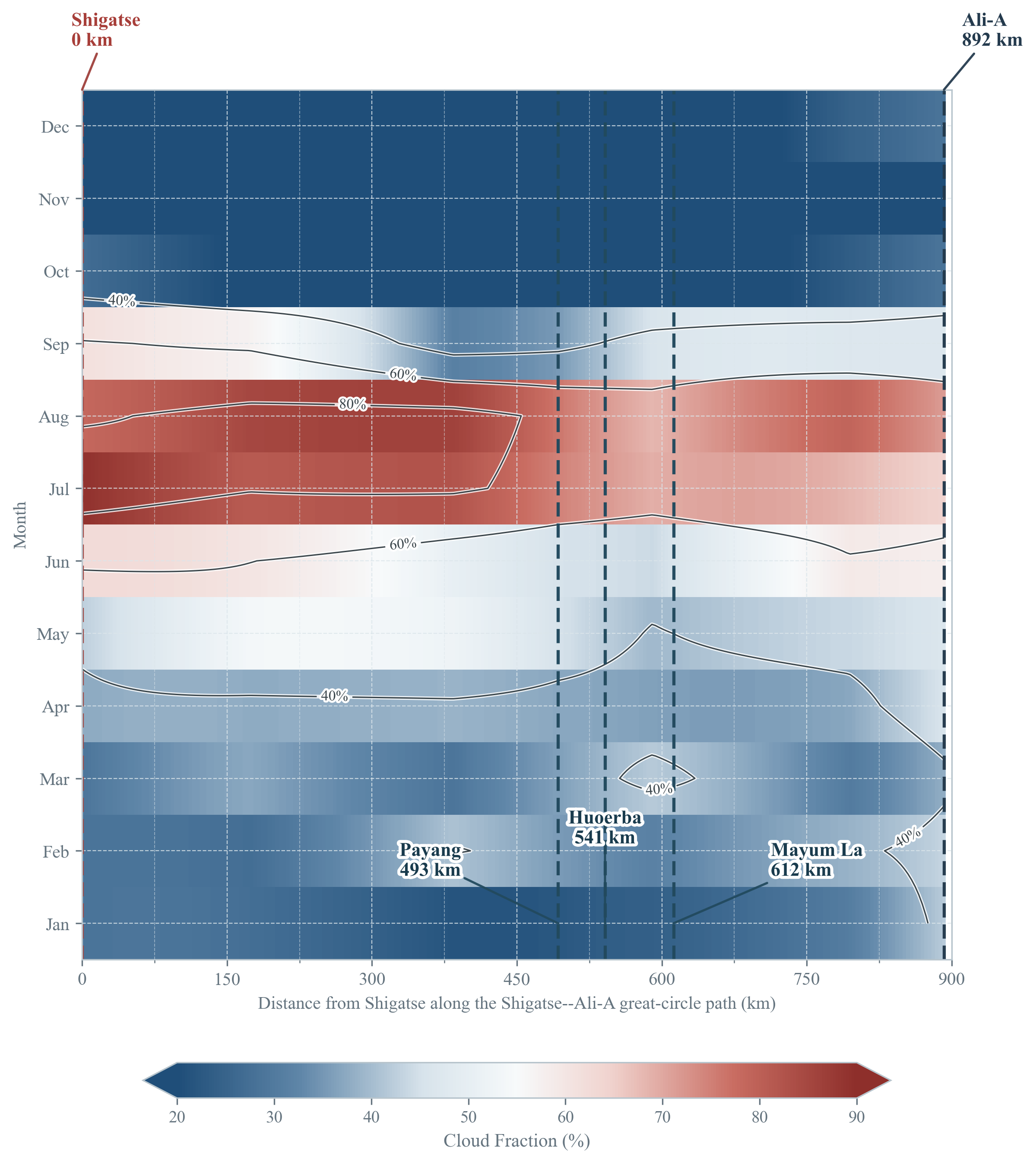}
\caption{GOCCP distance--month cloud-fraction section along the Shigatse--Ali-A great-circle path. The horizontal axis shows geodesic distance from Shigatse, and the vertical markers give the projected positions of Shigatse, Payang, Huoerba, Mayum La, and Ali-A along the same path. The field is based on bilinear interpolation of the $2^\circ \times 2^\circ$ GOCCP monthly climatology and represents satellite-product-scale gradients along the transect.}
\label{fig:7-1}
\end{figure}

The Shigatse--Ali-A great-circle section provides a direct view of this westward cloud field (Figure~\ref{fig:7-1}). The section is constructed by bilinearly interpolating the monthly GOCCP climatology along the geodesic from the Shigatse site to the Ali-A extraction point. At the \(2^\circ \times 2^\circ\) GOCCP support, it describes the product-scale cloud field along a reference transect for regional cloud-climatology assessment. The intermediate landmarks Payang, Huoerba, and Mayum La are projected onto this same geodesic for orientation; their perpendicular offsets from the line are about 116, 115, and 93 km, respectively, which is comparable to the spatial support of the GOCCP product.

The distance--month section clarifies what improves westward from Shigatse. The November--January low-cloud core is already strong at the Shigatse end of the path, and remains low along almost the full Shigatse--Ali-A section. In the interpolated GOCCP section, the annual mean cloud fraction decreases from about 42.4\% at Shigatse to a minimum of about 36.6\% near 590 km from Shigatse, close to \(31.3^\circ\) N and \(83.0^\circ\) E, before increasing again toward the Ali-A end. The October--May mean changes only modestly along the section, from about 27.1\% at Shigatse to about 26.5\% near the minimum. The larger difference occurs during the June--September monsoon interval: the Shigatse end has a mean cloud fraction of about 72.9\%, whereas the minimum zone has about 56.8\%. Thus the westward low-cloud signal is not primarily a further strengthening of the winter core, but rather a reduction in monsoon-season cloudiness and annual cloud burden around the Payang--Huoerba--Mayum La region. This is the path-scale expression of the transition discussed by \citet{Cao2020ClearNights}: this part of western Xizang lies near the inferred boundary between monsoon-dominated cloud formation to the east and a drier regime influenced more strongly by westerly circulation.

\begin{figure}[htbp]
\centering
\includegraphics[width=0.98\linewidth,keepaspectratio]{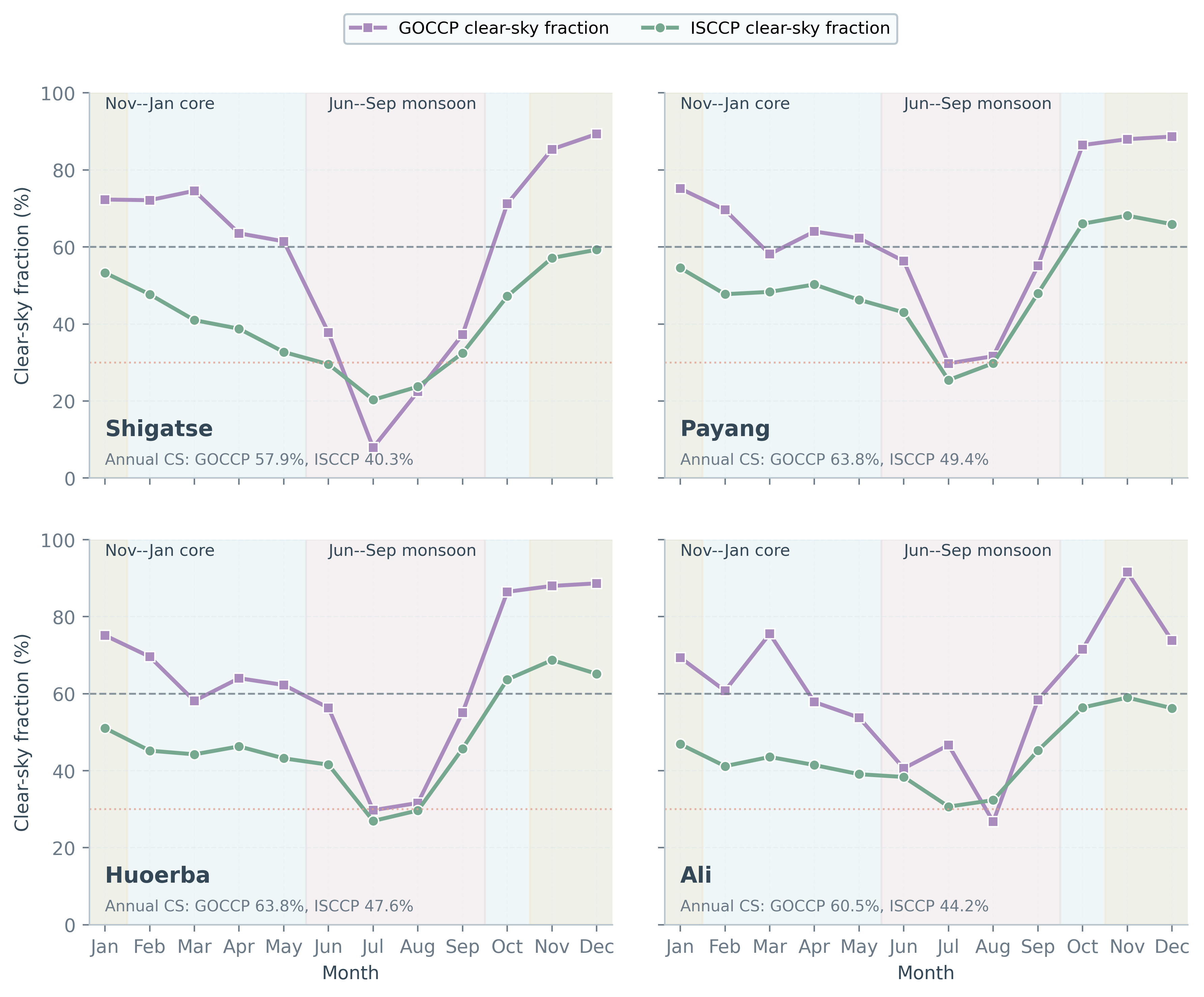}
\caption{Monthly satellite clear-sky fraction for Shigatse, Payang, Huoerba, and Ali. Clear-sky fraction is computed as $100\%-{\rm CF}$ for each product and used here as a product-scale low-cloud-season diagnostic. The background shading follows the month grouping used in Figure~\ref{fig:6-3}: pale blue marks the broader October--May low-cloud season, pale apricot marks the November--January core, and pale rose marks the June--September monsoon interval.}
\label{fig:7-2}
\end{figure}

The named extraction points provide a monthly clear-sky view of the same westward structure (Figure~\ref{fig:7-2}). At the satellite-product scale, the monthly structure separates Shigatse, Payang, Huoerba, and Ali by annual clear-sky level and monsoon-season amplitude; Mayum La is treated in the following spatial comparison as a topographic reference within the same low-cloud transition zone. In GOCCP, Payang and Huoerba share the same nearest grid cell and have an annual clear-sky fraction of 63.8\%, compared with 57.9\% at Shigatse and 60.5\% at Ali. In ISCCP, the westward points are separated at finer grid spacing: the annual clear-sky fractions are 49.4\% for Payang, 47.6\% for Huoerba, 44.2\% for Ali, and 40.3\% for Shigatse. The strongest difference again occurs in June--September. The westward points retain a clear late-autumn-to-winter peak while reducing the monsoon-season cloud penalty, which makes the Shigatse--Ali low-cloud corridor a useful framework for staged regional site testing.

\newpage

\begin{figure}[htbp]
\centering
\includegraphics[width=\linewidth,keepaspectratio]{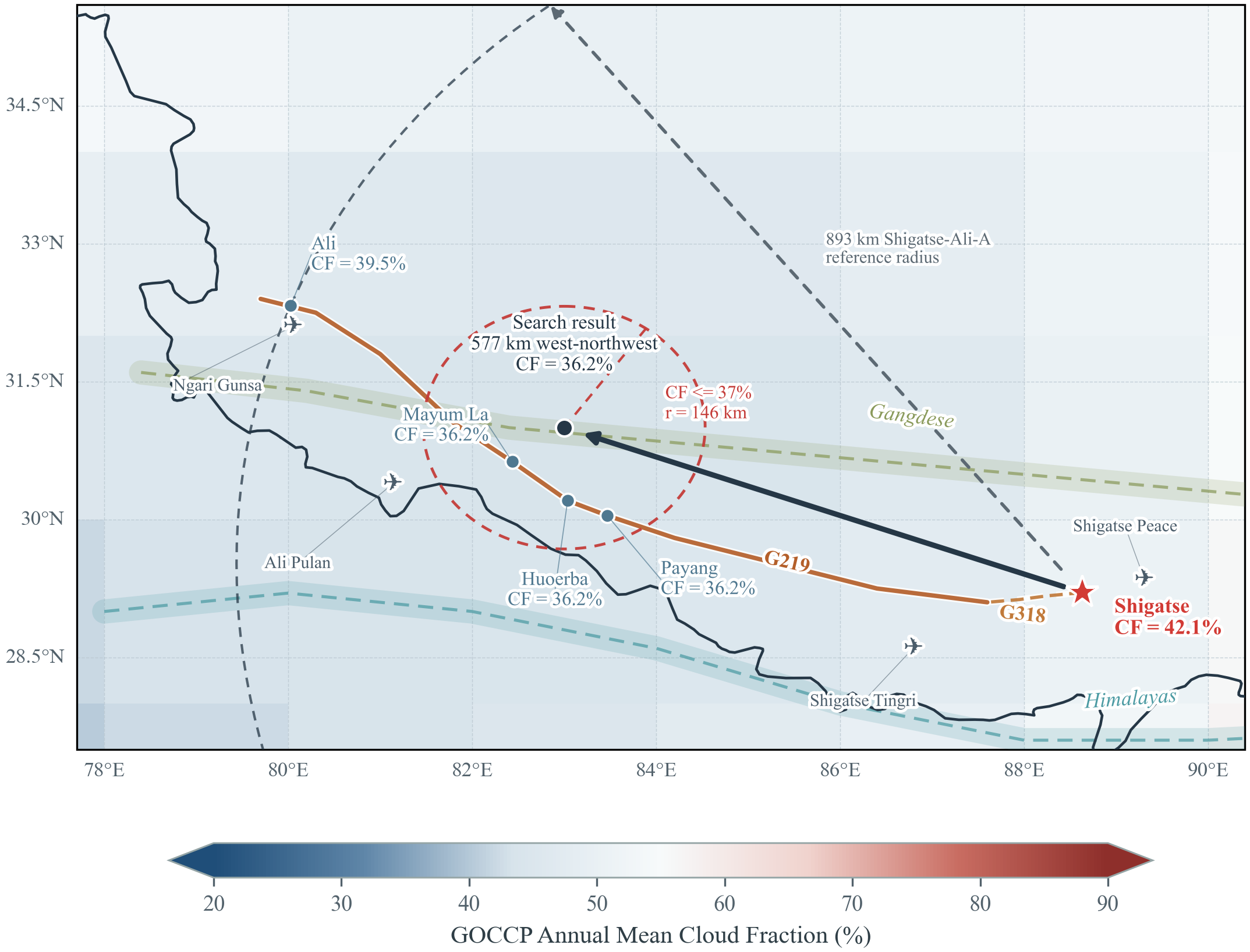}
\caption{GOCCP-based Shigatse--Ali low-cloud corridor. Shading: mean GOCCP CF. Red star: Shigatse; blue circles: Ali, Huoerba, Payang, and Mayum La; dark circle: lowest-CF grid cell within the 893 km radius. Grey/red dashed circles: reference-radius domain and CF $\leq 37\%$ low-cloud area. Dark arrow: west-northwest testing direction. G219/G318 lines, airport symbols, and terrain belts: field-campaign context.}
\label{fig:7-3}
\end{figure}

The GOCCP annual cloud-fraction field provides a spatial form of the westward comparison (Figure~\ref{fig:7-3}). The Shigatse-centered reference radius is set to 893 km, corresponding to the geodesic separation from the Shigatse site to the Ali-A extraction point used in the western-China site-testing framework \citep{Cao2020ClearNights}. This radius gives the search a defined cloud-climatology scale: it tests whether the region between the existing Shigatse facility and the established Ali-A high-plateau reference contains a lower-cloud area at the GOCCP product scale. Previous Chinese site-testing work has shown that competitive domestic optical candidates are associated with high, dry western or northern plateau environments, while satellite clear-night analysis is an appropriate first layer for prioritizing field campaigns \citep{Cao2020ClearNights,Deng2021Lenghu,Li2024LenghuCloud,Qian2024AliCloud}. Within this China-constrained search domain, the lowest-cloud GOCCP grid cell lies west-northwest of Shigatse, near \(31.0^\circ\) N and \(83.0^\circ\) E, with an annual cloud fraction of 36.2\%. It is about 577 km from Shigatse. The data therefore support the Shigatse--Ali low-cloud corridor as the westward cloud-climatology framework, with the Payang--Huoerba--Mayum La region as the main product-scale target.

The map annotations in Figure~\ref{fig:7-3} identify the cloud-climatology target and the feasibility of repeat field visits. The red dashed circle centered on the lowest-cloud grid cell marks the low-cloud area defined by GOCCP CF \(\leq 37\%\), and the CF labels beside Shigatse, Ali, Mayum La, Huoerba, Payang, and the search-result grid show nearest-grid annual GOCCP values. The G219 and G318 national highway corridors are based on the national highway route plan \citep{NDRCMOT2022NationalHighwayPlan}, and the airport symbols are transport nodes listed by the Civil Aviation Administration of China \citep{CAAC2025XizangAirports}. Boundary lines are used for regional orientation, with the standard-map source cited for reference \citep{MNR2019StandardMapService}.

The west-northwest low-cloud pattern is consistent with the western-Xizang cloud-regime interpretation of \citet{Cao2020ClearNights}. Figure~\ref{fig:7-3} marks the Ali, Payang, Huoerba, and Mayum La reference points discussed in that work; the Ali marker corresponds to the Ali-A extraction point. In that analysis, the Ali--Payang region was interpreted as lying north of the Himalayas and south of the Gangdese, within the longitude range where MODIS clear-night statistics separate group B and group C behavior. The authors inferred a change in cloud formation across the neighborhood of \(82^\circ\) E and identified Mayum La as the most important local landmark near that longitude. Mayum La is located at \(82.43429^\circ\) E, \(30.63046^\circ\) N, and an elevation of about 5285 m; it was described as the watershed between the sources of the Brahmaputra and Indus rivers and as a high local divide that appears to weaken the westward penetration of monsoon influence, leaving cloud formation west of Mayum La more affected by the westerlies and the east side more dominated by the monsoon. In the present cloud-climatology analysis, Mayum La provides the topographic and cloud-regime context, while Payang and Huoerba provide neighboring candidate-site reference points. The GOCCP field places this reference region within the west-northwest extension from Shigatse, with the low-cloud search grid near Mayum La, Huoerba, and Payang.

The point values in Figure~\ref{fig:7-3} are satellite-product values. GOCCP has a \(2^\circ \times 2^\circ\) grid in this analysis, so the Mayum La, Huoerba, and Payang markers fall in the same nearest GOCCP grid cell and therefore share the same GOCCP annual cloud-fraction value in Table~\ref{tab:7-3}. This grid-scale result identifies the Payang--Huoerba--Mayum La region, together with the nearby Ali reference, as part of a coherent lower-cloud regional zone relative to the Shigatse grid cell. The red dashed circle in Figure~\ref{fig:7-3} is centered on this search-result grid cell and encloses the low-cloud area defined by GOCCP CF \(\leq 37\%\) at the GOCCP grid scale. Together, Figure~\ref{fig:7-3} and Table~\ref{tab:7-3} identify the Payang--Huoerba--Mayum La region as a priority target for follow-up site testing within the Shigatse--Ali low-cloud corridor.

Table~\ref{tab:7-2} separates the reference site, search-result grid, and landmark entries used in the regional comparison. Shigatse is the existing 40 m radio site and monitoring reference site. The search-result grid is the lowest annual GOCCP cloud-fraction grid cell found inside China within the Shigatse--Ali-A reference-radius domain. Ali, plotted as Ali-A, defines the reference radius and connects this study to the western-Xizang site-testing framework. Mayum La is retained as the topographic reference for the transition in cloud formation near \(82^\circ\) E. Huoerba and Payang are the neighboring group C reference points from the same study. In the table, \(^{a}\) marks landmarks and sites following \citet{Cao2020ClearNights}, and \(^{b}\) marks the Ali-A extraction point and data-product context of \citet{Cao2020DataProducts}. The GOCCP and ISCCP entries are nearest product grid cells. Table~\ref{tab:7-3} gives the corresponding cloud-related quantities.

\begin{table}[htbp]
\centering
\caption{Geographic Coordinates and Satellite-grid Matching for the Shigatse West-northwest Region. Latitudes and longitudes are in degrees north and degrees east, respectively; satellite-grid coordinates are listed as (latitude, longitude), and distances are geodesic values from the Shigatse site.}
\label{tab:7-2}
\raatablestyle
\setlength{\tabcolsep}{2.8pt}
\begin{tabular*}{\linewidth}{@{\extracolsep{\fill}}lccccc@{}}
\hline
Point & Lat. & Lon. & Distance & GOCCP grid & ISCCP grid \\
\hline
Shigatse site & 29.20560 & 88.63220 & 0.0 km & (29.00, 89.00) & (29.25, 88.65) \\
Search-result grid & 31.00000 & 83.00000 & 577.3 km & (31.00, 83.00) & (30.95, 82.95) \\
Ali (Ali-A)$^{a,b}$ & 32.32573 & 80.02671 & 893.2 km & (33.00, 81.00) & (32.35, 80.05) \\
Mayum La$^{a}$ & 30.63046 & 82.43429 & 618.9 km & (31.00, 83.00) & (30.65, 82.45) \\
Huoerba$^{a}$ & 30.20617 & 83.03725 & 552.6 km & (31.00, 83.00) & (30.25, 83.05) \\
Payang$^{a}$ & 30.04289 & 83.46912 & 508.5 km & (31.00, 83.00) & (30.05, 83.45) \\
\hline
\end{tabular*}
\end{table}

Table~\ref{tab:7-3} summarizes the annual cloud-related quantities associated with the rows in Table~\ref{tab:7-2} and the points marked in Figures~\ref{fig:7-1}--\ref{fig:7-3}. The GOCCP and ISCCP columns are the annual mean cloud fractions extracted in this work at the nearest product grid cells. The published MODIS values from the western-Xizang clear-night analysis have a different definition: that work reports annual average percentages of clear nights, so Table~\ref{tab:7-3} lists the complementary cloudy-night percentage, \(100\%-{\rm clear\ nights}\), as a literature reference under the original clear-night definition \citep{Cao2020ClearNights}. For Shigatse, the independent local number is the weighted mean total cloud amount from the aligned conventional meteorological-station data used in Section~\ref{ground-based-cloud-amount-analysis-at-shigatse}. The earlier total-sky-image study of the Shigatse area provides seasonal support for the same autumn--winter low-cloud regime, while the table focuses on annual values that can be traced to explicit product or literature definitions \citep{Yang2018ShigatseSkyImages}.

\begin{table}[htbp]
\centering
\caption{Multi-source Annual Cloud-related Values for the Points Marked in Figures~\ref{fig:7-1}--\ref{fig:7-3}. CF denotes annual mean cloud fraction for GOCCP and ISCCP; the last column lists the independent annual metric and source used for comparison.}
\label{tab:7-3}
\raatablestyle
\setlength{\tabcolsep}{3.0pt}
\begin{tabular*}{\linewidth}{@{\extracolsep{\fill}}p{0.12\linewidth}ccp{0.60\linewidth}@{}}
\hline
Point & GOCCP (\%) & ISCCP (\%) & Independent annual cloud-related metric and source \\
\hline
\shortstack[l]{Shigatse\\site} & 42.1 & 59.7 & 29.7\% weighted mean total cloud amount from the Shigatse Meteorological Station common-period data in this work; the station is about 30 km east of the Shigatse site. \\
Payang & 36.2 & 50.6 & 13.4\% (NMC MODIS) and 18.0\% (CAS MODIS), expressed as cloudy-night complements of the published group C clear-night percentages. \\
Huoerba & 36.2 & 52.4 & 11.0\% (NMC MODIS), expressed as the cloudy-night complement of the published group C clear-night percentage. \\
Mayum La & 36.2 & 52.9 & Not reported as an independent annual cloud statistic; Mayum La is used in the western-Xizang analysis as a high watershed and regional divide near $82^\circ$ E. \\
\shortstack[l]{Ali\\(Ali-A)} & 39.5 & 55.8 & 13.2\% (NMC MODIS) and 17.4\% (CAS MODIS), expressed as cloudy-night complements of the published Ali-A clear-night percentages. \\
\hline
\end{tabular*}
\end{table}

GOCCP, ISCCP, and the published MODIS clear-night statistics consistently indicate lower cloud amount or lower cloudy-night occurrence in the Payang--Huoerba--Mayum La region than in the Shigatse grid cell \citep{Cao2020ClearNights}. Because these estimates are based on different sensors, retrieval methods, spatial supports, and cloud definitions, their agreement supports a consistent westward ordering of cloud amount. Together with the Mayum La topographic reference discussed above, the cloud minima shown in Figures~\ref{fig:7-1} and \ref{fig:7-3} identify the Payang--Huoerba--Mayum La region as the main follow-up target within the Shigatse--Ali low-cloud corridor. Subsequent site-testing campaigns should use this satellite- and literature-based target to select accessible ridges and passes for direct measurements of local cloud continuity, sky background, seeing, PWV, near-surface meteorology, radio-frequency environment, and logistical feasibility.

The transportation and airport layers in Figure~\ref{fig:7-3} evaluate whether the low-cloud target identified above can support repeatable site-testing campaigns. The official National Highway Network Plan identifies G219 as the Kanas--Dongxing national highway through western and southern Xizang and G318 as the Shanghai--Nyalam national highway through the Shigatse region \citep{NDRCMOT2022NationalHighwayPlan}. CAAC sources list the relevant Xizang transport-airport nodes, including Ali Pulan, Ali Gunsa, Shigatse Peace, Shigatse Tingri, and Lhasa Gonggar, and describe the recent expansion of the high-plateau airport network \citep{CAAC2025XizangAirports,CAAC2025XizangAviationDevelopment}. In the staged site-testing framework used by major optical/infrared programs, satellite cloud assessment identifies candidate cloud regimes, and ground campaigns then measure cloud continuity, seeing, sky background, water vapor, local meteorology, safety, and operational constraints before final site characterization \citep{Schoeck2009TMTOverview,Skidmore2008TMTASC,Varela2014ELTGroundMeteorology}. The existing Shigatse 40 m facility therefore provides a monitored eastern reference point for repeatable westward campaigns into the Payang--Huoerba--Mayum La region.

\subsection{Artificial Sky-brightness Trend along the Shigatse--Ali Low-cloud Corridor}\label{dark-sky-outlook-for-the-shigatseali-context}

The Shigatse--Ali low-cloud corridor is defined above from cloud-cover diagnostics. For optical site testing, the value of these low-cloud or cloud-free periods also depends on the long-term artificial sky-brightness background. The VIIRS/DNB-derived artificial sky-brightness model therefore extends the cloud-climatology analysis from transparent-sky occurrence to optical background, following the general framework of satellite-based night-sky-brightness mapping \citep{Cinzano2001WorldAtlas,Falchi2016WorldAtlas,Elvidge2021AnnualVIIRS,GoogleEarthEngineVIIRSAnnualV22,LightPollutionMapHelp2026}. This extension is physically relevant because clouds can modify the scattering and reflection of upward artificial light, and ground-based sky-brightness measurements have been used together with satellite night-light products to interpret nocturnal cloudiness and sky-background variability \citep{Hanel2018NightSkyMethods,Cavazzani2020SQMSatellite}. Here, the 2012--2025 annual sequence is used as an ancillary trend record to test whether Shigatse and the Shigatse--Ali low-cloud corridor also retain a stable dark-sky background. For the very dark high-plateau points considered here, the sequence shows that the western-Xizang points remain in a stable, very dark model regime over the available record.

\begin{figure}[htbp]
\centering
\includegraphics[width=\linewidth,keepaspectratio]{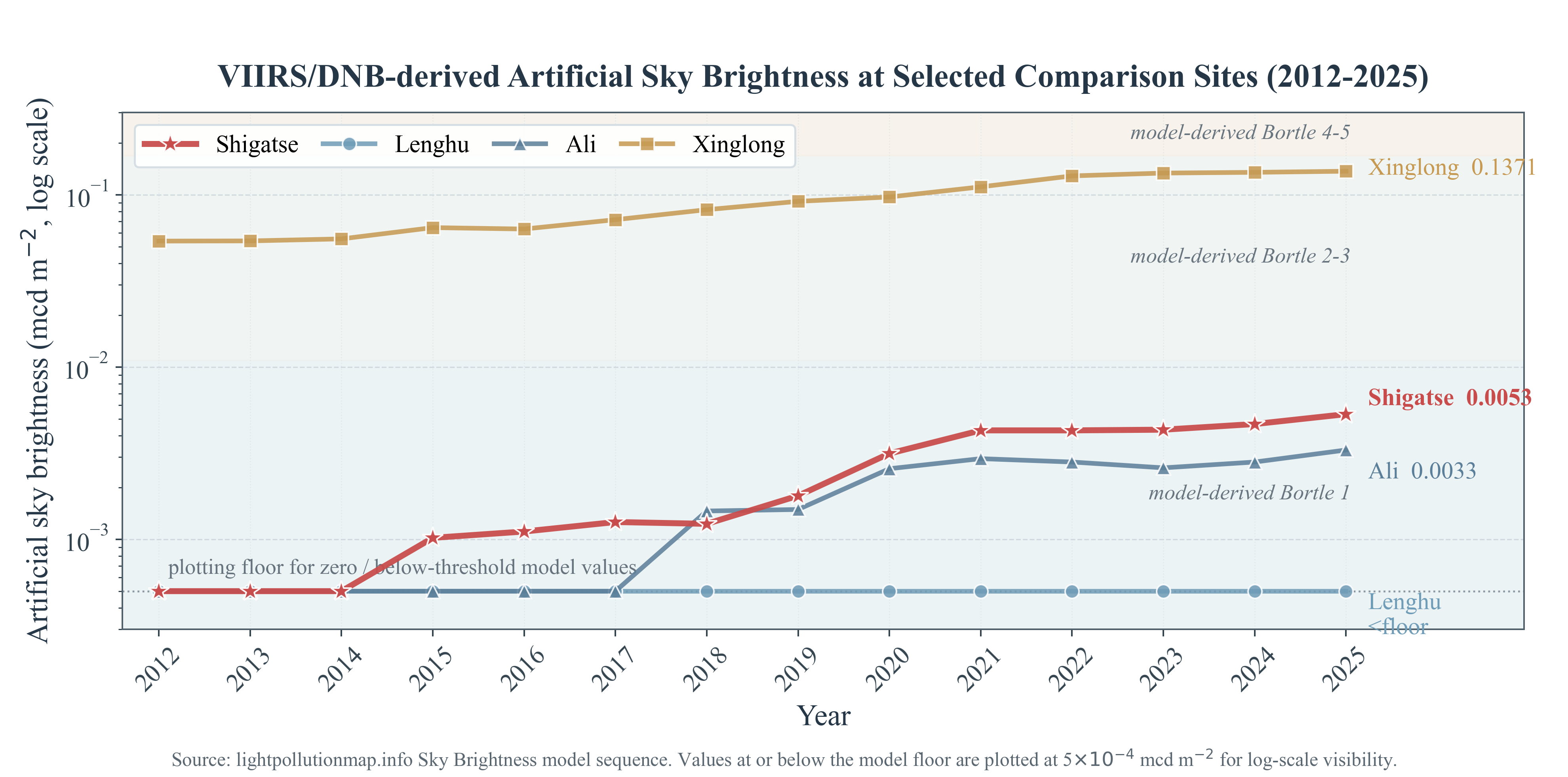}
\caption{VIIRS/DNB-derived artificial sky-brightness model sequence for Shigatse and selected comparison sites during 2012--2025.}
\label{fig:7-4}
\end{figure}

The broad comparison-site trend is clear (Figure~\ref{fig:7-4}). Shigatse remains in model-derived Bortle class 1 throughout the 2012--2025 sequence, with an SQM-equivalent value close to 22 mag arcsec\(^{-2}\) and artificial brightness at the \(10^{-3}\) mcd m\(^{-2}\) level. Here mcd m\(^{-2}\) denotes millcandela per square metre, a photometric luminance unit used for the model-derived artificial sky-brightness component. For orientation, a total dark-sky brightness of \(22.0~{\rm mag~arcsec^{-2}}\) corresponds to about \(0.17~{\rm mcd~m^{-2}}\). The model values are therefore used here as dark-regime trend indicators rather than calibrated in-situ SQM measurements. Ali and Lenghu show the same high-plateau dark-sky behavior: Ali remains close to 22 mag arcsec\(^{-2}\), and Lenghu stays at the model floor. By contrast, Xinglong declines from 20.77 to 20.24 mag arcsec\(^{-2}\), equivalent to about 63\% brightening over the same interval, and remains in a much brighter model-derived Bortle class 4 environment. Shigatse therefore follows the dark western high-plateau group rather than the brightening eastern-observatory example.

\begin{figure}[htbp]
\centering
\includegraphics[width=\linewidth,keepaspectratio]{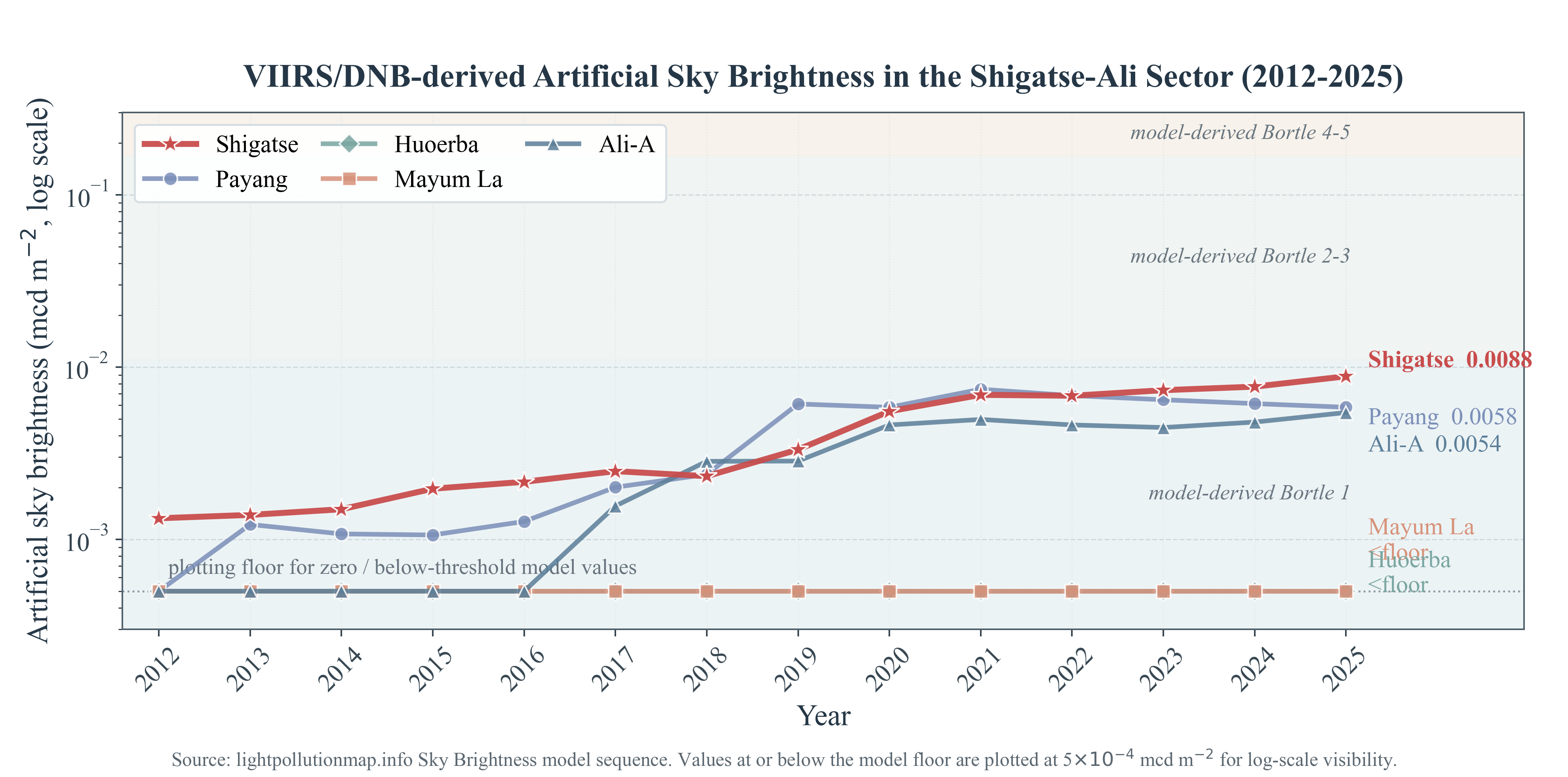}
\caption{VIIRS/DNB-derived artificial sky-brightness model sequence for Shigatse and four western-Xizang reference points during 2012--2025. Values are point-query extractions from the lightpollutionmap.info sky-brightness model; floor-level values are plotted with a small positive value for log-scale display.}
\label{fig:7-5}
\end{figure}

The same model was sampled at reference points along the Shigatse--Ali low-cloud corridor to examine whether the westward cloud target is also embedded in a stable artificial-light background (Figure~\ref{fig:7-5}). Mayum La and Huoerba remain at the model floor throughout 2012--2025. Ali-A is also at the floor through 2016 and then rises only to \(5.45\times10^{-3}\) mcd m\(^{-2}\) by 2025. Payang shows weak non-zero model values after 2013, with a 2025 value of \(5.84\times10^{-3}\) mcd m\(^{-2}\) and a sequence maximum of \(7.44\times10^{-3}\) mcd m\(^{-2}\). Shigatse increases from \(1.32\times10^{-3}\) to \(8.84\times10^{-3}\) mcd m\(^{-2}\), still within the same very dark model regime. In SQM-equivalent terms, all five points remain approximately in the 21.95--22.00 mag arcsec\(^{-2}\) range in 2025. The Shigatse--Ali low-cloud corridor therefore remains in a very dark modeled sky-brightness regime in the available 2012--2025 sequence: the monitored Shigatse 40 m site forms the eastern reference point, while Payang, Huoerba, Mayum La, and Ali-A define a westward target in which the modeled artificial-light contribution remains at the floor level or only a few \(10^{-3}\) mcd m\(^{-2}\).

Combining the comparison-site result and the four westward reference points, Shigatse and the Shigatse--Ali low-cloud corridor belong to the dark high-plateau group represented by Ali and Lenghu, while Xinglong represents the brighter eastern-observatory comparison case. The low-cloud months and westward low-cloud region retain a very low modeled artificial-light background through 2025. Published ground-based sky-background studies show that moonless-night statistics, instrument quality control, and calibrated sky-background distributions require local, higher-cadence SQM/SQM-LE-class monitoring, and such measurements have also been combined with satellite data in nocturnal cloud studies \citep{Hanel2018NightSkyMethods,Plauchu2017SPM,Cavazzani2020SQMSatellite}. The present trend indicates that dark-sky preservation should be included in follow-up optical site testing within the Payang--Huoerba--Mayum La low-cloud region: roads, power infrastructure, settlement growth, engineering lighting, and observatory operations should be managed so that the current dark high-plateau background is retained during future site-testing campaigns.

\subsection{Multi-decadal Stability of Ground-based Cloud Amount}\label{long-term-cloud-data-stability-assessment}

The 1988--2019 conventional cloud-amount sequence provides a multi-decadal stability check on the Shigatse cloud environment. It extends beyond the main 1988--2013 common-period statistics and follows the long-term trend-analysis convention, so it is used as an independent monitoring diagnostic alongside the aligned four-time analysis in Section~\ref{ground-based-cloud-amount-analysis-at-shigatse}. The reproduced annual diagnostic shows a weak downward first-order trend, with a Mann--Kendall coefficient of \(\tau=-0.25\), \(p=0.046\), and an ordinary least-squares slope of about \(-0.8\) percentage points per decade \citep{Mann1945Trend,Kendall1975RankCorrelation}. This sequence is consistent with multi-decadal recurrence and no obvious deterioration in the local cloud-amount diagnostic, supporting continued use of Shigatse as a cloud-cover monitoring reference.

The corresponding spectral diagnostic is retained in Appendix~\ref{appendix-long-term-cloud-amount-diagnostics} as a variability check. Multi-year power in the 2--7 yr band is physically plausible in the context of Tibetan Plateau cloud and monsoon variability, because cloudiness over the plateau is affected by monsoon moisture transport, westerly circulation, local topography, and regional climate variability \citep{DuanWu2006TPCloudClimate,Zhang2007TPCloud,Wang2001SouthAsianMonsoon,Li2008TibetanPlateauMonsoon,Xu2009EasternTPPrecipitation,Wu2024TPCloudReview,Peng2024ChinaSummerCloudCycles}. The spectral result is used only as a monitoring diagnostic for multi-year variability scales, while the main cloud-climatology conclusions rely on the monthly satellite products and the aligned ground-cloud statistics.

\subsection{Scale Translation from Satellite Cloud Climatology to Site-level Observing Statistics}\label{scale-translation-from-satellite-cloud-climatology-to-site-level-observing-statistics}

The Shigatse results identify the timing and regional placement of the low-cloud regime from satellite products, fixed-time ground cloud observations, and local surface meteorology. A useful external example for interpreting such product-scale cloud fractions is Mauna Kea, where both satellite cloud estimates and site-level all-sky-camera statistics have been studied. The TMT 13N site is located near the Mauna Kea summit region at an elevation of about 4050 m, and its cloud environment is strongly influenced by the Hawaiian trade-wind inversion. MODIS observations over Hawaii show that cloud-cover frequency increases up to the inversion layer and then decreases sharply at higher elevations, with the lowest cloud frequencies over the high-elevation parts of Maui and Hawaii above the inversion \citep{Barnes2016HawaiiCloud}. The TMT all-sky camera study reported a manually classified clear fraction of 70.9\% for Mauna Kea 13N \citep{Skidmore2011TMTASCA}. A CALIPSO/CALIOP analysis of cirrus and aerosol occurrence at astronomical sites reported an overall usable fraction of about 81\% for Mauna Kea 13N, compared with 71.4\% from the earlier Erasmus satellite study \citep{OtarolaHickson2017Cirrus,ErasmusVanStaden2003MaunaKeaSatellite}. These published values illustrate a scale and definition dependence that is common in site testing: gridded satellite cloud metrics and summit-level clear-time statistics are complementary, but they are not interchangeable.

\begin{figure}[htbp]
\centering
\includegraphics[width=\linewidth,keepaspectratio]{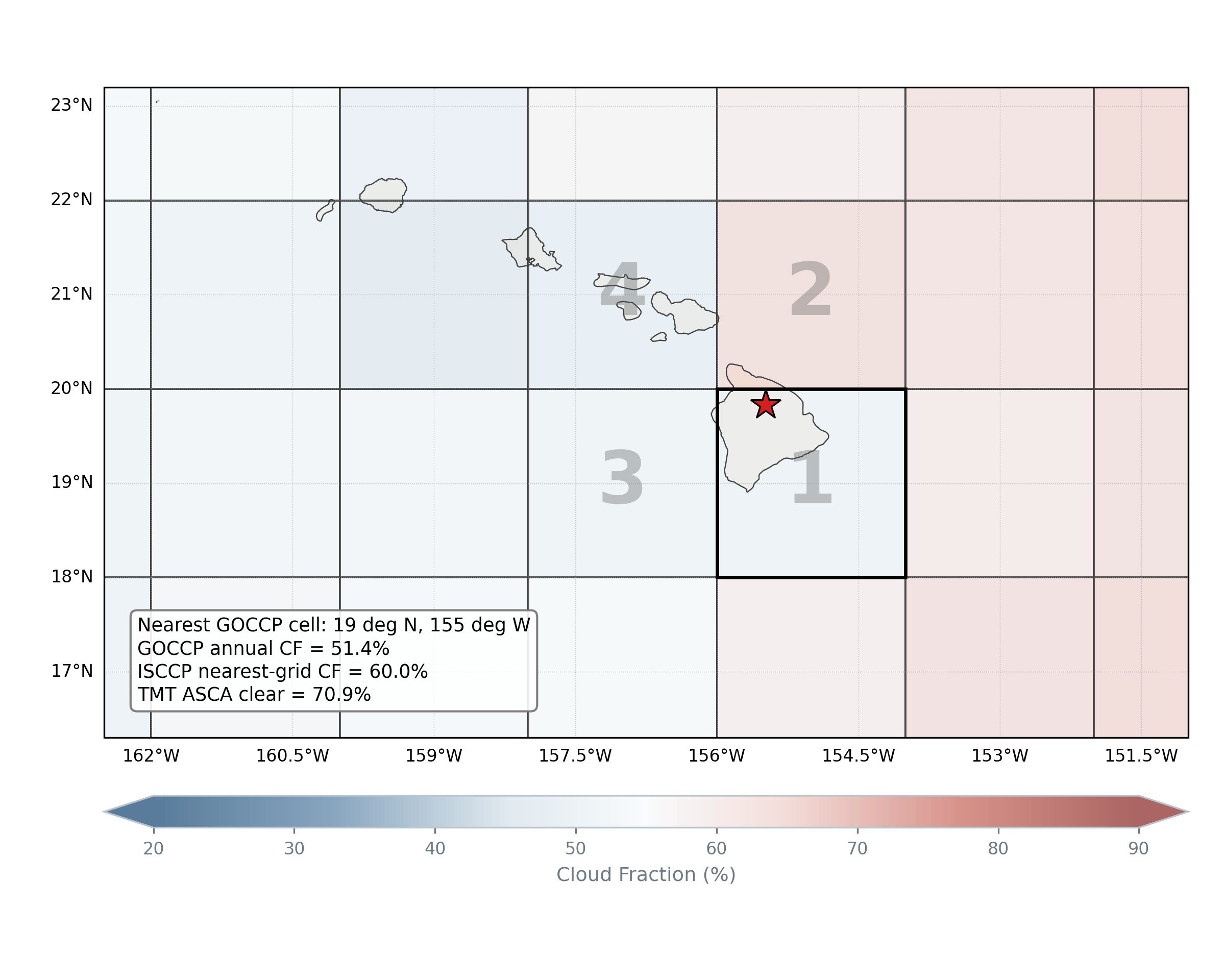}
\caption{Annual mean GOCCP total cloud fraction around Mauna Kea 13N, shown with the same cloud-fraction scale used for the GOCCP annual cloud-fraction maps in this paper. The red star marks the TMT 13N site near the Mauna Kea summit, and the black square marks the nearest \(2^\circ \times 2^\circ\) GOCCP grid cell used for the point extraction. The large grey numbers identify the four GOCCP cells listed in Table~\ref{tab:7-4}; cell 1 is the Mauna Kea 13N cell.}
\label{fig:7-6}
\end{figure}

Applying the same 2007--2016 satellite products used in the Shigatse analysis gives higher product-scale total cloud fractions at Mauna Kea (Figure~\ref{fig:7-6}). The nearest \(2^\circ \times 2^\circ\) GOCCP grid cell to Mauna Kea 13N has an annual cloud fraction of 51.4\%, while the nearest ISCCP grid point gives 60.0\%. The four numbered GOCCP cells around the site range from 50.1\% to 62.9\% (Table~\ref{tab:7-4}), showing appreciable product-scale heterogeneity around the island. These values describe grid-cell cloud occurrence over the surrounding island and adjacent ocean, and total cloud fraction can include cloud layers that do not correspond to opaque summit-level cloud cover along the telescope line of sight. The Mauna Kea comparison therefore provides a well-studied example for the interpretation used here: satellite cloud products define the regional cloud regime, and local high-cadence instruments convert that regime into summit-, ridge-, or facility-scale observing statistics.

\begin{table}[htbp]
\centering
\caption{Numbered GOCCP Grid Cells around Mauna Kea 13N.}
\label{tab:7-4}
\begingroup
\raatablestyle
\setlength{\tabcolsep}{4pt}
\begin{tabular*}{0.90\linewidth}{@{\extracolsep{\fill}}cllr@{}}
\hline
No. & Relation to Mauna Kea 13N & GOCCP grid-cell center & Annual CF \\
\hline
1 & Mauna Kea 13N cell & \(19^\circ\) N, \(155^\circ\) W & 51.4\% \\
2 & Northern adjacent cell & \(21^\circ\) N, \(155^\circ\) W & 62.9\% \\
3 & Western adjacent cell & \(19^\circ\) N, \(157^\circ\) W & 51.8\% \\
4 & Northwestern adjacent cell & \(21^\circ\) N, \(157^\circ\) W & 50.1\% \\
\hline
\end{tabular*}
\endgroup
\end{table}

The methodological implication for Shigatse is direct. GOCCP and ISCCP define the satellite-product-scale cloud regime over the southern Tibetan Plateau and along the Shigatse--Ali low-cloud corridor, while the multi-decadal ground cloud record provides an independent local check of the same seasonal phase at Shigatse. Their agreement supports a high-confidence multi-source estimate of the recurrent low-cloud period at Shigatse and of the westward low-cloud corridor at product scale. High-cadence all-sky monitoring and co-located site instruments can now translate this monthly cloud-climatology result into 24 h and night-by-night observing statistics.

\section{Conclusions}\label{conclusions}

Active-lidar satellite data, passive-satellite cloud fields, long fixed-time meteorological-station cloud observations, and on-site Weather Station measurements lead to five conclusions for Shigatse site testing. Together they identify a moderate-to-low annual GOCCP cloud fraction and a strongly organized monthly cycle: cloudiness is concentrated in the summer monsoon months, while October--May repeatedly provides the low-cloud observing period.

\begin{enumerate}
\def\labelenumi{\arabic{enumi}.}
\item
  CALIPSO-GOCCP places Shigatse in a southern-plateau cloud regime with a moderate annual cloud fraction and a well-organized monthly cycle. The extracted GOCCP grid cell has an annual mean cloud fraction of 42.1\% and an annual clear-sky fraction of 57.9\%. Cloud fraction is lowest in late autumn and winter, with monthly values of 14.6\% in November and 10.7\% in December, and highest during the summer monsoon, reaching 92.1\% in July. In the GOCCP climatology, October--May forms the broader low-cloud season and November--January is the clearest core.
\item
  ISCCP HXG gives higher absolute cloud fractions than GOCCP, but it supports the same seasonal phase and local spatial placement at finer grid spacing. The annual ISCCP cloud fraction at Shigatse is 59.7\%, with October--May and June--September means of 52.9\% and 73.5\%, respectively. The two satellite products differ in absolute cloud fraction because of sensor type, retrieval physics, sampling, spatial support, and high-terrain retrieval uncertainty. Their shared monthly phase is a recurrent non-monsoon low-cloud period and a monsoon-limited summer.
\item
  The conventional meteorological-station cloud data provide an independent, long-term consistency check on the satellite seasonal phase. After aligning the four observing times to the 1988--2013 common period and expressing valid 0--10 tenths values as 0--100\% cloud amount, the ground observations reproduce the late-autumn/winter low-cloud regime and the summer maximum. Using total cloud amount \(\leq 40\%\) as the ground-data low-cloud criterion, the fraction of fixed-time observations satisfying this criterion is 90.7\% during November--January and 80.7\% during October--May, compared with 39.9\% during June--September. The four fixed observing times constrain monthly recurrence and low-order diurnal behavior; the follow-up all-sky-camera paper in this series will extend this result to continuous 24 h cloud evolution.
\item
  The Weather Station data show that the satellite-defined low-cloud season also has high Good-condition fractions in the 2024--2025 surface-meteorological archive. For the 20:00--06:00 night-time proxy, valid 10-minute samples satisfying the Good-condition class account for 96.2\% in November--January and 92.6\% in October--May, compared with 62.5\% in June--September. The corresponding 24 h fractions are 97.1\%, 94.6\%, and 73.9\%. Sustained wind \(>18\) m/s is absent in the valid samples and measurable hourly precipitation is rare in the low-cloud season; the main limiting term is high relative humidity during the monsoon months.
\item
  Shigatse provides a lower-latitude southern-plateau cloud-cover reference within China's site-testing network. Its cloud-cover cycle is complementary to established reference sites such as Lenghu and Ali. The persistent non-monsoon low-cloud season defines a low-cloud observing period for subsequent local monitoring and provides the basis for later westward regional site testing.
\end{enumerate}

This study identifies a non-monsoon low-cloud observing period spanning October--May, with a November--January core, and delineates the Shigatse--Ali low-cloud corridor for subsequent westward site testing. The follow-up all-sky-camera paper in this series, led by Qihang He at Xizang University, will test this monthly cloud-cover framework with continuous 24 h cloud evolution, night-time cloud-duration, and cloud-gap statistics. Continued measurements of local weather, sky brightness, PWV, seeing, and operational conditions will then derive direct site-testing statistics for specific facilities from the low-cloud observing period and the Payang--Huoerba--Mayum La region identified here.

\appendix

\section{Spatial Correspondence Diagnostics}\label{appendix-spatial-correspondence-diagnostics}

The GOCCP--ISCCP spatial correspondence test uses the same western-China map domain as Figure~\ref{fig:6-2}, \(65^\circ\)--\(125^\circ\) E and \(20^\circ\)--\(50^\circ\) N. ISCCP monthly cloud fraction is area-averaged inside each GOCCP \(2^\circ \times 2^\circ\) grid cell, giving 465 paired cells for each month. Spearman rank correlation is then calculated across the paired cells. The diagnostic emphasizes spatial ordering rather than equality of absolute cloud fraction.

The low-cloud-area overlap is computed from product-specific lower-quantile masks. For a quantile \(q\), \(G_q\) denotes the set of GOCCP grid cells in the lower \(q\) part of the GOCCP cloud-fraction distribution, and \(I_q\) denotes the corresponding ISCCP set after aggregation to the GOCCP grid. The Jaccard overlap is
\[
J_q=\frac{|G_q\cap I_q|}{|G_q\cup I_q|},
\]
and the GOCCP low-cloud recovery fraction is
\[
R_q=\frac{|G_q\cap I_q|}{|G_q|}.
\]
The lower-tercile masks (\(q=1/3\)) are used as the main low-cloud-area diagnostic, and lower-quartile masks (\(q=1/4\)) are retained as a sensitivity check.

\begin{table}[htbp]
\centering
\caption{Monthly GOCCP--ISCCP Spatial Correspondence Diagnostics.}
\label{tab:appendix-spatial-correspondence}
\raatablestyle
\begin{tabular}{@{}lrrrrr@{}}
\hline
Month & Spearman \(\rho\) & \(J_{1/3}\) & \(R_{1/3}\) & \(J_{1/4}\) & \(R_{1/4}\) \\
\hline
Jan & 0.892 & 0.632 & 0.774 & 0.560 & 0.718 \\
Feb & 0.905 & 0.722 & 0.839 & 0.625 & 0.769 \\
Mar & 0.807 & 0.632 & 0.774 & 0.581 & 0.735 \\
Apr & 0.768 & 0.498 & 0.665 & 0.539 & 0.701 \\
May & 0.681 & 0.566 & 0.723 & 0.560 & 0.718 \\
Jun & 0.734 & 0.455 & 0.626 & 0.444 & 0.615 \\
Jul & 0.791 & 0.469 & 0.639 & 0.436 & 0.607 \\
Aug & 0.834 & 0.520 & 0.684 & 0.510 & 0.675 \\
Sep & 0.793 & 0.505 & 0.671 & 0.453 & 0.624 \\
Oct & 0.825 & 0.658 & 0.794 & 0.636 & 0.778 \\
Nov & 0.871 & 0.598 & 0.748 & 0.636 & 0.778 \\
Dec & 0.868 & 0.623 & 0.768 & 0.550 & 0.709 \\
\hline
\end{tabular}

\vspace{2mm}
\begin{minipage}{0.93\linewidth}
\footnotesize
Note. All diagnostics are calculated over the \(65^\circ\)--\(125^\circ\) E, \(20^\circ\)--\(50^\circ\) N map domain after ISCCP is area-averaged to the GOCCP grid. Each monthly comparison uses 465 paired grid cells. \(J\) is the Jaccard overlap of the two product-specific low-cloud masks, and \(R\) is the fraction of GOCCP low-cloud cells also selected by ISCCP.
\end{minipage}
\end{table}

\clearpage

\section{Long-term Cloud-amount Diagnostics}\label{appendix-long-term-cloud-amount-diagnostics}

The long-term cloud-amount analysis examined 1988--2019 conventional cloud observations from the Shigatse Meteorological Station. These diagnostics document the data provenance and provide a long-term stability comparison for future monitoring alongside the main 1988--2013 aligned cloud-amount statistics adopted in Section~\ref{ground-based-cloud-amount-analysis-at-shigatse}.

\begin{figure}[htbp]
\centering
\includegraphics[width=0.98\linewidth,keepaspectratio]{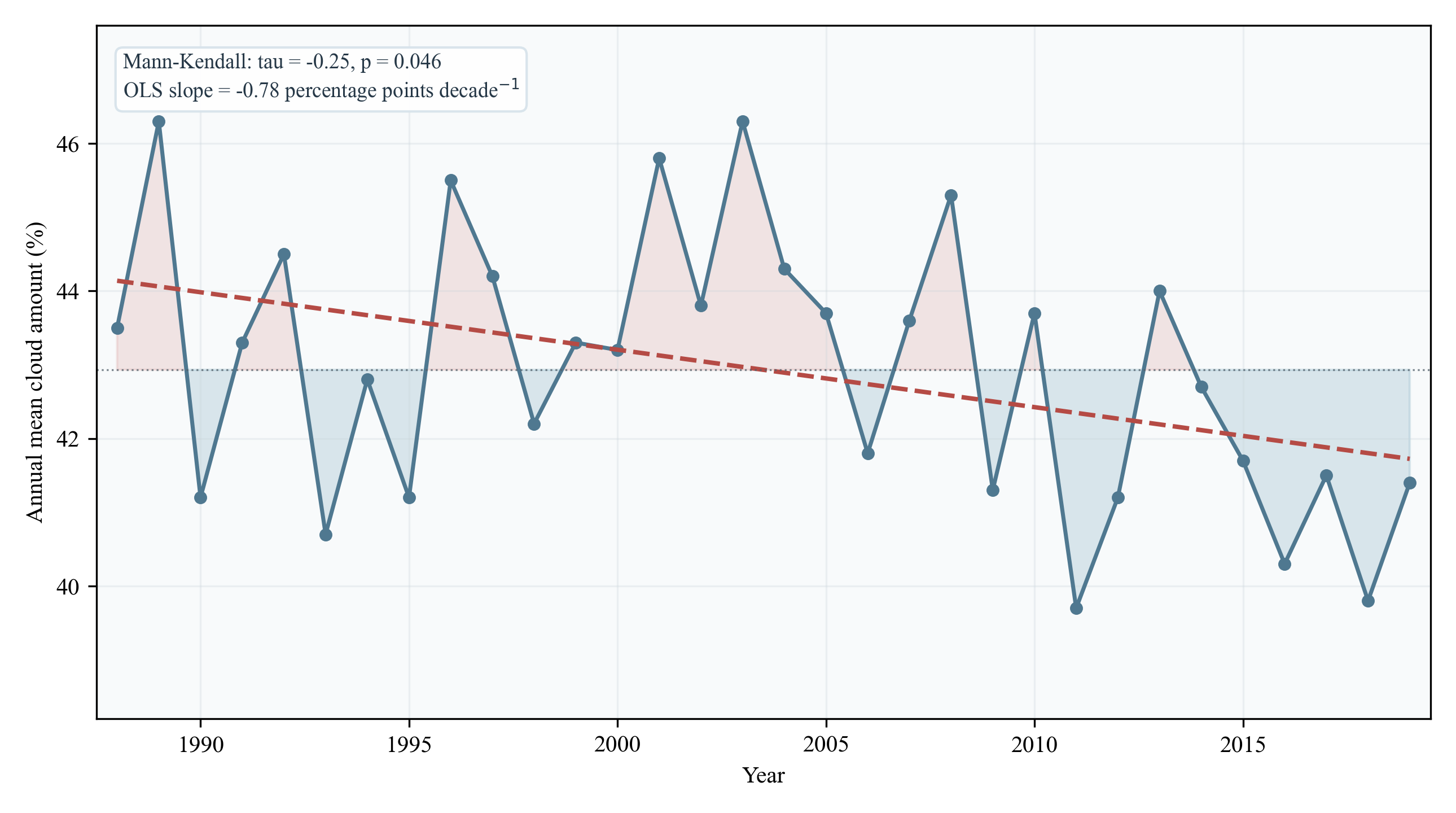}
\caption{Annual cloud-amount trend diagnostic for the Shigatse Meteorological Station. The figure uses the annual mean sequence extracted from the long-term trend analysis.}
\label{fig:appendix-cloud-amount-trend}
\end{figure}

The reproduced annual sequence shows a weak downward tendency in mean cloud amount (Figure~\ref{fig:appendix-cloud-amount-trend}). A Mann--Kendall test \citep{Mann1945Trend,Kendall1975RankCorrelation} applied to that sequence yields \(\tau=-0.25\) and \(p=0.046\), while an ordinary least-squares fit gives a slope of about \(-0.8\) percentage points per decade. This interannual diagnostic uses a longer 1988--2019 annual sequence with a treatment of value 11 and later-time samples that differs from the main 1988--2013 valid-entry analysis. Under this convention, the figure is used as a monitoring diagnostic for recurrence and long-term cloud-amount stability, while future homogenized cloud-value analyses can refine the formal trend estimate for the Shigatse region.

\begin{figure}[htbp]
\centering
\includegraphics[width=0.98\linewidth,keepaspectratio]{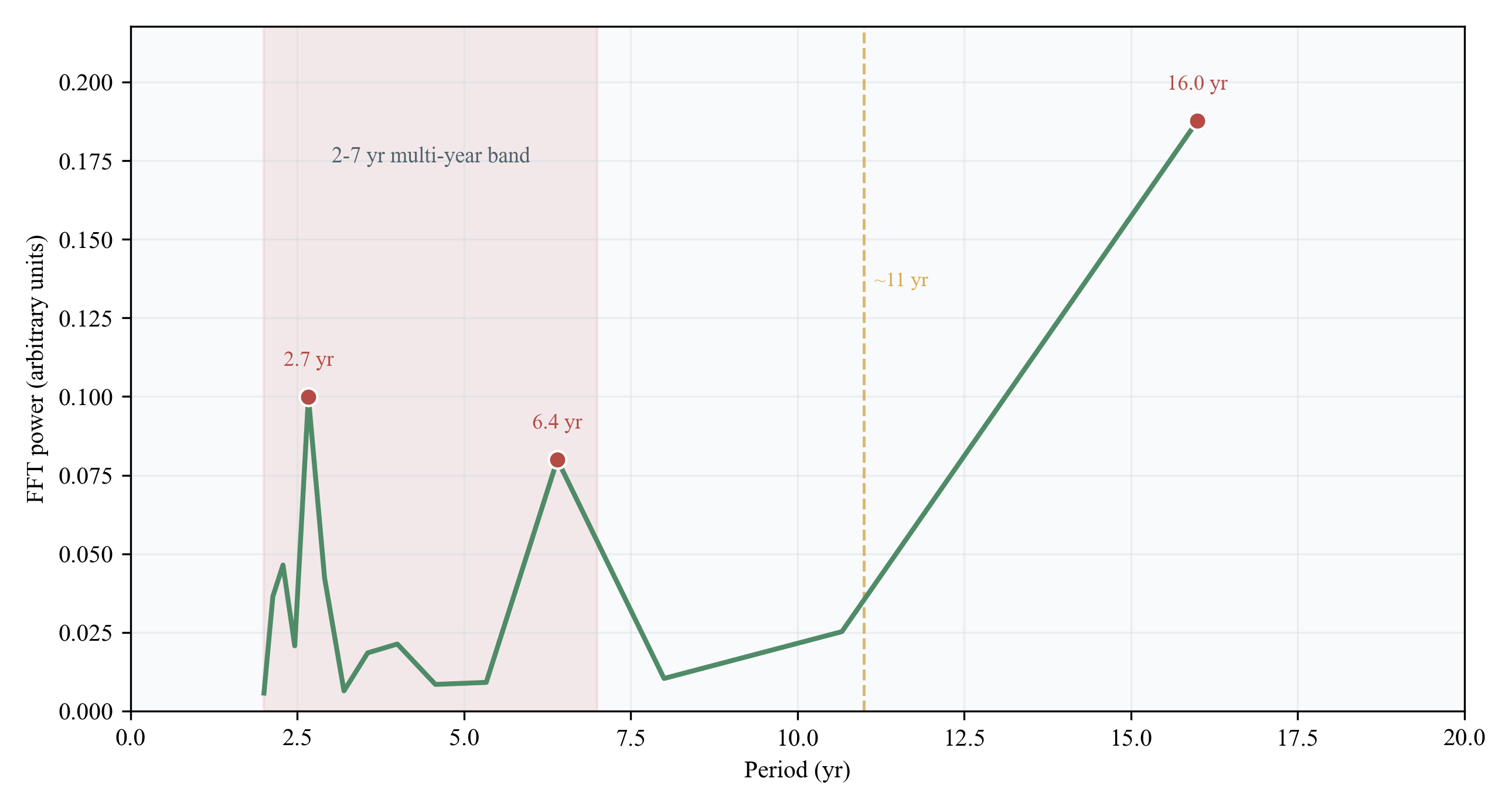}
\caption{FFT periodogram of the 1988--2019 Shigatse Meteorological Station annual cloud-amount sequence. The shaded band marks the 2--7 yr multi-year variability range, and the dashed vertical line marks an approximate 11 yr reference period. Peaks identify candidate variability scales for long-term monitoring.}
\label{fig:appendix-cloud-amount-fft}
\end{figure}

The corresponding fast Fourier transform diagnostic is retained as a variability check (Figure~\ref{fig:appendix-cloud-amount-fft}). The reproduced periodogram shows power within a 2--7 yr multi-year variability band, with local features near 2.7 yr and 6.4 yr, and a stronger low-frequency component near 16 yr. Such multi-year variability is physically plausible in the Tibetan Plateau cloud context because Chinese cloud studies have reported cloud-amount variability in similar bands, including 2--4 yr and 5--7 yr components \citep{Peng2024ChinaSummerCloudCycles}. For Shigatse, South Asian monsoon variability and Tibetan Plateau circulation provide a possible physical context for year-to-year cloud differences through changes in moisture transport and precipitation \citep{Wang2001SouthAsianMonsoon,Li2008TibetanPlateauMonsoon,Xu2009EasternTPPrecipitation}. Attribution of these variability scales to specific climate modes would require a dedicated climate-dynamics analysis with homogenized cloud records, climate indices, red-noise significance tests, lag tests, and time-localized methods. Here, the periodogram identifies variability scales that should be monitored in future Shigatse cloud records.

\clearpage

\begin{acknowledgements}
We thank China Manned Space Engineering and the Technology and Engineering Center for Space Utilization, Chinese Academy of Sciences, for providing data used in this study. This work was supported by the National Key Research and Development Program of China (grant Nos. 2025YFF0511000 and 2025YFF0510602), the National Natural Science Foundation of China (grant Nos. 12273075 and 12473070), the China Manned Space Project (grant Nos. CMS-CSST-2025-A19, CMS-CSST-2021-A12, and CMS-CSST-2021-B10), the International Partnership Program of the Chinese Academy of Sciences (grant No. 018GJHZ2023110GC), the Talent Plan of the Shanghai Branch, Chinese Academy of Sciences (No. CASSHB-QNPD-2023-016), the Tianshan Innovation Team Program of Xinjiang Uygur Autonomous Region (No. 2024D14015), and the Tian-shan Talent Training Program (No. 2023TSYCLJ0053). This work was also sponsored by the Natural Science Foundation of Xinjiang Uygur Autonomous Region (No. 2023D01A13). We thank Hongli Li and Qiang Zhang for sharing field-route reconnaissance information along the G318 and G219 national-highway corridors. We also acknowledge Zhejiang Lab, Zhejiang Province, China, for support of this work.
\end{acknowledgements}

\bibliographystyle{raa}
\bibliography{references}

\end{document}